\documentclass[prd,aps,showpacs,preprintnumbers,amsmath,amssymb,nofootinbib,preprintnumbers]{revtex4}
\usepackage{graphicx}
\usepackage{epsf}
\usepackage{amsmath}
\usepackage{amssymb,latexsym}
\usepackage{epstopdf}
\usepackage{bm}
\usepackage{color}
\usepackage{tabularx}
\usepackage{enumitem}
\usepackage{float}
\usepackage{array,booktabs}
\usepackage{footnote}
\usepackage{threeparttable}
\usepackage{graphicx}
\usepackage{hyperref}
\usepackage{amssymb,epsf}
\usepackage{latexsym}
\usepackage{epstopdf}
\usepackage{epsfig}
\usepackage{subfigure}
\usepackage{mathrsfs}
\usepackage{lmodern}

\begin{document}
     \title{A Comprehensive Review of Geometrical Thermodynamics: From Fluctuations to Black Holes}
     \author{
        S. Mahmoudi$^{1,2}$\footnote{email address: S.mahmoudi@shirazu.ac.ir},
        Kh. Jafarzade$^{1,2}$\footnote{email address: khadije.jafarzade@gmail.com} and
        S. H. Hendi$^{1,2,3}$\footnote{email address: hendi@shirazu.ac.ir}}
     \affiliation{
        $^1$Department of Physics, School of Science, Shiraz University, Shiraz 71454, Iran \\
        $^2$Biruni Observatory, School of Science, Shiraz University, Shiraz 71454, Iran \\
        $^3$Canadian Quantum Research Center 204-3002 32 Ave Vernon, BC V1T 2L7 Canada}

\begin{abstract}
This paper presents a comprehensive review of geometrical
thermodynamics, which employs geometric concepts to study the
thermodynamic properties of physical systems. The review covers
key topics such as thermodynamic fluctuation theory, proposed
thermodynamic metrics in various coordinate systems, and
thermodynamic curvature. Additionally, the paper discusses the
geometrical approach to black hole thermodynamics and provides an
overview of recent research in this field.
\end{abstract}
\maketitle
\tableofcontents
\newpage
\section{Introduction}
Differential geometry is a crucial mathematical discipline that
has found numerous applications in physics, particularly in the
description of the geometric properties of physical systems. One
of its most significant applications is in characterizing the
curvature and topology of spacetime to describe the four known
interactions of nature. This has yielded significant insights into
the fundamental principles of nature and allowed physicists to
formulate theories of gravity, such as Einstein's general theory
of relativity. Differential geometry also finds use in other areas
of physics, including quantum field theory, string theory, and
condensed matter physics. By providing a powerful framework for
understanding the behavior of massive objects and their
interactions with each other in the presence of gravity,
differential geometry represents an essential tool for physicists
to understand and describe the physical world through the lens of
geometry. In this regard, Einstein was the first pioneer of this
idea who came up with the relation between the field strength and
curvature and proposed the astonishing principle
\begin{equation}
    \textit{field strength}\approx \textit{curvature of a Riemannian manifold}\nonumber
\end{equation}
to comprehend the physics of the gravitational field \cite{frankel}.
The idea behind this principle can be represented schematically as
\begin{equation}
{\rm\it metric}\rightarrow {\hbox{\rm\it Levi-Civita connection}}\rightarrow {\hbox{\rm\it Riemann curvature $\approx$ \it gravitational field strength}}.
\nonumber
\end{equation}
In 1953, Yang and Mills introduced a principal fiber bundle
associated with a geometric structure to the electromagnetic
field. This bundle had the Minkowski spacetime as the base
manifold and $U(1)$, representing the internal symmetry of
electromagnetism, as the standard fiber \cite{ym53}. The
connection across the fibers was a local cross-section taking
values in the algebra of $U(1)$, and the Faraday tensor
represented the curvature of this fiber bundle. Their idea was
extended to non-abelian gauge theories by using different
connections as local cross-sections. They found that the weak and
strong interactions could be represented as the curvature of a
principal fiber bundle having a Minkowski base manifold and the
standard fiber $SU(2)$ and $SU(3)$, respectively \cite{frankel}.
This construction can be represented schematically as follows
\begin{equation}
\begin{array}[c]{ccccc}
    &         \Rsh       & \it U(1)-{\rm\it connection} & \rightarrow &\,\,\,\,\,\,\,\,\, \it U(1)-{\rm \it curvature} \approx {\hbox{\rm\it electromagnetic interaction strength}}  \\
    &    &  \\
    {\hbox{\rm\it Minkowski metric}}  & \rightarrow    &   \it SU(2)-{\rm\it connection} & \rightarrow & \it SU(2)-{\rm\it curvature} \approx {\hbox{\rm\it weak interaction strength}}\,\,\,\,\,\,\,\,\,\,\,\,\,\,\,\,\\
    &       & \\
    &            \rotatebox[origin=c]{180}{$\Lsh$}     & \it SU(3)-{\rm \it connection} & \rightarrow &\it SU(3)-{\rm\it curvature} \approx {\hbox{\rm\it strong interaction strength.}}\,\,\,\,\,\,\,\,\,\,\,\, \nonumber
\end{array}
\end{equation}

This line of thought continued until a few decades ago when some
physicists, inspired by the central role of curvature in
describing field interactions, attempted to extrapolate these
ideas to thermodynamic systems and make statements of the
following form
\begin{equation}
\textit{thermodynamic interaction}\approx \textit{curvature.}\nonumber
\end{equation}
Indeed, this idea led to the challenge of finding an intrinsic
geometric formulation for thermodynamics. The starting point is to
introduce an equilibrium space, an abstract space whose points can
be interpreted as representing the equilibrium states of the
system, which can be uplifted to a differential manifold endowed
with a metric structure. The existence of such a metric endows
spaces of thermodynamic equilibrium states with the notion of a
length between the states such that, shorter the distance between
a pair of thermodynamic states, the more probable is a fluctuation
between them. It is notable that significant information is
revealed through the invariants of the geometry, like the
curvature scalar. For instance, it has been turned out that the
magnitude of the scalar curvature invariant, known as
thermodynamic curvature,  is proportional to the correlation
volume of ordinary thermodynamic system and thus is a measure of
interaction strength, and yields fundamental information about
inter-particles interaction \cite{Ruppeiner1995, Rup91}. More
specifically, a negative (positive) sign of the thermodynamic
curvature is an indicator of the attractive (repulsive) nature of
the interaction between particles, whereas a zero value for the
thermodynamic curvature means there is no interaction between
particles \cite{Rup2,Rup3,Rup4}. Furthermore, it has been shown
that in addition to describing the critical point through its
singularity, the scalar curvature also encodes first-order phase
transitions in simple fluids, uniquely determines the Widom line
in different regimes, and identifies the experimentally determined
solid-like patches in the liquid phases \cite{Ruppeiner:2011gm,
Rup2015}.

Since the original works by Gibbs \cite{gibbs} and Caratheodory
\cite{car}, different geometric representations of thermodynamics,
depending on the potential chosen, have been explored until now,
the most important of which is as follows:

 \begin{itemize}
    \item{{\bf{Representation based on the Riemannian geometry:}} This representation was first applied in thermodynamics and statistical physics by Rao \cite{Rao}, in 1945, using the entropy as thermodynamic
potential. This approach which comes from information theory, leads to the Fisher-Rao metric in the thermodynamic limit. In fact,
 Rao introduced a metric whose components in local
        coordinates coincide with Fisher's information matrix \cite{Rao}. Later, his initial work was extended by a number of authors (for a review, see, e.g. \cite{amari85}). In the continuation of this path, Hessian metrics were postulated to describe the geometric properties of the equilibrium space, commonly known as thermodynamic geometry. In particular, metric structures as the Hessian of the internal energy and the entropy were proposed by Weinhold \cite{Wein1975, Wein75} and Ruppeiner \cite{Ruppeiner1995, Rupp79}, respectively. It is important to note that the Ruppeiner's and the Weinhold's metrics are conformally equivalent, with the inverse of the temperature as the conformal factor. Although
                both metrics have been widely used to study the geometry of the equilibrium space
        of ordinary systems \cite{nul85,san04,san05a,san05b,san05c,jjk03,jan04}, they suffer from the major drawback of not being invariant under Legendre transformations, which means that a given thermodynamical system has different geometrical properties, depending on
        the thermodynamical potential used.
        }

    \item{{\bf{Representation based on the contact geometry:}} Contact geometry was first introduced by Hermann \cite{Hermann} into the thermodynamic phase
        space and later explored by Mrugala \cite{Mrugala1, Mrugala2}. This approach aims to formulate the geometric version of the laws of thermodynamics in a consistent manner. Indeed, contact geometry allows us to consider a Legendre transformation as a coordinate transformation
        in the phase space \cite{Arnold}. This issue was used by Quevedo \cite{Quevedo} to propose the formalism
        of geometrothermodynamics whose purpose is to incorporate the property of the Legendre invariance in Riemannian structures at the level of the
phase space and the equilibrium space. This property guarantees that the thermodynamic characteristics of a system do
not depend on the thermodynamic potential used for its description. This geometric setting is widely used to study thermodynamics \cite{RT1, RT3, RT2}, mechanical systems with Rayleigh dissipation \cite{CM1, CM2}, statistical mechanics \cite{SM1}  as well as black hole thermodynamics \cite{Ghosh:2019100}.}
 \end{itemize}
 
\begin{figure}[h!]
    \centering\includegraphics[scale=1.6]{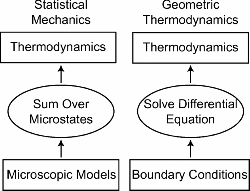}
    \caption{Contrasting philosophies of statistical mechanics and the
        geometric theory (adapted from \cite{Rupp05}).}
    \label{Figg1}
\end{figure}

 The geometric theory of thermodynamics offers a robust framework
for calculating equations of state without the explicit reliance
on underlying microscopic models. Consequently, it can serve as a
valuable tool to investigate systems whose microscopic nature is
unknown, precluding the use of statistical mechanics to study
their properties. Figure \ref{Figg1} provides a visual comparison
of the approaches employed by the geometric theory and statistical
mechanics in this regard.

Since the advent of the geometric representation of
thermodynamics, it has proven to be an invaluable tool for
uncovering the quantum nature of black holes and gravity. Although
a complete understanding of their microscopic degrees of freedom
remains elusive, black hole thermodynamics was first explored in
the early 1970s by Hawking and Bekenstein. Their pioneering work
revealed that black holes are not merely gravity systems but also
possess thermal properties. Subsequently, the laws of black hole
mechanics were formulated, which correspond to conventional
thermodynamics if suitable quantities such as temperature,
entropy, energy, etc., are appropriately identified. Based on
these discoveries, it was found that black holes radiate as black
bodies, possessing characteristic temperatures and entropies
\cite{Bardeen,Bekenstein:1973, Hawking:1975vcx,FN,Wald-book}
\begin{align}
kT_{\scriptscriptstyle\mathrm{H}} = \frac{\hbar\kappa}{2\pi} ,\qquad
S_{\scriptscriptstyle\mathrm{BH}}
= \frac{\ A_{\mathrm{\scriptstyle H}}}{4\hbar G}  ,
\label{Carlipa1}
\end{align}
where $\kappa$ is the surface gravity and $A_{\mathrm{\scriptstyle
H}}$ stands for the area of the horizon, $\hbar$ denotes Planck's
constant and $G$ is the Newton's constant. However, although it is
believed that a black hole does possess thermodynamic quantities
and extremely interesting phase structure and a vast amount of
studies have been done concerning this issue \cite{Davies:1978zz,
Hut:1977zx, Sokolowski:1980uva, Cai:1997cs, Shen:2005nu, David
Kubiznak12,Rong-Gen Cai13,Sharmila Gunasekaran12,Antonia M.
Frassino14,Natacha Altamirano14,Natacha Altamirano13,Joy Das
Bairagya20,X. H. Ge15,Rong-Gen Cai16,Jia-Lin Zhang15b,Ren
Zhao13,Ren Zhao13b,Shao-Wen Wei09,S H Hendi17,A. Dehghani20,Rabin
Banerjee20,M. Chabab2019,M. Chabab2016,Meng-Sen
Ma17b,Dehyadegari2020,Daniela
Mago2020,Sajadi2019,Hendi2019,Hendi2018,Eslam,Hendi2016a,Hendi2016b,Zou17,Jianfei15,
Mahmoudi:2022hqq, Hendi:2022opt, Hendi:2021yii, Dehghani:2020kbn,
Hendi:2020mhg, Hendi:2020knv, Yuichi-2006,
Miho-2009,Saoussen-2019,David-2016,
Simovic-2019,Sumarna-2020,Simovic-202020,Chabab-202020,Brian-2013,Bhattacharya-2016,James-2016,
Kanti-2017,Romans-1992,zhang-2016,zhang-2019, ma-2020,
guoxiong-2020}, the statistical description of the black hole
microstates has not yet been fully understood. Even though a
complete quantum gravity theory is still absent, there have been
some attempts to understand the microscopic structure of a black
hole
\cite{IN-Strominger:1996sh,IN-Callan:1996dv,IN-Emparan:2006it,
wei-2019,wei-2020,zou-2020,wei-202011,miao-2018,
miao-2019,guo-2020,guo-2019,mann-2021,Volovik-2021}. In this
regard, the thermodynamic geometry method has led to many insights
into microstructure of a black hole. In this paper, we aim to
provide a review of the studies have been done on the black hole
thermodynamics on the view point of geometry.\par

The aim of this paper is to provide a comprehensive review of
geometrical thermodynamics. We begin in section \ref{2} by
reviewing the fundamentals of thermodynamic fluctuation theory,
which serves as the basis for identifying appropriate
thermodynamic geometric representations. In section \ref{3}, we
explore some essential and fundamental geometries proposed in the
context of thermodynamics using the language of differential
geometry. This section includes an introduction to Hessian
thermodynamic metrics (section \ref{3A}) and Legendre invariant
metrics (section \ref{3B}). Moreover, we review the application of
these metrics to some ordinary thermodynamic systems in section
\ref{3C}. We then discuss and review the concept of thermodynamic
curvature, which arises from thermodynamic fluctuation theory, in
section \ref{4}. After introducing the geometric perspective of
thermodynamics, we shift our focus to studying black holes using
geometric thermodynamic approaches. To accomplish this, we first
review the laws of black hole mechanics in section \ref{5-1},
followed by the laws of black hole thermodynamics in section
\ref{5-2}. Subsequently, we discuss and review the applications of
geometric thermodynamics in investigating the nature of black
holes in section \ref{5-3}. Finally, we end the paper with a summary in section \ref{5-4a}.

\section{Thermodynamic fluctuation theory}\label{2}

The macroscopic physical observables of a system are rooted in
microscopic quantities at equilibrium, which typically remain
close to their average values. The random deviations from these
values that describe the natural behavior of a system are referred
to as thermodynamic fluctuations, which are studied within the
context of thermodynamic fluctuation theory.

\begin{figure}[h!]
    \centering\includegraphics[scale=0.3]{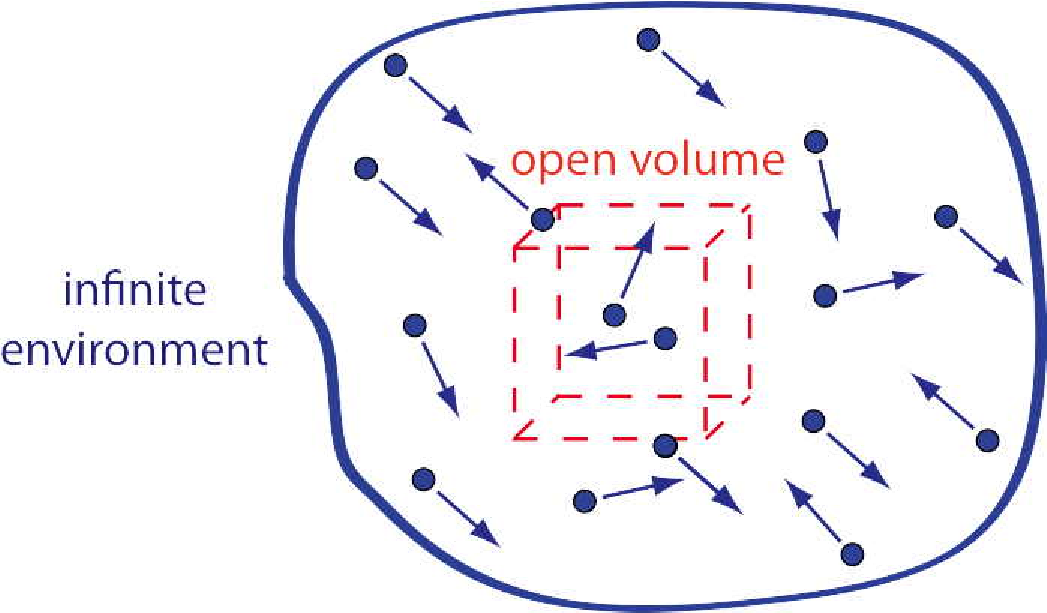}
    \caption{An infinite environment of particles and an open volume, with fixed volume V ,
        into which particles fluctuate in and out (adapted from \cite{Ruppeiner:2013yca}).}
    \label{fig2}
\end{figure}

In most statistical mechanical contexts, the Gaussian
approximation of thermodynamic fluctuations theory yields the
probability of finding a system in a particular thermodynamic
state \cite{Mishin}. To illustrate, consider an infinite system of
particles and an imaginary open volume with a fixed volume $V$,
into which the particles can move freely in and out, as depicted
in Figure \ref{fig2}. This theory provides a suitable framework
for determining the probability of finding a specific energy $U$
and a particular number of particles $N$ within the open volume.

However, it has been demonstrated that moving beyond the Gaussian
approximation by taking into account covariance, conservation, and
consistency results in a fundamentally novel entity: the
thermodynamic Riemannian curvature $\mathcal{R}$. This quantity is a
thermodynamic invariant that reveals information about
interparticle interactions \cite{Ruppeiner1995}.
In statistical mechanics,  Boltzmann's expression for the entropy of a system in the microcanonical ensemble is defined as follows \cite{Landau}
\begin{equation}\label{entropy1}
    S=k_{B}\ln\Omega,
\end{equation}
where $k_{B}$ is the Boltzmann constant and $\Omega$ indicates the
number of microstates of the system and is the function of
extensive quantities like energy, volume and particle numbers,
i.e. $\Omega=\Omega(U,V,N)$. In 1907, Einstein proposed a theory
which based on the relation \eqref{entropy1} is inverted  to
express the number of states in terms of entropy \cite{Landau}:
\begin{equation}\label{omega1}
    \Omega=\, \exp(\frac{S}{k_{B}}).
\end{equation}
This relation was used as the basis of thermodynamic fluctuation theory.\par

In what follows we begin by studying the thermodynamic
fluctuations with a single variable, and explain the problems that
lead to efforts to go beyond the Gaussian approximation. We will
show that the corrected theory is obtained by introducing a second
independent fluctuating variable which results in the appearance
of the thermodynamic curvature concept.

\subsection{One fluctuating variable}

Consider a thermodynamic system that exhibits equilibrium
properties described by a set of $n$ additive and invariant
parameters denoted as $X_1, X_2, ..., X_n$. In statistical
mechanics, the understanding of a system's thermodynamic
properties involves studying its fundamental equation as a
function of the conserved parameters. For instance, in the entropy
representation, the fundamental equation can be expressed as
\begin{equation}
    S=S(X_{1}, X_{2}, ..., X_{n}).
\end{equation}

For a simple fluid, the fundamental equation takes the form
$S=S(U,V)$ where $U$ represents the energy of the fluid, and $V$
denotes the volume of space that the fluid occupies. Since $U$,
$V$, and $S$ are additive quantities, the fundamental equation
can be expressed as
\begin{equation}
    S=V\,s(u),
\end{equation}
where $s$ and $u$ are the entropy and energy per volume, respectively.
\begin{figure}[h!]
    \centering\includegraphics[scale=0.6]{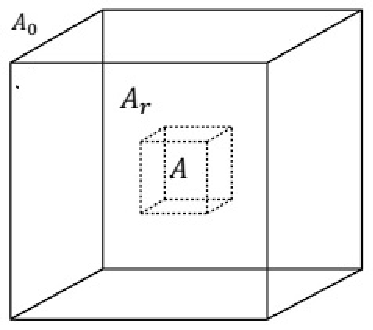}
    \caption{A closed system $A_0$ consists of an enclosed, open system $A$ and reservoir $A_r$.}
    \label{fig1}
\end{figure}

Now, we consider a very large closed thermodynamic system $A_0$ in
thermodynamic equilibrium, with fixed volume $V_0$ and fixed
energy per volume $u_0$. Suppose that the mentioned thermodynamic
system is divided into two subsystems by a fixed and open
partition: a finite system $A$, with fluctuating internal energy
per volume $u$ and constant volume $V$, and the reservoir $A_r$,
with fluctuating internal energy per volume $u_r$ and constant
volume $V_r$ as depicted in Fig. \ref{fig1}. It is important to
mention that due to being open the boundary between $A$ and $A_r$,
these two subsystems exchange energy through fluctuations, which
can be described by thermodynamic fluctuation theory. The basic
postulate of the microcanonical ensemble is that the occurrence
probability of all accessible microstates of $A_0$ is equal
\cite{Landau}. Hence, the probability of finding the internal
energy per volume of $A$ between $u$ and $u+du$ will be
proportioned to the microstates' number of $A_0$ existing to this
range:
\begin{equation}
P(u,V)du=C \Omega_{0}(u,V)du, \label{-25}
\end{equation}
where $\Omega_0(u,V)$ indicates the density of states, and $C$
stands for a normalization factor. Making use of
Eq.\eqref{omega1}, the Einstein's famous relation reduces to
\begin{equation}
P(u,V)du=C\exp\left[\frac{S_{0}(u,V)}{k_{B}}\right]du, \label{omega2}
\end{equation}
where $S_0(u,V)$ is the entropy of $A_0$. Statistical mechanics
asserts that for a thermodynamic system in the equilibrium state,
energy distribution between subsystems occurs in such a way that
the system's entropy is maximized. In this context, the maximum
value of $S_0(u,V)$ corresponds to the state in which the two
parts $A$ and $A_r$ admit the energy $u=u^*$ and $u_r=u_r^*$,
respectively. Using the fact that the entropy is additive
\begin{equation}
S_{0}(u,V) = V s(u) + V_{r} \,s(u_{r}), \label{-40}
\end{equation}
and expanding each of entropy densities $s(u)$ and $s(u_r)$ around their maximum energy values, we get
\begin{eqnarray}\label{S zero}
S_{0}(u,V) &=& \tilde{S}_{0} + V s'(u^*) \Delta u + V_{r} s'(u_{r}^*)\Delta u_{r} + V\frac{1}{2!} s''(u^*)(\Delta u)^2 \nonumber \\ &&{} + V_{r}\frac{1}{2!} s''(u_{r}^*)(\Delta u_{r})^2 + V\frac{1}{3!} s'''(u^*)(\Delta u)^3 + \cdots, \label{-50}
\end{eqnarray}
where $\Delta u=u-u^*$ and $\Delta u_{r}=u_{r}-u_r^{*}$. Moreover, the
prime denotes differentiation with respect to $u$, and $
\tilde{S}_0$ indicates the entropy of $A_0$ in equilibrium. Since
the first order expansion of the entropy must be zero at the
maximum points
 \begin{equation}
 V s^{\prime}(u^{*}) \Delta u + V_{r} s^{\prime}(u_r^{*})\Delta u_{r}=0, \label{first expand}
 \end{equation}
applying the conservation law of energy, $V\Delta u=-V_{r}\Delta
u_{r}$, results in the relation $s'(u^*) = s'(u_r^*)$ which forces
$u^*= u_{r}^{*}$. In addition, additivity of energy requires that $V
u^{*}+V_{r} u_{r}^{*}=V_{0} u_{0}$, leading to $u^{*}= u_{r}^{*}=u_{0}$. Therefore,
 \begin{equation}
    \Delta u=u-u_{0},\,\,\,\,\,\,\,\,\,\,\,\,\,\,\,\,\,\,\,\,\Delta u_{r}=u_{r}-u_{0}.
 \end{equation}
According to the energy conservation law, higher-order terms in
the expansion of $s(u_{r})$ can often be neglected when compared to
the corresponding terms in the expansion of $s(u)$. By truncating
the series of equation Eq. \eqref{S zero} at second order and
considering the resulting Gaussian distribution, we can
approximate fluctuations in thermodynamic fluctuation theory. This
approximation is commonly used in statistical mechanics and has
been described by authors such as \cite{Landau,Pathria}
\begin{equation} P_{G}(u,V)du = C \exp\left[-\frac{V}{2}g(u_{0}) (\Delta u)^{2}\right]\,du, \label{Gauss}
\end{equation}
where $ g(u_{0})\equiv \frac{-s^{\prime\prime}(u_{0})}{k_{B}}$ and due to being
proportional to the heat capacity, it is always positive. In order
to obtain the  normalization factor, one can use the normalization
condition $\int P_{G}(u,V)du =1 $ which results in
\begin{equation}
    C= \sqrt{\frac{V\,g(u_{0})}{2\pi}}.
\end{equation}

\subsubsection{Some shortcomings of thermodynamic fluctuation theory }

While Eq. \eqref{Gauss} is effective, there are limitations when
we go beyond the Gaussian approximation. One problem occurs when
we change the thermodynamic parameter $u$ to a general parameter
$x(u)$. Since the counting of microstates resulting in Eq.
\eqref{omega2} can be performed equally well with the parameter
$x$, we should get the same expression except that $u$ is replaced
by $x$:
\begin{equation}
P(x,V)dx={C}\exp\left[\frac{S_{0}(x,V)}{k_{B}}\right]dx. \label{transform1}
\end{equation}

However, we obtain a different expression if we make a
straightforward coordinate transformation on the left hand side of
Eq. \eqref{omega2}
\begin{equation} P(u,V)du = \left[P(u,V)\left(\frac{du}{dx}\right)\right]dx \equiv P(x,V)dx, \label{transform2}
\end{equation}
which gives the probability of finding the new parameter in the
range from $x$ to $x+dx$. Since entropy is a function of state,
then $S_{0}(u,V)=S_{0}(x,V)$. Therefore, we end up with
\begin{equation} P(x,V)dx = C \left(\frac{du}{dx}\right) \exp\left[\frac{S_{0}(x,V)}{k_{B}}\right]dx, \label{transform3}
\end{equation}
which is inconsistent with Eq. (\ref{transform1}) because the
factor  $du=(du/dx)dx$ is not generally a constant and cannot
simply be absorbed into the normalization factor. Hence, it can be
concluded that the formulation of the thermodynamic fluctuation
theory depends on the coordinates used and its equation is not
covariant. It is important to mention that this problem does not
arise in the Gaussian approximation. One can easily find that the
Gaussian approximation will be covariant if the following
transformation rule
  \begin{equation}
     g(x_{0})=g(u_{0})\left(\frac{du}{dx}\right)^2, \label{transform rule}
  \end{equation}
is used for the function $g(x_{0})$. Indeed, since the fluctuations
are small in the regime of validity of the Gaussian approximation,
the derivative of $u$ with respect to $x$ is constant over this
range. Therefore, small fluctuations guarantee the covariance
validity of the theory in the  Gaussian approximation. \par

Another class of problems is concerned with the fact that this
theory does not provide correct values for the average value of
the standard densities in the entropy expansion beyond the
Gaussian approximation. To explain more, we define the average
value of a thermodynamic function as
\begin{equation}
     \left<f\right> = \int f(u)\,P(u,V) du,\label{average1}
     \end{equation}
and hence,
\begin{equation}
     \left<\Delta u\right> = \int \Delta u\,P(u,V) du .\label{average2}
\end{equation}

From physical point of view, we expect that $\left< \Delta
u\right>=0$, which results in a conservation rule, for any
physically correct probability density $P(u,V)$ and for all $V$.
Although the mentioned conservation rule, i.e. $\left< \Delta
u\right>=0$, obtains in the Gaussian approximation because the
probability density is an even function of $u-u_{0}$, thermodynamic
fluctuation theory violates this rule beyond that approximation.
In fact, if the fluctuations are not small enough, we should
consider the third-order term in the entropy expansion which leads
to $\left<\Delta u\right>\not= 0$.
\subsubsection{Toward a consistent and covariant theory}
Now, we try to find an appropriate solution to the problems that
have been mentioned. To this end, we take a different approach
from Eq. \eqref{omega2} because this equation is fundamentally
incorrect beyond the Gaussian approximation.

Since the form of the Gaussian approximation Eq. (\ref{Gauss}) is
similar to the solution of a diffusion equation, with the role of
``time'' played by the inverse volume i.e. $t=1/V$, and with a
Dirac delta function peaked at $\Delta u=0$, one can be guided to
find a diffusion-like equation. The most general form of the
diffusion equation which maintains normalization can be expressed
as \cite{Ruppeiner1995,Graham}
\begin{equation}
\frac{\partial P}{\partial t}=-\frac{\partial}{\partial x}[K(x)P] + \frac{1}{2}\frac{\partial^2}{\partial x^2}\left[g^{-1}(x)P\right], \label{partial1}
\end{equation}
where $K(x)$ and $g(x)$ are drift term which should be determined.
It is important to mention that $t$ is simply a measure of volume
and nothing is diffusing in time. Indeed, the {\it{diffusion}}
reflects the fact that the thermodynamic state of $A$ gets
increasingly undetermined by decreasing $V$. By specifying $K(x)$
and $g(x)$, along with the initial condition and appropriate
boundary conditions, one can find the covariant theory
\cite{Ruppeiner1995}.\par

Now we can examine some general consequences of this equation:

\vspace{0.3cm} $\bullet$ For constant functions $K(x)=K$ and
$g(x)=g$ which is justified for small $t$, this has the exact
normalized solution of the form of Gaussian expression
\begin{equation} P_{G}(x,t)=\sqrt{\frac{g}{2\pi\, t}} \exp\left[-\frac{1}{2t}\,g(\Delta x-K\,t)^2\right], \label{Gauss4}
\end{equation}
where $\Delta x\equiv x-x_{0}$, and $x_{0}$ is a constant. This solution has the Dirac delta function $\delta(x-x_{0})$ start at $t=0$.

\vspace{0.3cm} $\bullet$ If $x_{a}$ and $x_{b}$ are the limits of
thermodynamic space and hence the smallest and largest values of
$x$, we will arrive at the following relations by successive
multiplications by $x$ and integration by parts
 \begin{equation}\label{itegrate1}
    \frac{d}{dt} \int_{x_{a}}^{x_{b}} P\, dx= -(K\,P)\bigg|_{x_{a}}^{x_{b}}+\frac{1}{2}\frac{\partial}{\partial x}(g^{-1}P)\bigg|_{x_{a}}^{x_{b}},
 \end{equation}
\begin{equation}\label{integrat2}
    \frac{d}{dt} \left< x\right>\,=\,\left<K\right>- (xKP)\bigg|_{x_{a}}^{x_{b}}+\frac{1}{2}x\frac{\partial}{\partial x}(g^{-1}P)\bigg|_{x_{a}}^{x_{b}}-\frac{1}{2}(g^{-1}P)\bigg|_{x_{a}}^{x_{b}},
\end{equation}
and
\begin{equation}\label{integrate3}
\frac{d}{dt} \left< x^2\right>=\left<g^{-1}\right>+2\left<x\,K\right> + boundry\,\,\, terms.
\end{equation}
 Taking $t$ to be small, and switching back to the $u$ coordinate we find that Eqs. \eqref{integrat2} and \eqref{integrate3} reduce to
\begin{equation} \label{integrate5}
\left<\Delta x\right> \,=\, \int \Delta x\,P_{G}(x,t) dx = K\,t,
\end{equation}
and
\begin{equation}\label{integrate6}
 \left<\left(\Delta x\right)^2\right> = \int \left(\Delta x\right)^2 P_{G}(x,t) dx=\frac{t(1+g\,K^2\,t)}{g}.
 \end{equation}
Comparing Eqs. (\ref{Gauss4}) and (\ref{integrate5}) with the
corresponding ones from the thermodynamic Gaussian expression Eqs.
(\ref{Gauss}), (\ref{average2}) shows that $K=0$ and $t=\frac{1}{V}$.
Also, $g$ is fixed by $ g(u_{0})\equiv \frac{-s^{\prime\prime}(u_{0})}{k_{B}}$,
specifying the width of the fluctuations. It may seem that the
function $K(x)$ in Eq. (\ref{partial1}) is irrelevant due to the
fact that $K(u)=0$ in $u$ coordinates. However, the  importance of
its non-zero value becomes clear in the topic of  coordinate
transformation.\par

\vspace{0.3cm}
 $\bullet$
To study about how to transform Eq. (\ref{partial1}) to another
coordinate $\tilde{x}=\tilde{x}(x)$, we first investigate the
possibility of transforming $P(x,t)$, $K(x)$, and $g(x)$ such that
the form of the partial differential equation is unchanged
  \begin{equation}
  \frac{\partial \tilde{P}}{\partial\,t}\,=\,-\frac{\partial}{\partial \tilde{x}}\left[\tilde{K}(\tilde{x})\tilde{P}\right] + \frac{1}{2}\frac{\partial^2}{\partial \tilde{x}^2}\left[\tilde{g}^{-1}(\tilde{x})\tilde{P}\right]. \label{partial eq}
  \end{equation}
  
Since the probability is a scalar quantity independent of the
choice of coordinate, it requires $\tilde{P} d\tilde{x}=P dx$, and
thus
 \begin{equation}
 \tilde{P}\,=\,P\,\left(\frac{dx}{d\tilde{x}}\right).
 \end{equation}
Substituting this expression into Eq. (\ref{partial eq}) and after
some straightforward calculation we find that the coefficients of
the corresponding derivatives of $P$ is equal to Eq.
(\ref{partial1})
 if and only if the functions $K$ and $g$ transform as
 \begin{equation}
 \tilde{g} =g\left(\frac{dx}{d\tilde{x}}\right)^2, \label{condition1}
 \end{equation}
 \noindent and
 \begin{equation}
 \tilde{K}\,=\,K\,\frac{d\tilde{x}}{dx}\,+\,\frac{1}{2}\,g^{-1}\frac{d^2 \tilde{x}}{dx^2}. \label{condition2}
 \end{equation}
Hence, by using the appropriate transformation rules, the partial
differential equation Eq. (\ref{partial1}) will be covariant and
in this way, we have solved the problem of fluctuation theory with
one variable. \par Before ending this subsection, it is worth
mentioning that Eq. \eqref{condition1} which is matched Eq.
\eqref{transform rule} is exactly the metric transformation in
Riemannian geometry. Therefore, this geometry arises naturally
from our demand for a covariance fluctuation theory under
coordinate transformations. Based on this issue, a thermodynamic
Riemannian geometry can be built and system's fluctuations can be
investigated according to their geometrical properties.

\subsection{Two fluctuating variables}
Now, we discuss fluctuations with two independent variables. This
situation can occur if the subsystem and its environment exchange
particles as well as energy. We will see that the two independent
fluctuating variables allow a Riemannian geometry with non-zero
thermodynamic curvature.\par

Consider a pure fluid system of $N$ identical particles in a volume $V$  whose the fundamental equation is \cite{Call}
\begin{equation} S=S(U,N,V).\label{fundamential eq2}
\end{equation}
Since $S$, $U$, $N$, and $V$ are all additive thermodynamics parameters, we can write
\begin{equation}
S=V s(u,\rho),
\end{equation}
where $\{s,u,\rho\}\equiv\{\frac{S}{V}, \frac{U}{V}, \frac{N}{V}\}$ are quantities per
volume. Similar to the one variable fluctuating case, we divide a
closed, infinite system $A_{0}$ into two parts: a finite $A$ and an
infinite reservoir $A_{r}$, by an imaginary, immovable partition as
shown in Fig. 1. Both the internal energy $U$ and the number of
particles $N$ are allowed to fluctuate \cite{Landau}. Here, we
define the extensive parameters as
$\{\mathcal{E}^1,\mathcal{E}^2\} \equiv\{u,\rho\}$, and the
corresponding intensive parameters get as
\begin{equation}\label{intensive}
 \mathcal{I}_\alpha \equiv \frac{\partial\,s}{\partial \mathcal{E}^{\alpha}},
\end{equation}
where $\alpha=1,2$. Basic thermodynamics gives
$\{\mathcal{I}_{1},\mathcal{I}_{2}\}=\{\frac{1}{T},-\frac{\mu}{T}\}$, where $T$ is the
temperature, and $\mu$ is the chemical potential. The properties
of $A_{0}$ and $A_{r}$ are denoted by subscripts $0$ and $r$,
respectively. \par

The probability of finding the state of $A$ in the range $(u,\rho)$ to $(u+du,\rho+d\rho)$ is
\begin{equation} Pdu d\rho=C \exp\left(\frac{S_{0}}{k_{B}}\right)du d\rho,\label{probablity}
\end{equation}
where $C$ is a normalization factor, and $S_0$ can be written as
\begin{equation}
S_{0} = V s(u,\rho) + V_{r} s(u_{r},\rho_{r}).\label{entropy2}
\end{equation}
There will be a maximum value for $S_{0}$ in the state where
$\mathcal{E}^{\alpha}=
\mathcal{E}^{\alpha}_{r}=\mathcal{E}^{\alpha}_{0}$. By applying the
same argument as used in Sec. II.A, and expanding the entropies in
Eq. (\ref{entropy2}) around this maximum in powers of the
differences $\Delta
\mathcal{E}^{\alpha}=\mathcal{E}^{\alpha}-\mathcal{E}^{\alpha}_{0}$
and $\Delta
\mathcal{E}^{\alpha}_{r}=\mathcal{E}^{\alpha}_{r}-\mathcal{E}^{\alpha}_{0}$,
we get
\begin{eqnarray}
\Delta S_{0} &=& V \mathcal{I}_{\mu} \Delta \mathcal{E}^{\mu} + V_{r} \mathcal{I}_{r\mu} \Delta \mathcal{E}_{r}^{\mu} + V\frac{1}{2!}\,\frac{\partial \mathcal{I}_{\mu}}{\partial \mathcal{E}^\nu}\Delta \mathcal{E}^\mu\Delta \mathcal{E}^{\nu} \nonumber \\ &&{}+ V_{r}\,\frac{1}{2!}\,\frac{\partial \mathcal{I}_{r\mu}}{\partial \mathcal{E}_{r}^{\nu}}\Delta  \mathcal{E}_{r}^{\mu}\Delta \mathcal{E}_{r}^{\nu} + \cdots, \label{delta entropy}
\end{eqnarray}
where $\Delta S_{0}$ is the difference between $S_{0}$ and its maximum
value, and summation over repeated indices is assumed. A necessary
condition for maximum entropy, i.e. $\mathcal{I}_{\alpha} =
\mathcal{I}_{r\alpha}$, is given by the conservation of both
energy and particles which requires that $V \Delta
\mathcal{E}^{\alpha}=-V_{r}\Delta \mathcal{E}_{r}^{\alpha}$. Also, by
neglecting the second quadratic term of Eq. (\ref{delta entropy})
compared with the first one, this equation reduces to
\begin{equation}
\frac{\Delta S_{0}}{k_{B}} =- \frac{V}{2}g_{\mu\nu}\Delta \mathcal{E}^{\mu} \Delta \mathcal{E}^{\nu},\label{140}
\end{equation}
where the symmetric intensive matrix
\begin{equation}
g_{\mu\nu} \equiv -\frac{1}{k_{B}}\frac{\partial^2\,s}{\partial \mathcal{E}^{\mu}\partial \mathcal{E}^{\nu}}, \label{metric}
\end{equation}
must be positive-definite because the entropy has a maximum value
in equilibrium. By substituting the above equation into Eq.
(\ref{probablity}) and calculating the normalization factor, we
arrive at the Gaussian approximation  as follows\cite{Landau}
\begin{equation}
 P_{G} \, d\mathcal{E}^1d\mathcal{E}^{2} = \Big(\frac{V}{2\pi} \Big) \exp \Big(-\frac{V}{2}g_{\mu\nu}\Delta \mathcal{E}^{\mu} \Delta \mathcal{E}^{\nu}\Big)\sqrt{g}\,d\mathcal{E}^1d\mathcal{E}^{2}, \label{probablity Gauss}
\end{equation}
where $g$ is the determinant of the matrix $g_{\alpha\beta}$. Regarding the transformation law for a first rank contravariant tensor
\begin{equation}
    \Delta \mathcal{E}^{\mu}=\frac{\partial \mathcal{E}^{\mu}}{\partial x^{\alpha}}\,\Delta x^{\alpha},\label{transform law}
\end{equation}
it can easily be shown that the exponential part of
Eq.\eqref{probablity Gauss} behaves like a scalar quantity under
the coordinate transformation. This invariant treat should be
maintained because the probability of fluctuation between two
states and the entropy difference should not depend on the
coordinates used to describe the system. Since the part of
$g_{\mu\nu}\Delta \mathcal{E}^{\mu} \Delta \mathcal{E}^{\nu}$ has the
look of a distance between thermodynamic states in the form of a
Riemannian metric, it leads us to define a new and useful
component for our discussion named {\it{thermodynamic length}}
\begin{equation}
 \Delta \ell^2\equiv g_{\mu\nu}\Delta \mathcal{E}^{\mu} \Delta \mathcal{E}^\nu, \label{thermo length}
\end{equation}
which is a positive-definite quantity. Indeed, the above relation
represents a well-defined Riemannian metric in the space of
thermodynamic state, as described in the Appendix \ref{one}. From a physical
standpoint, the interpretation of the distance between two
thermodynamic states in this space is clear: {\it{the distance
represents the measure of probability in thermodynamics}}. The
greater the probability of a fluctuation occurring between two
states, the closer together they are considered to be. It's worth
noting that the dimension of the squared thermodynamic length is
inversely proportional to volume, which differs from Riemannian
geometry in general relativity (GR) where the dimension of the squared
length is typically measured in meters squared.\par

\section{Thermodynamic metrics in different coordinate systems}\label{3}

After familiarizing ourselves with the concept of geometrical
thermodynamics, we will now examine some important and fundamental
geometries proposed in the context of thermodynamics using the
language of differential geometry. Differential geometry has been
extensively applied in physics, chemistry, and engineering, and
has found broad applications in thermodynamics, offering an
alternative way to describe thermodynamic systems. To begin, we
introduce the equilibrium space, which is an abstract space whose
points can be interpreted as representing equilibrium states of
the system. Various attempts have been made to provide a suitable
thermodynamic metric, some of which we will review below.

\subsection{Hessian thermodynamic metrics}\label{3A}

In the previous section, we discussed the origin of the concept of
geometrical thermodynamics in the context of thermal fluctuations.
Drawing on the fluctuation theory of equilibrium states of
physical systems, we define the appropriate thermodynamic metric
as the Hessian metric, which is based on the Hessian of the
thermodynamic potentials. Here, we will examine the issue that
different metrics have been proposed depending on the parameters
of the thermodynamic phase space and introduce some of the
provided metrics:
\subsubsection{\textbf{Ruppeiner metric}}

The Ruppeiner metric naturally appears in thermodynamic
fluctuation theory and is based on the line element given in Eq.
\eqref{thermo length} \cite{Rupp79, Ruppeiner1995}
\begin{equation}
    \Delta \ell^{2} = g_{\mu\nu}^{R}\Delta \mathcal{E}^{\mu} \Delta \mathcal{E}^{\nu}. \label{thermo length Rup}
\end{equation}
An important point that should be noted is that we are not
constrained to express this metric in a specific coordinate and
different metrics can be defined with different choices of
thermodynamic coordinates. Indeed, using the fact that the value
of $\Delta \ell^2$ must be independent of the coordinates used to
calculate one can explore the form of metric in other coordinate
systems. Keeping this point in mind, we will find the form of the
Ruppeiner metric in some thermodynamic coordinate systems:

\vspace{0.3cm}
$\blacktriangleright{\textbf{$(u, \mathcal{E}^i)$ fluctuation coordinate }}$
\vspace{0.3cm}

As explained earlier, working in the systems whose coordinates are
extensive parameters, i.e. $(u, \mathcal{E}^1, \mathcal{E}^2, ...,
\mathcal{E}^r)$ \footnote{The zeroth component is internal energy
and other components denote other extensive parameters of the
system}, leads to Eq.
\begin{equation}
g_{\mu\nu}^{R} = -\frac{1}{k_B}\frac{\partial^2 s}{\partial \mathcal{E}^\mu\partial \mathcal{E}^\nu}, \label{Ruppiener}
\end{equation}
which indeed is the Hessian matrix of the thermodynamic entropy
\eqref{metric} \cite{Rupp79, Ruppeiner1995}. Moreover, the
thermodynamic stability necessitates the definite positivity of
$g_{\mu\nu}^{R}$.

 \vspace{0.3cm}
 $\blacktriangleright{ \textbf{$(T,\mathcal{E}^i)$ fluctuation coordinate}}$
 \vspace{0.3cm}

Another coordinate system is defined based on the intensive as
well as extensive parameters. In this case, Making use of Eqs.
\eqref{metric} and \eqref{thermo length} along with
Eq.\eqref{intensive} and the tarnsforation law \eqref{transform
law} we get
\begin{equation}
\Delta \ell^2= -\frac{1}{k_{B}}\Delta \mathcal{I}^\mu \Delta \mathcal{E}_\mu . \label{thermo length1}
\end{equation}
Moreover, to express the metric completely in terms of the intensive parameters, we use
\begin{equation}
\Delta \mathcal{E}^{\mu}=\frac{\partial \mathcal{E}^{\mu}}{\partial \mathcal{I}^{\alpha}}\,\Delta \mathcal{I}^{\alpha},\label{transform law2}
\end{equation}
and substituting it into Eq. \eqref{thermo length1} yields
\begin{equation}
\Delta \ell^2= \frac{1}{k_{B}} \frac{\partial^2 \phi}{\partial \mathcal{I}^\mu\partial \mathcal{I}^\nu}\Delta \mathcal{I}^\mu \Delta \mathcal{I}^\nu, \label{thermo length2}
\end{equation}
where $\phi$ is the Legendre transformation of the entropy relative to all extensive parameters of the system (except volume) and reads as
\begin{equation}
    \phi(\mathcal{I}^{0}, \mathcal{I}^{1}, ..., \mathcal{I}^{r})=s-\mathcal{I}^{\mu}\mathcal{E}_{\mu}.
\end{equation}
Working in the entropy representation, the intensive parameters,
given by Eq.\eqref{intensive}, can be obtained as follows
\begin{equation}
    \{\mathcal{I}^0,\mathcal{I}^i\}=\{\frac{1}{T},-\frac{\mu^i}{T}\},
\end{equation}
in which $\mu^i$ is the chemical potential of the $i^{\text{th}}$ fluid
component. Using the above relation and doing the following
substitution into Eq. \eqref{thermo length1}
\begin{eqnarray}
    \Delta \mathcal{E}^{0}&=&\Delta u=\,T\,\Delta s+\sum_{i=1}^{r}\mu^{i}\Delta \mathcal{E}^i,\\
    \Delta \mathcal{I}^{0}&=&-\frac{1}{T^2}\Delta T,\\
    \Delta \mathcal{I}^{i}&=&\frac{\mu^i}{T^2}\Delta T-\frac{1}{T}\Delta\mu^i,\,\,\,\,\,\,\,\,\,\,\,\,\,\,\,\,\,\,\,\,\,\,\,\,1\leq i \leq r
\end{eqnarray}
we obtain
\begin{equation}
\Delta \ell^2= \frac{1}{k_{B}T}\Delta T\Delta s+\frac{1}{k_{B}T}\sum_{i=1}^{r}\Delta \mu^{i}\Delta \mathcal{E}^i. \label{thermo length3}
\end{equation}
Let us find the form of the metric in $(T,\mathcal{E}^1,
\mathcal{E}^2, ..., \mathcal{E}^r)$ coordinate system. To this
end, we use the following relation
\begin{eqnarray}
    \Delta s&=&\frac{\partial s}{\partial T}\Delta T+ \sum_{i=1}^{r}\frac{ \partial s}{\partial \mathcal{E}^i}\Delta \mathcal{E}^i,\\
    \Delta \mu^i&=&\frac{\partial \mu^i}{\partial T}\Delta T+ \sum_{i=1}^{r}\frac{ \partial \mu^i}{\partial \mathcal{E}^i}\Delta \mathcal{E}^i,
\end{eqnarray}
as well as the Maxwell relation
\begin{equation}
    \frac{\partial s}{\partial \rho^i}=-\frac{\partial \mu^i}{\partial T},\label{55}
\end{equation}
and substituting Eq.\eqref{55} into Eq. \eqref{thermo length3}, we get
\begin{equation}
\Delta \ell^2= \frac{1}{k_{B}T}\frac{\partial s}{\partial T}(\Delta T)^2+\frac{1}{k_{B}T}\sum_{i,j=1}^{r}\frac{\partial \mu^i}{\partial \mathcal{E}^j}\Delta \mathcal{E}^i \Delta \mathcal{E}^j. \label{thermo length4}
\end{equation}

\vspace{0.3cm}
$\blacktriangleright{ \textbf{$(T,\mathfrak{I}^i)$ fluctuation coordinate}}$
\vspace{0.3cm}

As another example, we try to express the metric in
$\mathfrak{I}=(T,\mu^1, \mu^2, ..., \mu^r)$ coordinate which are
the intensive parameters in the energy representation, i.e.
$\mathfrak{I}^0=\frac{\partial u}{\partial s}=T$ and
$\mathfrak{I}^i=\frac{\partial u}{\partial \mathcal{E}^i}=\mu^i$.
Making use of the following relations
\begin{eqnarray}
        \Delta s&=&\frac{\partial s}{\partial T}\Delta T+ \sum_{i=1}^{r}\frac{ \partial s}{\partial \mu^i}\Delta \mu^i,\\
        \Delta \mathcal{E}^i&=&\frac{\partial \mathcal{E}^i}{\partial T}\Delta T+ \sum_{j=1}^{r}\frac{ \partial \mathcal{E}^i}{\partial \mu^j}\Delta \mu^j,
\end{eqnarray}
and substituting them into Eq. \eqref{thermo length3} leads to
\begin{equation}
\Delta \ell^2= -\frac{1}{k_{B}T} \frac{\partial^2 \omega}{\partial \mathfrak{I}^\mu\partial \mathfrak{I}^\nu}\Delta \mathfrak{I}^\mu \Delta \mathfrak{I}^\nu ,\label{thermo length5}
\end{equation}
where
\begin{eqnarray}
\omega(\mathfrak{I}^{0}, \mathfrak{I}^{1}, ..., \mathfrak{I}^{r})&=&u-Ts-\sum_{i=1}^{r}\mu^i\mathcal{E}^i\\
&=&-\phi(\mathcal{I}^{0}, \mathcal{I}^{1}, ..., \mathcal{I}^{r}) T.
\end{eqnarray}
A summary of the mentioned metrics is tabulated in the Table
\ref{table}. It should be noted that in the all selected
frameworks, the volume of the system is considered as a quantity
whose value is fixed and does not fluctuate. Moreover, as a
several examples,  the thermodynamic metric elements
$g_{\alpha\beta}$ are evaluated in the state $x^\alpha=x^\alpha_0$
for a number of coordinate systems which are listed in Table
\ref{table2}.

\begin{table}
    \begin{center}
        \begin{tabular}{|l| c| l|}
            \hline
            \hline\ & & \\
        \,\,\,\,\,\,\,\,\,\,\,\,\,  {{\it{\textbf{Coordinates}}}} & {\it{\textbf{Potential}}} & \,\,\,\,\,\,\,\,\,\,\,\,\, {\it{\textbf{Line element $(\Delta \ell)^2$}}} \\ & & \\
        \hline\ & & \\
        $\mathcal{E}=(u,\mathcal{E}^1, \mathcal{E}^2, ..., \mathcal{E}^r)$ & $s$ & $ -\frac{1}{k_B}\frac{\partial^2 s}{\partial \mathcal{E}^\alpha\partial \mathcal{E}^\beta}$ \\ & &\\
            $\mathcal{I}=\{\frac{1}{T},-\frac{\mu^1}{T}, -\frac{\mu^2}{T},..., -\frac{\mu^r}{T}\}$ & $\phi(\mathcal{I})=s-\mathcal{I}^{\mu}\mathcal{E}_{\mu}$ & $\frac{1}{k_{B}} \frac{\partial^2 \phi}{\partial \mathcal{I}^\mu\partial \mathcal{I}^\nu}\Delta \mathcal{I}^\mu \Delta \mathcal{I}^\nu $ \\ & & \\
            $\mathfrak{I}=(T,\mu^1, \mu^2, ..., \mu^r)$ & $\omega(\mathfrak{I})=u-Ts-\sum_{i=1}^{r}\mu^i\mathcal{E}^i$ & $-\frac{1}{k_{B}T} \frac{\partial^2 \omega}{\partial \mathfrak{I}^\mu\partial \mathfrak{I}^\nu}\Delta \mathfrak{I}^\mu \Delta \mathfrak{I}^\nu $ \\ & & \\
            $(T,\mathcal{E}^1, \mathcal{E}^2, ..., \mathcal{E}^r)$ & $f=u-Ts$ & $\frac{1}{k_{B}T}\frac{\partial s}{\partial T}(\Delta T)^2+\frac{1}{k_{B}T}\sum_{i,j=1}^{r}\frac{\partial \mu^i}{\partial \mathcal{E}^j}\Delta \mathcal{E}^i \Delta \mathcal{E}^j$ \\ & & \\
                        \hline
        \end{tabular}\label{table}
    \end{center}
    \caption {Thermodynamic potentials and the Riemannian line elements in four coordinate systems.}
\end{table}

\begin{table}
    \begin{center}
        \begin{tabular}{|l|c|l|}
            \hline
            \hline  & & \\
            {{\it{\textbf{Coordinates}}}} & \it {\textbf{Potential}} & $\{g_{11},g_{12},g_{22}\}$ \\
            \hline  & & \\
            \,\,\,\,\,\,\,\,$\{u,\rho\}$ & $s(u,\rho)$ & ${\displaystyle -\frac{1}{k_B } \left\{\frac{\partial^2 s}{\partial u^2},\frac{\partial^2 s}{\partial u\,\partial \rho},\frac{\partial^2 s}{\partial \rho^2}\right\}}$ \\  & & \\
            \,\,\,\,\,\,\,  $\{s,\rho\}$ &   $u(s,\rho)$ & ${\displaystyle \frac{1}{k_B T } \left\{\frac{\partial^2 u}{\partial s^2},\frac{\partial^2 u}{\partial s\,\partial \rho},\frac{\partial^2 u}{\partial \rho^2}\right\}}$ \\  & & \\
            \,\,\,\,\,\,\,\,    $\{T,\rho\}$ & $f(T,\rho) = u - Ts $ & ${\displaystyle \frac{1}{k_B T } \left\{-\frac{\partial^2 f}{\partial T^2}, 0, \frac{\partial^2 f}{\partial \rho^2}\right\}}$ \\  & & \\
            \,\,\,\,\,\,\,\,\,\,$\{T,\mu\}$ & $\Omega(T,\mu) = u - T s - \mu \rho$ & ${\displaystyle -\frac{1}{k_B T } \left\{\frac{\partial^2 \Omega}{\partial T^2},\frac{\partial^2 \Omega}{\partial T\,\partial\mu},\frac{\partial^2 \Omega}{\partial \mu^2}\right\}}$ \\
            \hline
        \end{tabular}\label{table2}
    \end{center}
    \caption {Thermodynamic potentials and the thermodynamic metric elements in four coordinate systems. (adapted from \cite{Rup3})}
\end{table}

\subsubsection{ \textbf{Weinhold metric}}
An alternate geometric method, using the idea of conformal mapping
from the Riemannian space to thermodynamic space,  was proposed by
Weinhold in 1975 \cite{Wein1975} in which a metric is
introduced in the space of equilibrium states of thermodynamic
systems as the Hessian of the internal energy $u$
\begin{equation}
g_{\mu\nu}^{W} = \frac{1}{k_B}\frac{\partial^2 u}{\partial \mathscr{E}^\mu\partial \mathscr{E}^\nu}, \label{Weinhold}
\end{equation}
where $\mathscr{E}=(s, \mathcal{E}^1, \mathcal{E}^2, ...,
\mathcal{E}^r)$. It is notable that this metric can also be
obtained from the Ruppeiner metric via a transformation of
fluctuation coordinates. In addition, the Weinhold geometry is
conformaly related to the Ruppeiner geometry with the temperature
being the conformal factor \cite{Rupp79, Ruppeiner1995,
conformal, Liu:2010sz},
\begin{equation}
ds^2_{(R)}=\frac{1}{T}\,ds^2_{(W)}\,.
\end{equation}
Weinhold showed that the laws of thermodynamics assure that $
g^{W}_{\mu \nu} $ has the positivity required of a metric or first
fundamental form on the surface of thermodynamics states. This
property enabled him to re-derive thermodynamic relations using
simple geometric arguments \cite{Salamon;1980}. Such a metric
gives us a way to define distances and angles and, therefore, it
enables us to study the geometry of the surface
\cite{san05a}. This metric turns out to be positive as a
consequence of the second law of thermodynamics.

Salamon et al., considered Weinhold metric to study the physical
significance of thermodynamic length \cite{Salamon;1980}. They
found that the local meaning of $ g^{W}_{\mu \nu} $ is the
distance  between the energy surface and the linear space tangent
to this surface at some point where $ g^{W}_{\mu \nu} $ is
evaluated. In Ref. \cite{san05b}, the authors generalized
the physical interpretation of thermodynamic length to a
two-dimensional thermodynamic system with constant heat capacity
and showed that in an isochoric thermodynamic system with two
degrees of freedom, thermodynamic length is related to the heat
flux of a quasi-static process at a constant mole number. A
relation between thermodynamic length and work for an isentropic
Ideal and quasi-Ideal Gas along isotherms was found in Ref.
\cite{san04}. In fact, the thermodynamic length was
considered as a measure of the amount of work done by the system
along isotherms. After that in Ref.\cite{san05a} a
generalization of this relation was provided for un-fixed
temperature and also found that the thermodynamic length of an
isentropic Ideal or quasi-Ideal Gas measures the difference of the
square roots of the energies of two given states. Naturally, if
there is no work received or done by such a system then the length
of the path is zero. The Weinhold approach has been intensively used
to study, from a geometrical point of view, the properties of the
space generated by Weinhold's metric
\cite{Gilmore;1981,Feldman;1985}, the chemical and physical
properties of various two-dimensional thermodynamic systems
\cite{san05b,san04,san05a,nul85,san05c},
and the associated Riemannian structure
\cite{Rupp79,Torres;1993,RT3}. Despite the
efforts conducted in the context of the Weinhold metric, the
geometry based on this metric seems physically meaningless in the
context of purely equilibrium thermodynamics.

\subsubsection{ \textbf{Shortcomings of Hessian metrics}}

Weinhold and Ruppeiner metrics have been widely used to describe
the phase structure of condensed matter systems \cite{Rupp79,
condensed1, condensed2, condensed3, jjk03, condensed5,
condensed6, condensed7, condensed8, condensed9, condensed10}.
However, in some cases, these metrics have shown inconsistency
with each other
\cite{Quevedo;2008,Mirza:059,Medved:2149,Aman:2003,Quevedo;2008cd}.
One reason for this discrepancy is due to their dependence on the
choice of thermodynamic potential. Specifically, Hessian metrics
lack Legendre invariance, which results in the loss of geometric
structure when a different thermodynamic potential is employed to
describe equilibrium states. To address this problem, Quevedo
{\it{et al.}} proposed a Legendre invariant metric formalism
\cite{Quevedo} that preserves geometric structure regardless of
the chosen thermodynamic potential. This formalism will be further
discussed in the following section.

\subsection{Legendre invariant metric} \label{3B}
Invariance under Legendre transformations is a fundamental
property of thermodynamics, and it follows that any
thermo-geometric metric should maintain this invariance.
Legendre-invariant metrics are important because they convey the
essential message that the thermodynamic properties of physical
systems do not depend on the choice of thermodynamic potential
from a geometric standpoint. To devise a thermo-geometric metric
in a Legendre-invariant manner, one must incorporate concepts from
differential geometry, such as contact manifolds, metrics, and
Riemannian geometry, as well as Legendre transformations. Each of
these concepts is explained in detail in the appendix. By taking a
rigorous and systematic approach to the geometrization of
thermodynamic systems, researchers can develop Legendre-invariant
metrics that provide critical insights into the behavior of
thermodynamical systems.

\subsubsection{\textbf{Quevedo Metric}}
As it was already mentioned, neither Weinhold nor Ruppeiner metric
is formulated in a Legendre-invariant way which makes them
inappropriate for describing the geometry of thermodynamic
systems. By using purely mathematical considerations and geometric
objects, Quevedo put forward the idea of geometric thermodynamic
and showed that both Weinhold and Ruppeiner formalisms can be
unified into a single approach called geometrothermodynamics (GTD)
\cite{Quevedo}. The main motivation for introducing the formalism
of GTD was to formulate a geometric approach which takes into
account the fact that in ordinary thermodynamics the description
of a system does not depend on the choice of the thermodynamic
potential, i. e., it is invariant with respect to Legendre
transformations.\par

One of the primary geometric components of GTD is the {\it{phase
manifold}}  which is essential in order to have Legendre
transformations as diffeomorphisms. The triad $({\cal{T}}, \theta,
G)$ is called phase manifold, or generally a {\it{Riemannian
contact manifold}}, if the following conditions are satisfied (see
appendixes \ref{two} and \ref{three} for more detail):
\begin{itemize}
    \item{$ \mathcal{T} $ is a (2n+1)-dimensional manifold which coordinatized by the set $Z^A=\{\Phi, E^{a}, I^{a}\}$ with $A=0, ....., 2n$, and $a=1,...., n$. Here $\Phi$ is the thermodynamic potential, and $E^{a}$ and $I^{a}$ denote the extensive and intensive variables, respectively.}
    \item{$\theta$ is a non-vanishing differential contact one-form defined on the cotangent manifold $T^*({\cal T})$, satisfying the condition $\theta \wedge (d \theta)^n \neq 0$, that according to Darboux theorem always exists in any odd
dimensional manifold and can be expressed as follows
        \begin{equation}
        \theta= d\Phi - I_{a} d E^{a}. \,\,\,\,\,\,\,\,\text{(Darboux theorem)}
        \label{1form}
        \end{equation}
        This expression for the contact one-form $\theta$ is clearly invariant under the Legendre transformations given in (\ref{Legendre1} - \ref{Legendre3}), i. e.
        \begin{equation}
        \theta\rightarrow \bar\theta =  d\bar\Phi - \bar I_{a} d \bar E^{b}.
        \end{equation}
    }
        \item{$G$ is a Legendre invariant Riemannian metric on $\cal T$ , i.e. $G = G_{AB} dZ^AdZ^B$. Indeed, demanding Legendre invariance of the metric $G$ results in a set of algebraic equations for the components $G_{AB}$, whose solutions can be split into three different metrics \cite{Belgiorno:2002iw}, namely
            \begin{equation}
                G^{{I}} = (d\Phi - I_{a} d E^{a})^2 + (\xi_{ab} E^{a} I^{b}) (\delta_{cd} dE^{c} dI^{d}),\label{G1}
            \end{equation}
            \begin{equation}
                G^{{II}} = (d\Phi - I_{a} d E^{a})^2 + (\xi_{ab} E^{a}\, I^{b}) (\eta_{cd} dE^{c}, dI^{d}),\label{G2}
            \end{equation}
            \begin{equation}
                G^{{III}}  =(d\Phi - I_{a}\, d E^{a})^{2}  + \sum_{a=1}^{n}\,E_{a} I_{a}\, d E^{a} \,d I^{a},\label{G3}
            \end{equation}
            which the first and the second relations are invariant under total Legendre transformations while the third one is invariant with respect to partial Legendre transformations.
            Here $\xi_{ab}$ is a diagonal constant $(n\times n)$-matrix, $\delta_{ab}= {\rm diag}(1,\cdots,1)$ and $\eta_{ab}= {\rm diag}(-1,\cdots,1)$. It is notable that  $G^{I}$ and $G^{{II}}$ are used to describe systems with first order and second order phase transitions, respectively.}
\end{itemize}
 In the context of thermodynamics,
any Riemannian contact manifold $({\cal T}, \theta,G)$ whose components are Legendre invariant is
called a {\it thermodynamic phase space} (phase manifold) that can be considered as the starting point for a description
of thermodynamic systems using geometric concepts. \par

On the other hand, the space of {\it{thermodynamic equilibrium states}} (equilibrium manifold) is an $n$-dimensional Riemannian submanifold
${\cal E} \subset {\cal T}$, with a non-degenerate metric $g$ and
the extensive variables ${E^a}$ as the coordinates, defined by
means of a smooth embedding mapping $ \varphi : \   {\mathcal E} \
\longmapsto {\mathcal T}$, i.e. $ \varphi :  (E^a) \longmapsto
(Z^{A}(E^a))=(\Phi(E^a), E^a, I^a(E^a))$, if the condition
\begin{equation}
    \varphi^*(\theta) =\varphi^{\star}(d\Phi\,-\,I_{a}\,d E^{a})=0,\label{condition33}
\end{equation}
is satisfied, where $\varphi^{\star}$ indicates the pullback of
$\varphi$. Moreover, a thermodynamic metric $g$ is induced in the
equilibrium manifold $\mathcal E$, which we demand to be
compatible with the metric $G$ on $\cal T$, by means of
$g=\varphi^{*}(G)$. Therefore, regarding \eqref{G1}-\eqref{G3},
the following relation can be obtained for the thermodynamic
metric $g$
\begin{equation}
g^{{I/II}}_{ab} =   \beta_{\Phi} \Phi  \xi_{a}^{\ c}
\frac{\partial^2\Phi}{\partial E^{b} \partial E^{c}}   ,
\label{gdownf}
\end{equation}
where $\xi_a^{\ c}=\delta_a^{\ c}={\rm diag}(1,\cdots,1)$ for
$g^{I}$ and $\xi_a^{\ c}=\eta_a^{\ c}={\rm diag}(-1,1,\cdots,1)$
for $g^{{II}}$. Besides, the constant $\beta_\Phi$ indicates the
degree of homogeneity of the thermodynamic potential $\Phi$
\cite{Qeovedo1}. The third metric of the equilibrium space can  be
written as
\begin{equation}
g^{{III}} = \sum_{a=1} ^n \left(\delta_{ad} E^{d} \frac{\partial\Phi}{\partial E^{a}}\right) \delta^{ab} \frac{\partial ^{2} \Phi}{\partial E^{b} \partial E^{c}}
dE^{a} dE^{c} \ .
\label{gIII}
\end{equation}
Note that $\Phi(E^a)$ represents a fundamental thermodynamic
equation which must be known explicitly. $\Phi$ can be either the entropy or the internal energy of the system \cite{Call}.\par

Considering condition \eqref{condition33} leads us to the following relations
\begin{eqnarray}
d\Phi &=& \delta_{ab} I^{a} d E^{b} \ ,\label{condition11} \\
 \frac{\partial\Phi}{\partial E^a} &=&
\delta_{ab} I^{b} \label{condition22},
\end{eqnarray}
which correspond to the {\it{first law of thermodynamics}} and the
standard conditions of thermodynamic equilibrium, respectively
\cite{Call}. Eq.\eqref{condition22} also means that the intensive
thermodynamic variables are dual to the extensive ones. In this
construction, the {\it{second law of thermodynamics}} implies that
the fundamental equation satisfies the condition
\cite{Call,Burke;1987}
\begin{equation}
\pm\frac{\partial^2\Phi}
{\partial E^{a} \partial E^{b}}\geq 0,\,\,\quad\quad \,(\text{convexity condition})
\end{equation}
where the sign depends on the thermodynamic potential. For example, if $\Phi$ is identified as the entropy, the sign is positive whereas it should be negative in the case of $\Phi$ being the internal energy of the system \cite{Call}. \par

It is worth mentioning that as the thermodynamic potential is a
homogeneous function of its arguments, it satisfies the
homogeneity condition
\begin{equation}
    \Phi(\lambda E^{a}) =\lambda^{\beta} \Phi(\lambda E^{a}),
\end{equation}
for constant parameters $\lambda$ and $\beta$.  Making use of the
condition \eqref{condition22} along with differentiating the
homogeneity condition with respect to $\lambda$, one can easily
get the following expression for the Euler's identity
 \begin{eqnarray}
    \beta \Phi(E^{a}) = \delta_{ab}I^{b}E^{a},
 \end{eqnarray}
which the result has been evaluated at $\lambda=1$. In addition,
calculating the exterior derivative of Euler's identity and using
the first law of thermodynamics \eqref{condition11} gives rise to
the generalized Gibbs-Duhem relation
 \begin{eqnarray}
 (1-\beta)I^{a}dE^{b} \delta_{ab}+E^{a}dI^{b} \delta_{ab} = 0.
 \end{eqnarray}
It is important to note that the classical expressions for Euler's
identity and Gibbs-Duhem relation will be obtained by instituting
$\beta = 1$ into the above equations.
\subsection{Calculation of thermodynamic metrics for ordinary thermodynamic systems} \label{3C}
In order to apply the geometric representation, which was
explained throughout the previous parts, to some ordinary
thermodynamic systems, we consider ideal gas and real gas with
5-dim phase space with variables $(U,S,T,p,V)$, and study them in
more detail.\par

\subsubsection{\bf{Ideal gas}}
\par
First, we consider the ideal gas, as the simplest example of a
thermodynamic system with two degrees of freedom. In this case,
the equation of state is given by
\begin{equation}
    PV=Nk_{B}T,
\end{equation}
where $k_{B}$ indicates the Boltzmann constant and $N$ is
 the number of molecules. Its  internal energy is a function of the temperature and reads as
 \begin{equation}
    U_{\mathrm{ideal}}(T)=n\,C_{V}T,
 \end{equation}
in which $n\equiv\frac{N}{N_A}$ is defined as the number of moles
of the gas with $N_{A}$ denoting Avogadro's number. Besides, $C_V$
indicates the heat capacity at constant volume. Making use of the
first law of thermodynamics, $dQ=T dS =dU+PdV$, one can
immediately get
 \begin{equation}
     dS=nC_{V}\frac{dU}{U}+nR\frac{dV}{V},
 \end{equation}
 or equivalently
 \begin{equation}
    S_{\mathrm{ideal}}\left(  U\text{, }V\right)  =n\,C_{V}\ln\left(  U\right)
    +nR\ln\left( V\right)  +\mathcal{S}_{0}, \label{entropyideal}%
 \end{equation}
where $\mathcal{S}_0$ is an integration constant and
$R=N_{A}k_{B}$ denotes the gas constant. This relation is known as
the fundamental relation in the entropy representation. Working in
the energy representation we get
\begin{equation}
    U_{\mathrm{ideal}}(S,V)=V^{-\frac{R}{C_{V}}}e^{\frac{S-\mathcal{S}_{0}}{nC_{V}}}.
\end{equation}
 Having the above information, we try to examine thermodynamic metrics in different representations:

   \vspace{0.3cm}
\noindent  $\bullet$ {\it{\textbf{Ruppeiner metric:}}}\quad
  Recalling that the Ruppeiner metric is defined as $
 g_{\mu\nu}^{R} \overset{\text{def}}{=} -\frac{1}{k_B}\frac{\partial^2 S}{\partial{E}^\mu\partial {E}^\nu}$,
 with $E=\left(  E^{1}\text{, }E^{2}\right)  =\left(  U\text{,
 }V\right)  $, the metric components can be obtained as
   \begin{align}
    g_{11}^{{R}}\left(  U\text{, }V\right)   &  =\frac{n\,C_{V}}{k_{B}U^{2}}\text{,
    }\nonumber\\
    g_{12}^{{R}}\left(  U\text{, }V\right)   &  =g_{21}\left(  U\text{, }V\right)
    =0\text{,}\nonumber\\
    g_{22}^{{R}}\left(  U\text{, }V\right)   &  =\frac{nR}{k_{B}V^{2}}\text{.}
  \end{align}
  Hence
  \begin{equation}
  g^R_{\mathrm{ideal}}
  =\frac{nC_{V}%
  }{k_{B}U^{2}}dU^{2}+\frac{n\,R}{k_{B}V^{2}}dV^{2}\text{.}
  \end{equation}
  
  \vspace{0.3cm} \noindent $\bullet$ {\it{\textbf{Weinhold
metric:}}}\quad Working in the energy representation and using the
relation $g_{\mu\nu}^{W} \overset{\text{def}}{=}
\frac{1}{k_B}\frac{\partial^2 U}{\partial \theta^\mu\partial
\theta^\nu}$ with $\left(\theta^{1}\text{, }\theta^{2}\right)
=\left(  S\text{, }V\right) $, the components of Weinhold's metric
will be given as follows
\begin{align}
g_{11}^{W}\left(  S\text{, }V\right)   &  =\frac{1}{k_{B}n^2C_{V}^{2}}\,e^{\frac{S}{nC_{V}}}V^{-\frac{R}{C_{V}}}\text{,
}\nonumber\\
g_{12}^{W}\left(  S\text{, }V\right)   &  =g_{21}\left(  U\text{, }V\right)
=-\frac{RV}{nC_{V}^2}e^{\frac{S}{nC_{V}}}V^{-\frac{C_{P}}{C_{V}}}\text{,}\nonumber\\
g_{22}^{W}\left(  S\text{, }V\right)   &  =\frac{RC_{P}}{C_{V}^2}e^{\frac{S}{nC_{V}}}V^{-(1+\frac{C_{P}}{C_{V}})}\text{,}
\end{align}
and as a result
\begin{equation}
g^W_{\mathrm{ideal}}
=e^{\frac{S}{nC_{V}}}\Bigg[\frac{V^{-\frac{R}{C_{V}}}}{k_{B}n^2C_{V}^{2}}\,\,dS^{2}+\frac{RC_{P}}{C_{V}^2}V^{-(1+\frac{C_{P}}{C_{V}})}\,\,dV^{2}-\frac{RV}{nC_{V}^2}V^{-\frac{C_{P}}{C_{V}}}\,\,dSdV\Bigg]\text{.}
\end{equation}

\vspace{0.3cm}
\noindent $\bullet$ {\it{\textbf{Quevedo Metric:}}}\quad
In the case of GTD, by using the most general metric \eqref{gIII} and
working in the entropy representation, the components of Quevedo's metric will take the following form
    \begin{align}
    g_{11}^{Q}\left(  U\text{, }V\right)    & =\frac{(nC_{V})^{2}}{U^{2}%
    }\text{,}\\
    & \\
    g_{12}^{Q}\left(  U\text{, }V\right)    & =g_{21}^{ideal}\left(  U\text{,
    }V\right)  =0\text{,}\\
    & \\
    g_{22}^{Q}\left(  U\text{, }V\right)    & =\frac{\left(  nR\right)
        ^{2}}{V^{2}}\text{.}\\%
\end{align}
Therefore,
\begin{equation}
g_{\mathrm{ideal}}^{\text{Q}}=\frac{(nC_{V})^{2}%
}{U^{2}}dU^{2}+\frac{\left(  nR\right)  ^{2}}{V^{2}}dV^{2}\text{.}%
\end{equation}


\subsubsection{\bf{Van der Waals gas}}
Now, we consider a more realistic model of a gas known as the Van der Waals Gas whose equation of state is
given by
\begin{equation}
\left(  P+\frac{n^{2}}{V^{2}}a\right)  \left(  V-nb\right)=nRT,
\end{equation}
where $a$ and $b$ are positive constants, indicating the characteristic of the particular gas under consideration \footnote{Indeed, $a$ is a characteristic of molecule interaction in the gas and $b$ is characteristic of the part of the volume occupied by the molecules \cite{Call}.}. Moreover, its internal energy reads as
\begin{equation}
    U_{\mathrm{VdW}}(T)=n\,C_{V}T-n^{2}\frac{a}{V}+U_{0}.
\end{equation}
Using the first law of thermodynamics
\begin{equation}
    dS=\frac{n\,C_{V}}{U+n^{2}\frac{a}{V}}dU+\frac{n\,R}{V-n\,b}dV,
\end{equation}
the fundamental equation for the Van der Waals gas in the entropy representation is given by
\begin{equation}
S_{\mathrm{VdW}}\left(  U\text{, }V\right)  =nC_{V}\ln\left(  U+n^{2}\frac
{a}{V}\right)  +n\,R\ln\left(  V-n\,b\right)  +\mathcal{C}_{r},
\label{entropyreal}%
\end{equation}
with $\mathcal{C}_r$ an integration constant. Setting $\mathcal{C}_{r}=0$ and solving the above relation for $U$ we get
the internal energy of Van der Waals gas as a function of S and V
as follows
\par
\begin{equation}
U_{VdW}(S,V)=(V-n\,b)^{-\frac{R}{C_{V}}}e^{\frac{S}{n\,C_{V}}}-n^2\frac{a}{V}.
\end{equation}
\par
Considering the mentioned information for the Van der Waals gas as
well as Eqs. \eqref{Ruppiener}, \eqref{Weinhold} and \eqref{gIII},
some thermodynamic metrics will be obtained as follows:

\vspace{0.3cm}
 \noindent  $\bullet$ {\it{\textbf{Ruppeiner metric:}}}\quad
 \begin{align}
g_{\text{\tiny{UU}}}^{\text{\tiny{VdW(R)}}}\left(  U\text{, }V\right)   &  =n\,C_{V}\frac{V^{2}}{k_{B}\left(
    an^{2}+U\,V\right)  ^{2}}\text{, }\nonumber\\
g_{\text{\tiny{UV}}}^{\text{\tiny{VdW(R)}}}\left(  U\text{, }V\right)   &  =g_{\text{\tiny{VU}}}^{\text{\tiny{VdW}}}\left(  U\text{, }V\right)
=-\frac{an^{3}C_{V}}{k_{B}\left(  an^{2}+UV\right)  ^{2}}\text{,}\nonumber\\
g_{\text{\tiny{VV}}}^{\text{\tiny{VdW(R)}}}\left(  U\text{, }V\right)   &  =\frac{nR}{k_{B}V^{2}}\frac{V^{2}}{\left(
    V-bn\right)  ^{2}}-\frac{an^{3}C_{V}}{k_{B}V^{2}}\frac{\left(  an^{2}+2UV\right)
}{\left(  an^{2}+UV\right)  ^{2}}\text{.}
\end{align}

  \vspace{0.3cm}
 \noindent $\bullet$ {\it{\textbf{Weinhold metric:}}}\quad
  \begin{align}
 g_{\text{\tiny{SS}}}^{\text{\tiny{VdW(W)}}}\left(  S\text{, }V\right)   &  =\frac{1}{k_{B}n^2C^{2}_{V}}e^{\frac{S}{nC_{V}}}(V-nb)^{-\frac{R}{C_{V}}} \text{,
 }\nonumber\\
 g_{\text{\tiny{SV}}}^{\text{\tiny{VdW(W)}}}\left(  S\text{, }V\right)   &  =g_{\text{\tiny{VS}}}^{\text{\tiny{VdW(W)}}}\left(  U\text{, }V\right)
 =-\frac{R}{k_{B}nC^{2}_{V}}e^{\frac{S}{nC_{V}}}(V-nb)^{-(1+\frac{R}{C_{V}})}\text{,}\nonumber\\
 g_{\text{\tiny{VV}}}^{\text{\tiny{VdW(W)}}}\left(  S\text{, }V\right)   &  =\frac{R}{k_{B}C^{2}_{V}}e^{\frac{S}{nC_{V}}}(C_{V}+R)(V-nb)^{-(2+\frac{R}{C_{V}})}-n^2\frac{2a}{k_{B}V^{3}}\text{,}\label{wein}
 \end{align}
 
\vspace{0.3cm}
 \noindent $\bullet$ {\it{\textbf{Quevedo Metric:}}}\quad
  \begin{align}
\quad\quad\quad\quad\quad g_{\text{\tiny{UU}}}^{\text{\tiny{VdW(Q)}}}\left(  U\text{, }V\right) & =\left(  nC_{V}\frac{U}%
 {U+n^{2}\frac{a}{V}}\right)  ^{2}\frac{nC_{V}V^{2}}{\left(  an^{2}%
    +UV\right)  ^{2}}\text{,}\\\label{Q1}
 &\nonumber \\
 g_{\text{\tiny{UV}}}^{\text{\tiny{VdW(Q)}}}\left(  U\text{, }V\right) & =g_{\text{\tiny{VU}}}^{\text{\tiny{VdW(Q)}}}\left(  U\text{,
 }V\right)  =-\frac{1}{2}\frac{an^{3}C_{V}}{\left(
 an^{2}+UV\right)  ^{2}}\nonumber\\
&\nonumber \\
&\Bigg[  \left(  nC_{V}\frac{U}{U+n^{2}\frac{a}{V}%
}\right)  ^{2}+\left(  nR\frac{V}{V-nb}-nC_{V}\frac{n^{2}\frac{a}{V}%
}{U+n^{2}\frac{a}{V}}\right)  ^{2}\Bigg] \text{,}\\\label{Q2}
& \nonumber\\
g_{\text{\tiny{VV}}}^{\text{\tiny{VdW(Q)}}}\left(  U\text{, }V\right) & =-\left(  nR\frac{V}{V-nb}%
-nC_{V}\frac{n^{2}\frac{a}{V}}{U+n^{2}\frac{a}{V}}\right)  ^{2}\nonumber\\
&\nonumber\\
&\left[
-\frac{nR}{V^{2}}\frac{V^{2}}{\left(  V-bn\right)  ^{2}}+\frac{an^{3}C_{V}%
}{V^{2}}\frac{\left(  an^{2}+2UV\right)  }{\left(  an^{2}+UV\right)  ^{2}%
}\right]  \text{.}%
\end{align}
In the limit $a\rightarrow 0,\
b\rightarrow 0$, the obtained results for the Van der Waals gas will reduce to the ideal gas.

\section{Thermodynamic curvature}\label{4}
According to what was discussed, based on the thermal fluctuation
between equilibrium states, a new method was proposed to study the
thermodynamics of physical systems using Riemannian geometry.
Given a metric in  Riemannian geometry, one of the immediate
objects of interest is its Ricci curvature scalar $\mathcal{R}$
\cite{Laugwitz1965}, which is known as the thermodynamic
Riemannian curvature in the context of thermodynamic systems.
$\mathcal{R}$ is the only geometrical scalar invariant in
thermodynamics that has a volume dimension and as long as we are
dealing with a two-dimensional hypersurface, all the information
lies in this quantity and hence plays a fundamental role.\par

To go further, we examine the scalar curvature for some ordinary
thermodynamic systems and then discuss some interesting
connections between curvature and interactions in more detail.

{\vspace{0.3cm}
\noindent $\bullet$ {\it{\textbf{Ideal gas:}}}\quad}
 As
a first example, let's consider the ideal gas. Making use of the
results of thermodynamic metric for ideal gas as well as using
\eqref{Ricci} one can easily find that the thermodynamic curvature
is zero for a monoatomic ideal gas system. Since the ideal gas is
a non-interacting gas, this result suggests that $\mathcal{R}$ can
be some type of measure of interactions between particles.\par

	{\vspace{0.3cm}
	\noindent $\bullet$ {\it{\textbf{Van der Waals gas:}}}\quad}

As a second example, we investigate the scalar curvature of the
the Van der Waals gas in different geometrical representation:

\vspace{0.3cm}
 $\centerdot$ {{\textbf{Weinhold representation:}}}\quad
 
Using Eq. \eqref{wein} and \eqref{Ricci}, the scalar curvature in the Weinhold representation will be obtained as
following form
\begin{equation}
\mathcal{R}^{W}=\frac{aRV^{3}}{C_{V}(pV^{3}-aV+2ab)^{2}}.
\end{equation}
\par
As this relation shows the scalar curvature goes to zero when $a\rightarrow{0}$ or
$V\rightarrow{\infty}$. Since the quantity $\frac{a}{V^{2}}$
characterizes the attractive interaction within a system, scalar
curvature seems to be a measure of the attraction among particles
while its dependence on the parameter $b$ is more quantitative
then qualitative. Besides, the scalar curvature diverges at the critical points determined
by the algebraic equation $pV^3-aV + 2ab = 0$, which is exactly the equation that determines
the location of first order phase transitions of the van der Waals gas \cite{Call}. Consequently, a
first order phase transition can be interpreted geometrically as a curvature singularity. This
is in accordance with our intuitive interpretation of thermodynamic curvature.\par

\vspace{0.3cm}
$\centerdot$ {{\textbf{Ruppeiner representation:}}}\quad

In order to work in this geometry, we consider the fluctuation coordinates to be the temperature $T$ and the volume $V$. Therefore, the line element of the Ruppeiner geometry will be as follows \cite{Wei:2019yvs}
\begin{eqnarray}
dl^{2}
&=&\frac{3}{2T^{2}}dT^{2}+\frac{TV^{3}-2a(V-b)^{2}}{TV^{3}(V-b)^{2}}dV^{2}.
\end{eqnarray}
Employing the \eqref{Ricci}, we get the scalar curvature as
\begin{eqnarray}
\mathcal{R}^{R}=\frac{4a(V-b)^{2}\left(a(V-b)^{2}-TV^{3}\right)}{3\left(2a(V-b)^{2}-TV^{3}\right)^{2}}.
\end{eqnarray}
It is easy to check that the denominator of the scalar curvature $\mathcal{R}$ vanishes at the critical point. Clearly, the condition $V=b$ results in $\mathcal{R}=0$. Indeed, at this point, all the volume is taken up by the fluid molecules so that no extra room remains. Therefore, the total fluid becomes a rigid body and no interaction can exist between the fluid molecules, which is consistent with the result that $\mathcal{R}=0$ corresponds to vanishing interaction \cite{Oshima}.
\par

\vspace{0.3cm}
$\centerdot$ {{\textbf{Quevedo representation:}}}\quad

 Using the components of Quevedo metric for the Van der Waals gas as well as \eqref{Ricci}, the corresponding curvature scalar can be written as \cite{Quevedo:2012jg}
 \begin{equation}
 \mathcal{R}^{Q}=\frac{\mathcal{N}_{VdW}(U,V)}{\left( 3V(V-b)(U+PV)\right) ^{3}\left( \frac{3}{2}(V-b)(PV^3-aV+2ab)  \right) ^{2} 
 }\ ,
 \end{equation}
 where the function
 $\mathcal{N}_{vdW}(U,V)$ is a polynomial which is  different from zero at points where the denominator vanishes. Similar to the two previous representations, there exists curvature singularities at those points where the condition $PV^3-aV+2ab=0$ is satisfied, which as mentioned earlier, these are the points  where first order phase transitions occur in the van der Waals gas.
 
%
%

In addition to the mentioned results regarding the scalar
curvature in the cases of ideal gas and Van der Waals gas, some
calculations done on models with critical points which represent
that $|\mathcal{R}|$ diverges near these points in the same way as
the correlation volume $\xi^d$, where $d$ is the spatial dimension
of the system and $\xi$ is the correlation length
\cite{Rupp79, Ruppeiner1995, jjk03}. These
properties, i.e. the connection between curvature and
interactions, suggest that {\it{$\mathcal{R}$ can be considered as
a measure of interparticle interaction.}} Indeed, the calculations
indicate that the thermodynamic curvature yields a measure of the
smallest volume in which the thermal fluctuation theory is
effective. We expect this volume to be $\xi^d$ near criticality.
Therefore, these results leads us to interpret $|\mathcal{R}|$ as
being proportional to the correlation volume
\begin{equation}
|\mathcal{R}|\sim   \xi^d,
\end{equation}
which has been confirmed near critical points, where $\xi^d$ is
large enough to contain many particles, for several statistical mechanics models
\cite{Rupp81, Rupp90B, Brody95, condensed5, condensed6, Jan02, condensed7, Brody03, jjk03} \par

\begin{figure}
    \centering
    \includegraphics[height=3cm]{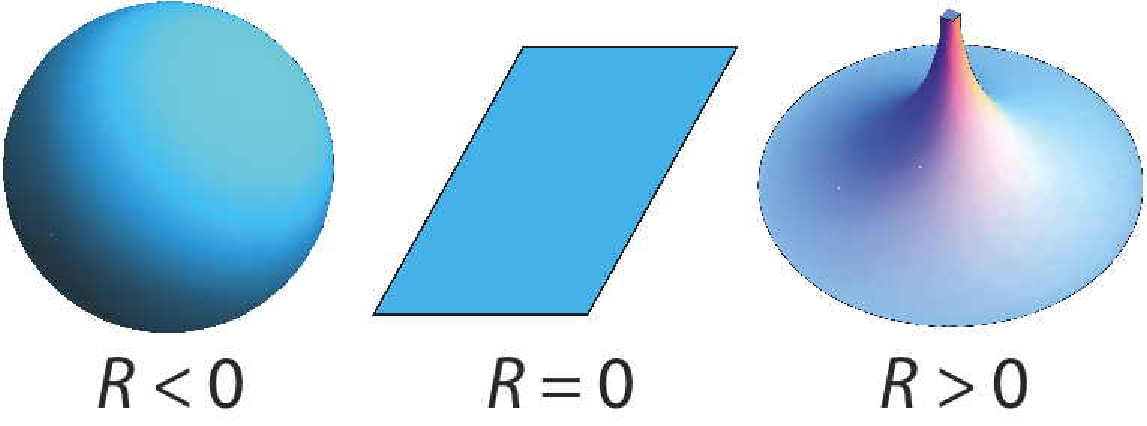}
    \caption{Surfaces related to constant Ricci curvature scalar R: the sphere, the plane, and the pseudosphere. , $R<0$ represents attractive interparticle interactions dominate. $R>0$ is related to repulsive interactions dominant. Regardless of the sign convention for $  R$, attractive and repulsive interactions correspond to the geometry of the sphere and pseudosphere, respectively \cite{Ruppeiner:2013yca}. }
    \label{fig:2}
\end{figure}

An important point that should be mentioned is that $\mathcal{R}$
is a signed quantity and various calculations reveal that the sign
of $\mathcal{R}$ corresponds to whether interactions are
effectively attractive or repulsive (The sign of $\mathcal{R}$ is
subject to convention and we use Weinberg's sign convention
\cite{Weinberg1972}). For fluid and solid systems, an overall
pattern is that $ \mathcal{R} $ is negative for systems where
attractive interparticle interactions dominate, and positive where
repulsive interactions dominate (see Fig. \ref{fig:2} for a better
understanding). For systems with no statistical mechanical
interactions (e.g. an ideal gas), the scalar curvature vanishes,
and consequently, the geometry of the associated two-dimensional
space is flat. Therefore, the sign of $ \mathcal{R} $ alone
provides direct information about the properties of the
interaction among particles. For this reason, the thermodynamic
length is considered a measure of statistical mechanical
interactions within a thermodynamic system. These results have
been reviewed in \cite{Ruppeiner1995,jjk03,jan04}.
\par

Here, we try to find the relation between the scalar curvature and
phase transitions. To this end we take a system whose fluctuating
coordinates are $T$ and some extensive quantities $\mathcal{E}$
and examine the scalar curvature concerning the associated metric.
Therefore, we work with metric \eqref{thermo length4} and we
consider the metric is two-dimensional and diagonal, for
simplicity,
\begin{equation}
d \ell^2= \frac{1}{k_{B}T}\frac{\partial s}{\partial T}dT^2+\frac{1}{k_{B}T}\frac{\partial \mu}{\partial \mathcal{E}}d \mathcal{E}^2.
\end{equation}
Since the heat capacity at constant $\mathcal{E}$ is
\begin{equation}
C_{\mathcal{E}}=T\left(\frac{\partial S}{\partial T}\right)_{\mathcal{E}},
\end{equation}
the line element can be expressed as
\begin{equation}
d \ell^{2}=\frac{C_{\mathcal{E}}}{k_BT^{2}}dT^{2}+\frac{(\partial_{\mathcal{E}}\mu)_{T}}{k_BT}d\mathcal{E}^{2},\label{xxy}
\end{equation}
and the scalar curvature is easily obtained as follows
\begin{eqnarray}
\mathcal{R}&=&\frac{1}{2C_{\mathcal{E}}^{2}(\partial_{\mathcal{E}}\mu)^{2}}
\bigg\{
T(\partial_{\mathcal{E}}\mu)\bigg[(\partial_{\mathcal{E}}C_{\mathcal{E}})^{2}+(\partial_{T}C_{\mathcal{E}})(\partial_{\mathcal{E}}\mu-T\partial_{T,\mathcal{E}}\mu)\bigg]\nonumber\\
&+&C_{\mathcal{E}}\bigg[(\partial_{\mathcal{E}}\mu)^{2}+T\left((\partial_{\mathcal{E}}C_{\mathcal{E}})(\partial_{\mathcal{E},\mathcal{E}}\mu)-T(\partial_{T,\mathcal{E}}\mu)^{2}\right)
+2T(\partial_{\mathcal{E}}\mu)(-(\partial_{\mathcal{E},\mathcal{E}}C_{\mathcal{E}})+T(\partial_{T,T,\mathcal{E}}\mu))\bigg]
\bigg\}. \nonumber\\\label{RR}
\end{eqnarray}
This relation represents that $\mathcal{R}$ can potentially
diverge at $(\partial_{\mathcal{E}}\mu)_{T} =0$, which is the
critical point related to a Van der Waals phase transition, or
$C_{\mathcal{E}}=0$. Hence, the scalar curvature of the Ruppeiner
geometry experiences divergent behaviour at the critical point of
the phase transition. This property provides a possible connection
between the scalar curvature and the correlation length, which
goes to infinity at the critical point according to phase
transition theory.


Before concluding this section, it is important to mention the
analysis of several studies conducted on this subject.

$\bullet$ In a captivating research conducted by Mirza et al.
\cite{Mirza1}, they investigated the thermodynamic curvature of
the Ising model on a kagome lattice under the influence of an
external magnetic field. The findings of the study demonstrated
that the curvature displays a singularity at the critical point,
consistent with the Fisher expression. Furthermore, it was
observed that the scalar curvature in this model deviates from the
conventional scaling $R$. \par

$\bullet$ In Ref. \cite{Mirza2}, the authors investigated the
thermodynamic geometry of an ideal $q-$deformed boson and fermion
gas. They observed that the statistical interaction between
$q-$deformed bosons is attractive, while for $q-$deformed fermions
it is repulsive. Additionally, they found that the $q-$deformed
fermion gas may be more stable than the $q-$deformed boson gas
regardless of the deformation parameter's value. The authors'
analysis suggests that a large value of the deformation parameter
($q<1$) makes the $q-$deformed boson gas more stable compared a
small value, indicating that an ideal boson gas is more stable
than its deformed counterpart. Furthermore, they examined the
singular point of the thermodynamic curvature and discovered new
insights into the condensation of $q-$deformed bosons. They
concluded that the singular point coincides with the critical
point, where phase transitions like Bose-Einstein condensation
occur. \par

$\bullet$ The thermodynamic geometry of an ideal gas, where
particles follow non-Abelian statistics, was investigated in Ref.
\cite{Mirza3}. It was discovered that the behavior of the gas is
dependent on the statistical parameter $\alpha$. When $\alpha$
exceeds $\frac{1}{4}$, condensation does not occur. However, for
$0 \leq \alpha \leq \frac{1}{4}$, the non-Abelian statistics
resemble bosonic behavior and do not exhibit exclusion properties.
At a specific fractional parameter value and given fugacity
values, the thermodynamic curvature exhibits a singularity,
indicating the occurrence of condensation. This observation
suggests the presence of Bose-Einstein condensation in non-Abelian
statistics. \par

$\bullet$ The study on the thermodynamic geometry of fractional
exclusion statistics has been expanded to higher dimensions and
other types of fractional statistics, such as ideal Haldane,
Gentile, and Polychronakos fractional exclusion statistics gas
\cite{Mirza4}. The findings indicate that gases become less stable
in higher dimensions. In the case of an ideal gas following
Gentile statistics, the thermodynamic curvature is predominantly
positive in the classical limit, reflecting the attractive
statistical interaction among Gentileons. However, beyond the
classical limit, particles obeying Haldane fractional statistics
and Gentile statistics exhibit a Fermi surface, indicating that
the repulsive statistical interaction becomes dominant at low
temperatures. While the definitions of Haldane and Polychronakos
fractional exclusion statistics coincide in the classical limit,
their differences become apparent as we move further away from
this limit. In the low-temperature regime, the exclusion property
associated with particles obeying Hald statistics takes
precedence, implying that Bose-Einstein condensation does not
occur for particles governed by Haldane fractional statistics.\par

$\bullet$ A thorough examination of the non-perturbative
thermodynamic curvature of a two-dimensional ideal anyon gas was
conducted in Ref. \cite{Mirza5}. The study revealed that, under
low temperature conditions and with a constant particle count, the
thermodynamic curvature converges to that of a fermion gas. This
implies that, at $T = 0$, particles with general exclusion
statistics display a Fermi surface.

$\bullet$ In Ref. \cite{Mirza6}, the thermodynamic curvature of
the anyon gas \footnote{The concept of "anyons" or particles with
fractional statistics in two-dimensional systems has applications
in the theory of fractional quantum Hall effect \cite{Halperin;1984}.} was investigated using the Ruppeiner approach in
both classical and quantum limits. The study explored the concept
of "anyons," which refers to particles with fractional statistics
in two-dimensional systems, and its relevance to the theory of
fractional quantum Hall effect. The results revealed that the
behavior of the ideal anyonic gas in the classical limit depends on
the value of the statistical parameter $\alpha$. When
$\alpha<\frac{1}{2}$ ($\alpha>\frac{1}{2}$), the scalar
thermodynamic curvature is positive (negative), indicating an
attractive (repulsive) statistical interaction. Consequently, the
ideal anyonic gas is referred to as "Bose-like" ("Fermi-like") for
$\alpha<\frac{1}{2}$ ($\alpha>\frac{1}{2}$). In the case of
$\alpha=\frac{1}{2}$, the equation of state resembles that of an
ideal classical gas, resulting in a thermodynamic curvature of
zero. In the quantum limit, the zero point of the thermodynamic
curvature shifts from $\alpha=\frac{1}{2}$ to lower values,
suggesting that quantum corrections alter the value of $\alpha$.
This implies that the anyon gas behaves as a free non-interacting
gas when quantum corrections are applied.

\section{Application to black holes } \label{5}

This section aims to review the application of thermodynamic
geometries to black hole issues.

Understanding the meaning of the event horizon of black holes
helps us gain a better understanding of black hole physics and
its connection to thermodynamics. Since gravity always attracts,
self-gravitating systems tend to grow rather than shrink. As we
know, nothing - not even light - can escape the event horizon of a
black hole, making the horizon akin to a one-way asymmetric
surface: objects can enter but not exit, leading to an increase in
the surface area. This aligns with the second law of
thermodynamics, which states that the total entropy of a system
will increase in a physical process and never decrease. In fact,
there is an asymmetric tendency for a one-way increase in entropy,
since the area of the event horizon is proportional to entropy.
Thus, the size of a black hole cannot decrease in any process.

The properties of a black hole are independent of the details of
the collapsing matter, and this universality arises from the fact
that black holes can be the thermodynamic limit of the underlying
quantum gravitational degrees of freedom. This suggests that the
classical and semi-classical features of black holes can provide
important information about the nature of quantum gravity. The
lack of experimental or observational results presents a major
challenge in constructing a theory of quantum gravity. Therefore,
understanding the fundamental laws of black hole mechanics may
serve as a necessary (if not sufficient) constraint on the theory
of quantum gravity.

\begin{figure}[!htb]
    \centering
    {\includegraphics[width=0.6\textwidth]{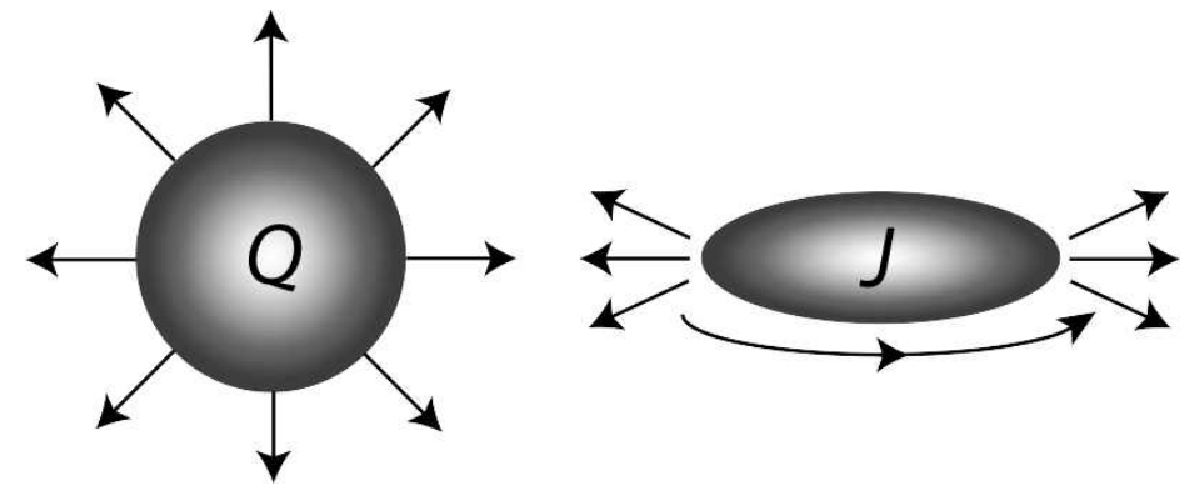}}
    \newline
    \caption{ The left graph shows a charged extremal black hole with a high charge Q which is on the verge of exploding under its own electrostatic repulsion. The graph on the right displays a rotating black hole with a very large angular momentum J on the verge of pulling apart \cite{Ruppeiner:2013yca}.}
    \label{FigQJ}
\end{figure}

The classical (non-quantum) characteristics of black holes go back
to the solution of Einstein's field equations by Schwarzschild
\cite{Misner:1973}. This solution is obtained by assuming a
static, uncharged, spherically symmetric point mass M, located at
a central singularity. Einstein's field equations can also be
solved by adding charge $Q$ (the Reissner Nordstr\"{o}m solution),
angular momentum $J$ (the Kerr solution), or with all three
quantities $(M, J, Q)$ (the Kerr-Newman solution). If the
collapsing matter is sufficiently dense, we inevitably approach
one of these solutions. A question arises: Is it possible to add
enough charge to a black hole to make it explode outward under its
own electrostatic repulsion? Or can enough angular momentum be
added to a black hole so that it tears apart under its own spin,
as in Fig. \ref{FigQJ}? According to cosmic censorship, both
scenarios - or any combination of them - are forbidden. Black
holes are considered extremal if they are as close as possible to
these mechanical limits.

A principle often cited in black hole physics is the "no-hair
theorem" \cite{Misner:1973}, which states that stationary
solutions in GR can be fully described by their
mass, spin, and electric charge. Other properties, such as
geometry and magnetic moment, are uniquely determined by these
three parameters, and all other information - referred to as
"hair" - about the matter that formed or falls into a black hole
disappears behind its event horizon, becoming permanently
inaccessible to external observers. After collapsing matter forms
a black hole, there is a short period of settling during which
information related to the black hole's creation disappears.
Remarkably, the final equilibrium state of a black hole depends
solely on mass, electric charge, and angular momentum. This
reduction in complexity is essential for black hole
thermodynamics.

The no-hair conjecture rules out any possibility of a distribution
of black hole equilibrium microstates. The distribution of
microstates is necessary for statistical mechanics and
thermodynamic fluctuations \cite{Lewis:1931}.

\subsection{Black Hole thermodynamics}\label{5-1}

When a system is in thermal equilibrium, it is possible to predict
its behavior by determining only a handful of known properties
such as internal energy, entropy, temperature, volume, angular
momentum, total mass, etc. Such systems which is in interaction with the
outside environment follow four laws, which are known as
thermodynamic laws \cite{Adamson:edu}:

 \begin{itemize}
    \item{{\textbf{Zeroth Law:}} the temperature T is constant for a system in thermal
        equilibrium.}
    \item{{\textbf{First Law:}} when an infinite amount of work W is performed  on a
        system at temperature T, the variation of the entropy S and energy
        E of the system is given by
        \begin{equation}
        \delta E=T \delta S+\delta W.
        \end{equation}}
    \item{{\textbf{Second Law:}} the entropy of an isolated system can never decrease
        over time.
        }
        \item{{\textbf{Third Law:}} it is not possible to cool a system to an absolute
            zero temperature \cite{Kittel:1980}.}
 \end{itemize}

Studies in black hole mechanics have shown that a black hole
behaves like a thermodynamic system, with the area of the event
horizon as the entropy and a geometric quantity called surface
gravity as the black hole's temperature. Black hole thermodynamics
appears when quantum fields propagate on a background black hole
spacetime. Thermodynamics of black holes suggest that there exists
an underlying microstructure of spacetime.

According to Hawking's area theorem the total surface area of the
event horizon of the black hole never decreases in any
"classical" processes \cite{Hawking;1972hk}. But a more physical
derivation of the theorem occurs by considering a quasistationary
process in which the second law can be shown to hold as a
consequence of the first law and the null energy condition. In
such a process, positive energy matter can be created by a black
hole independent of the details of the gravity action. From what
was expressed, energy can flow not only into black holes but also
out of them. In other words, a black hole can act as an
intermediary in energy exchange processes. For the unchanged
horizon area, energy extraction is maximally efficient, but
processes during which the horizon area increases are
irreversible. The analogy with thermodynamic behavior was noted
when Bekenstein employed the area theorem and proposed that the
area of the black hole is indeed related to the thermodynamic
entropy of the event horizon \cite{Bekenstein:1973}
\begin{equation}
S\equiv \frac{1}{4\hbar G}\times \text{Area of horizon}.
\end{equation}

This analogy was vigorously pursued as soon as it was recognized
at the beginning of the 1970s, although there were  several
obvious flaws at first \cite{Jacobson:2mn}:

a) the black hole temperature vanishes.

b) entropy is dimensionless, whereas horizon area is a length
squared.

c) the area of every black hole is separately non-decreasing,
while only the total entropy is nondecreasing in thermodynamics.

By 1975, it was understood that these flaws in black hole physics
could be addressed by incorporating quantum theory. As we know, the thermodynamics of black holes is based on applying quantum field theory in curved spacetime \cite{Hawking:1975vcx,Bardeen:1973abc,Gibbons:1977ab}.  The nature of black hole radiation is strongly related to the quantum gravity effects and pair creations. This suggests the black hole is a quantum mechanical system like any other quantum system.
In some respects, a black hole plays in gravitation the same role as an atom plays in quantum mechanics \cite{Bekenstein:92111}. This analogy suggests that contrary to classical general relativity in which the mass spectrum of black holes is a continuum, in quantum theory the black hole mass spectrum must be discrete. It means that the black hole horizon area has a discrete eigenvalue spectrum and is in fact quantized \cite{Hod:1998,Hod:1999,Bekenstein:124052}. The quantization of black holes was first proposed by Bekenstein in the early 1970s. According to his opinion, the non-extremal BH horizon area behaves as a classical adiabatic invariant and therefore it should exemplify a discrete eigenvalue spectrum with quantum transitions \cite{Hod:1998,Hod:1999}. This strongly suggests the quantum mechanical properties of black holes in different physical aspects. When a black hole captures or releases a massive point particle, its mass increases or decreases, respectively, which directly affects its horizon area. For an uncharged particle absorption process, the BH horizon area increase is as \cite{Hod:1998,Hod:1999}
\begin{equation}
\Delta A=8\pi \mu b,
\end{equation}
where $ \Delta A $ is the BH horizon area change. $  \mu$ and $ b $ are, respectively, the particle rest mass and particle finite proper radius. According to the uncertainty principle, the particle cannot be localized to better than a Compton wavelength. Therefore, the radial position for the particle's center of mass is subject to an uncertainty of $ b\geq \xi\hslash/ \mu$ \cite{Bekenstein:92111}. Hence, the smallest possible increase in
horizon area is \cite{Hod:1998,Hod:1999,Sakalli:2015,Maggiore:2008}
\begin{equation}
(\Delta A)_{min}=8\pi\xi\hslash=\alpha l_{p}^{2},
\end{equation}
where $ \xi$ is a number of order unity and $ l_{p}=\sqrt{G\hslash}/c^{3} $ is the Planck length in gravitational units $G=c=1$. For the BH's charged particle absorption process, the BH horizon area increase is defined as \cite{Hod:1998,Hod:1999}
\begin{equation}
\Delta A=4 l_{p}^{2}.
\end{equation}

From what was expressed, the quantization of horizon area in equal steps brings to mind a horizon formed by patches of equal area $ \alpha l_{p}^{2} $ which get added one at a time. In quantum theory, degrees of freedom independently reveal distinct states. Since the patches are all equivalent, each has the same number of quantum states, say $ k $. Therefore, the total number of quantum states on the horizon is \cite{Bekenstein:92111}
\begin{equation}
N=k^{A/\alpha l_{p}^{2}}.
\end{equation}

The statistical (Boltzmann) entropy associated with the horizon is $ln N$ or \cite{Bekenstein:92111}
\begin{equation}
S_{BH}=\frac{A ln k}{\alpha l_{p}^{2}}.
\end{equation}

Comparing the above equation with Hawking's coefficient in the black hole entropy calibrates the constant $ \alpha $ \cite{Bekenstein:124052}:
\begin{equation}
\alpha=4 ln k.
\end{equation}

In the present work, we would like to investigate the classical
aspects of the theory. Since these form the basis of quantum
black hole thermodynamics, their study is crucial. However, it is
also interesting to understand what can be inferred without
invoking quantum theory, as this may provide insight into the
deeper origins of gravity.

\subsection{The four laws of black hole mechanics}\label{5-2}

The surprising similarities between the geometrical properties of
black holes and classical thermodynamic variables suggest an
in-depth connection between the laws of black hole mechanics and
the laws of thermodynamics. Employing such a connection, one is
able to identify the geometrical features of black holes with
their thermodynamic counterparts. In fact, the four laws of black
hole mechanics are similar to the four laws of thermodynamics if
one considers a correspondence between the temperature $ T $ and
surface gravity $ \kappa $ of the black holes and also between
entropy S and the black hole area $ A $. Before formulating the
four laws of black hole mechanics, Bekenstein suggested that the
entropy of a black hole is proportional to the horizon area $ A $
($S=\eta A$).  Although he could not find the value of $  \eta$
exactly, he provided heuristic arguments to conjecture its value
to be $ \frac{\ln 2}{8\pi} $ \cite{Bekenstein:3292}. By discovering Hawking radiation and
the fact that the radiation has a thermal spectrum, Hawking found
that his remarkable discovery did make Bekenstein's idea
consistent, of a finite entropy proportional to black hole area,
though not conjectured value $\eta= \frac{\ln 2}{8} $, but $
\eta=\frac{1}{4} $ \cite{Page;2032005}. In the following, we review the four laws of black hole mechanics:
\subsubsection{Surface gravity and the zeroth law of black hole mechanics}

To explain the zeroth law of black hole mechanics, we need to know
the notions of stationary, static, and axisymmetric black holes
and also the notion of a Killing horizon. A black hole in
asymptotically flat spacetime $(M,g_{ab})$ is stationary if there
is a one-parameter group of isometries on $(M,g_{ab})$ generated
by a Killing vector field $ \chi^{a} $ which is unit timelike at
infinity \cite{Wald;62001}. A  stationary black hole is said to be
static if the Killing field $ \chi^{a}$ is hypersurface
orthogonal. The black hole is called axisymmetric if one can find
a one-parameter group of isometries that correspond to rotations
at infinity. A black hole is stationary-axisymmetric if it has
"$t-\varphi$ orthogonality property"\footnote{A black hole has the
"$t-\varphi$ orthogonality property" if the 2-planes defined by $
\chi^{a}$ and the rotational Killing field $ \varphi^{a} $ are
orthogonal to a family of 2-dimensional surfaces.}.  This property
holds for all stationary-axisymmetric BH solutions to the vacuum
Einstein or Einstein-Maxwell equations \cite{Heusler:1996}.

A null hypersurface $ \mathcal{K} $ is called Killing horizon if
Killing vector field $ \chi^{a} $ is normal to it  i.e. $ \chi^{a}
\chi_{a}=0 $. For stationary black holes, event horizon $
\mathcal{H} $ is a Killing horizon. For a static black hole, the
static Killing field $ t^{a} $ is normal to the horizon, while for
a stationary-axisymmetric black hole with the "$t-\varphi$
orthogonality property", Killing vector field $ \chi^{a} $ is
defined as $ \chi^{a}=t^{a}+\Omega \varphi^{a} $ which becomes
null on the $ \mathcal{H} $ \cite{Carter:1973a}. Here the constant $ \Omega $ is the
angular velocity of the horizon and $ \varphi^{a} $ denotes the
Killing vector associated with the axial symmetry. Since $
\nabla^{a}(\chi^{b}\chi_{b}) $ is normal to $ \mathcal{K} $, the
vector fields must be proportional at every point on $ \mathcal{K}
$. Therefore, one can introduce a function, $ \kappa $ known as the
surface gravity defined as \cite{Jacobson:2mn}
\begin{equation}
\kappa^{2}=-\frac{1}{2}(\nabla^{a}\chi^{b})(\nabla_{a}\chi_{b}).
\end{equation}

Here, one can provide an equivalent definition of the surface
gravity such that $ \kappa $ is the magnitude of the acceleration,
with respect to Killing time, of a stationary zero angular
momentum particle just outside the horizon \cite{Jacobson:2mn}. In fact, it is the
same force per unit mass that should be employed on a particle at
infinity to continue its movement on a path. Although $ \kappa $
is defined locally on the horizon, it turns out that it is always
constant over the entire event horizon of a stationary black hole.
Since the surface gravity is proportional to the temperature of
black holes, this resembles the zeroth law of thermodynamics which
states that temperature $T$ is constant throughout a system in
thermal equilibrium \cite{Bardeen:1973abc}. It is worth mentioning that there are two
independent versions of the zeroth law of black hole mechanics.
The first one was proposed by Carter, stating that the surface
gravity $ \kappa $, has to be constant over its event horizon $
\mathcal{H} $ for all static black holes or
stationary-axisymmetric with the $t-\varphi$ orthogonality
property. This result is purely geometrical without using field
equations.  The second one is related to Bardeen, Carter, and
Hawking's idea which expresses that if Einstein's equation holds
with the matter stress-energy tensor satisfying the dominant
energy condition, then $ \kappa $ is uniform on any Killing
horizon. Evidently, in the second version, the existence of the
$t-\varphi$ orthogonality property is omitted, instead, the field
equations of GR are used. A property of the surface gravity is
that it illustrates a proportionality factor between the affinely
parametrized null geodesics that generate the event horizon and
the Killing vector field parametrization. Then each component of
the affinely parametrized null geodesic generators $ k^{a} $ is
defined as $ \lambda k^{a}=\frac{1}{\kappa} \chi^{a}$, where $
\lambda $ is the affine parameter. It should be noted that near
the event horizon $ k^{a} $ will locally be parallel to the
Killing vector field $ \chi^{a} $, but that the two vector fields
will have different parametrizations \cite{Adamson:edu}.

\subsubsection{First law of black hole mechanics}

It would be interesting to study how the event horizon area is
related to the other properties of a stationary black hole such as
mass, angular momentum, and surface gravity. Also, it is very
important to investigate how a small change in one of these
properties will cause changes in the others. To do so, we first
introduce the Raychaudhuri equation, defined as \cite{Adamson:edu}
\begin{equation}
\frac{d\theta}{d\lambda}=-\frac{1}{2}\theta^{2}-\sigma_{ab}\sigma^{ab}+\omega_{ab}\omega^{ab}-R_{ab}k^{a}k^{b},
\label{dtheta}
\end{equation}
in which $ \theta $ is expansion/contraction of volume which is
given by the divergence of $ k^{a} $ defined as $
\theta=\nabla_{a}k^{a} $ and $ \sigma_{ab} $ is a shear tensor
that describes distortions in shape with no change in volume and
is represented by a symmetric tensor which is trace free as
\begin{equation}
\sigma_{ab}=\theta_{ab}-\frac{1}{3}\theta h_{ab},
\end{equation}
in which $ h_{ab}=g_{ab}-k_{a}k_{b}$. The vorticity tensor $
\omega_{ab} $, describing the rotation of an area element of the
null hypersurface is given by
\begin{equation}
\omega_{ab}=k_{[a;b]}-\dot{k}_{[_{a}k_{b}]},
\end{equation}
where $ k_{[a;b]}=\frac{1}{2} (k_{a;b}-k_{b;a})$. Using the
expression for the Ricci tensor contained in the Einstein
equation, the last term of Eq. (\ref{dtheta}) can be written as
\begin{equation}
R_{ab}k^{a}k^{b}=8\pi \left[ T_{ab}k^{a}k^{b}+\frac{1}{2} \right].
\end{equation}

Since the quantity $ T_{ab}k^{a}k^{b} $ is the local energy
density as measured by an observer with 4-velocity $ k^{a} $, it
is expected that this quantity is non-negative everywhere for
time-like and null $ k^{a} $. Although internal stresses in a
matter can have a negative contribution to $  T$, for a physically
reasonable matter, this contribution is always much smaller than
the mass and momentum terms which contribute positively to $
T_{ab}k^{a}k^{b} $. So, one can assume that $ T_{ab}k^{a}k^{b}
\geq -\frac{T}{2}$. This reveals the fact that $
R_{ab}k^{a}k^{b}\geq 0 $. The terms $ \sigma_{ab}\sigma^{ab}$ and
$ \omega_{ab}\omega^{ab}  $ are also greater than or equal to
zero. If a small amount of matter drops into a black hole, the
local value of $ T_{ab} $ near the black hole surface changes as $
\delta T_{ab} $. The resulting change in the black hole area is
governed by Eq. (\ref{dtheta}). Since changes $\theta^{2}$,
$\sigma_{ab}\sigma^{ab}$, and $\omega_{ab}\omega^{ab}$ can be
neglected to first order in $ \delta T_{ab} $, the Raychaudhuri
equation is simplified as
\begin{equation}
\frac{d\theta}{d\lambda}=-8\pi \delta T_{ab}k^{a}k^{b}.
\end{equation}

As was mentioned in the discussion of surface gravity, $
k^{a}=\frac{1}{\kappa \lambda} \chi^{a}=\frac{1}{\kappa
\lambda}\left( t^{a}+\Omega \varphi^{a} \right) $. Here,
evaluating the change $ \delta T_{ab} $ on the black hole is a
matter of integrating both sides of the Raychaudhuri equation over
the black hole event horizon surface and overall future values of
$ \lambda$ \cite{Adamson:edu}
\begin{eqnarray}
\label{integral} \kappa \int d^{2}S \int_{0} ^{\infty} \lambda
\frac{d\theta}{d\lambda}&=&-8\pi \int_{0} ^{\infty} \int d^{2}S
\delta T_{ab}k^{a}k^{b} \\ \nonumber \kappa \int
d^{2}S(\theta\lambda)\bigg\vert_{0} ^{\infty}-\kappa \int d^{2}S
\int_{0} ^{\infty}\theta d\lambda &=& -8\pi \left( \int_{0}
^{\infty} d\lambda \int d^{2}S \delta T_{ab}t^{a}k^{b}+\Omega
\int_{0} ^{\infty} dV \int d^{2}S \delta
T_{ab}\varphi^{a}k^{b}\right).
\end{eqnarray}

Now we need to evaluate these integrals exactly. The boundary term
on the left-hand side can be taken to be zero by taking the
surface out to a region of spacetime where expansion is
insignificant. Also, the second term is the integral of the
expansion of each infinitesimal area element of the event horizon
over the surface of the event horizon. This is the same as the
infinitesimally small change in event horizon surface area $
\delta A $ which is caused by $ \delta T_{ab} $. On the other
hand, the first integral on the right is an integral of the
$T_{00}$ component, which is called the mass $M$ in special
relativity (it comes from the fact that $t^{a}$ and $k^{a}$ both
point forward in time). In the second term of the right side, the
term $ T_{ab}\varphi^{a}k^{a} $ is a projection onto the time-$
\varphi $ component of $ T_{ab} $, which is the negative of the
angular momentum $  J$. Thus, for a far distant observer where space
is approximately flat, Eq. (\ref{integral}) can be written as
\begin{equation}
\kappa \delta A=8\pi(\delta M-\Omega \delta J). \label{first-law}
\end{equation}

Eq. (\ref{first-law}) is called the first law of black hole
mechanics which is analogous to the first law of thermodynamics
defined as
\begin{equation}
T \delta S=\delta E+\delta W \label{first-law1}.
\end{equation}

It is worth mentioning that the zeroth and first laws of black
hole mechanics are concerned with equilibrium or quasi-equilibrium
processes. That is, they concern stationary black holes, or
adiabatic changes from one stationary black hole to another.

\subsubsection{Second law of black hole mechanics}
The second law of black hole mechanics is Hawking’s area theorem
\cite{Hawking;1972hk}, which based on the horizon area $ A $ of a
black hole can never decrease assuming Cosmic Censorship and a
positive energy condition. The Cosmic Censor Conjecture states
that a complete gravitational collapse of a body results in a
singularity, covered by an event horizon \cite{Ohanian:923}. In
other words, every singularity must possess an event horizon that
hides the singularity from view.  Assuming the Weak energy
condition and the Cosmic Censor Conjecture, it has been found
\cite{Bardeen:170161} that the area of a future event horizon
never decreases in asymptotically flat spacetime. The second law
of black hole mechanics is similar to the second law of
thermodynamics, stating that the entropy $ S $ of a closed system
cannot decrease.

Now, we explore the relation of Hawking's area theorem to the
second law of thermodynamics. According to the area theorem, if
the spacetime on and outside the future event horizon is a regular
predictable space, and for arbitrary null $k^{a}$, the null energy
condition, $ T_{ab} k^{a} k^{b}$ is satisfied by the stress
tensor, then the area of spatial cross-sections of the event
horizon is non-decreasing. As was mentioned in the discussion of
the first law, the Raychaudhuri equation is defined by Eq.
(\ref{dtheta}). Also, it was discussed that $ R_{ab}k^{a}k^{b} $,
$ \sigma_{ab}\sigma^{ab}$ and $ \omega_{ab}\omega^{ab}  $ are
greater than or equal to zero. These constraints help us to arrive
at an important restriction on the expansion of geodesics, namely
\begin{equation}
\frac{d \theta}{d\lambda}\leq
-\frac{1}{2}\theta^{2}~~~~\Longrightarrow ~~~
\int_{\theta_{0}}^{\theta}\frac{d \theta}{\theta^{2}}\leq
-\int_{0}^{\lambda}\frac{1}{2} d\lambda ~~~~ \Longrightarrow ~~~
\frac{1}{\theta (\lambda)}\geq
\frac{1}{\theta_{0}}+\frac{1}{2}\lambda \label{thetalambda}.
\end{equation}

In the above relation, if one takes $ \theta_{0} $ to be negative,
then by increasing $ \lambda $,  moving forward along the null
geodesic, a point will be reached where the right side of the
inequality would become zero. Thus, to hold the inequality, the left
side must be zero at the same point. This reveals an infinite
expansion of nearby geodesics which is highly unphysical and is in
contradiction with some of the basic properties of black holes.
Hence, the only choice is to consider positive $ \theta_{0} $, in
which case $ \theta $ will be non-negative for all $ \lambda $.
Since increasing $ \lambda $ means moving forward in time to a
time-like observer, this restriction implies that the event
horizon area observed by an observer far from the black hole
should never decrease with time. This statement is known as the
second law of black hole mechanics.

 \vspace{0.3cm}
 \noindent  $\bullet$ {\it{\textbf{Generalized second law:}}}\quad

\vspace{0.2cm} Classically, there is a serious difficulty with the
ordinary second law of thermodynamics when a black hole is
present. If a black hole radiates energy it will lose mass and, in turn, its
event horizon surface area will decrease, thereby the area
increase theorem violates. The violation of the theorem is not
unexpected since the condition $ T_{ab}\geq 0 $ is not satisfied
on the event horizon by taking into account quantum effects. Therefore, it
seems that one can have Hawking radiation or the area increase
theorem, but not both. Similarly, when a matter drops into a black
hole it disappears into a spacetime singularity. So, the entropy
initially present in the matter is lost, and no compensating gain
of ordinary entropy occurs. This means that the total entropy of
matter in the universe decreases which violates the second law of
thermodynamics. Bekenstein proposed a way out of this difficulty
by defining the generalized entropy $ S^{\prime} $ to be the sum
of the entropy outside the black hole $ S $ and the entropy of the
black hole itself \cite{Wald;62001}
\begin{equation}
S^{\prime}\equiv S+S_{bh}=S+\frac{A}{4}.
\end{equation}

In fact, this means that dropping an object into a black hole
increases its surface area and that Hawking radiation creates
thermal particles outside the black hole with finite entropy. This
is known as  the generalized second law which is the statement
that the total entropy cannot decrease:
\begin{equation}
\Delta S^{\prime}\geq 0.
\end{equation}

\subsubsection{Third law of black hole mechanics}

The third law of black hole mechanics states that it is impossible
to achieve a surface gravity of zero within a finite number of
steps \cite{Israel:1971nm}. Since otherwise the black hole would
seem unattractive to a distant observer. Moreover, it will be more
unphysical if one imagines $ \kappa $ to be negative because in
this case, the black hole would seem repulsive to an observer far
from it. This law can also be proven by calculating $ \kappa $
explicitly for the Kerr metric. The non-negative nature of surface
gravity is then guaranteed by the requirement that the solution
does not have any closed-timelike curves \cite{Wald-book}. This
law corresponds to the weaker form of the third law of
thermodynamics which states that the temperature of a real
thermodynamic system is always greater than absolute zero.
However, the classical third law of black hole mechanics is not
analogous to the stronger form of the third law of thermodynamics
which asserts that $ S\rightarrow 0 $ as $ T\rightarrow 0 $.
Although the temperature of a black hole can be zero at the
extremal limit, black holes do not always have zero area in such a
limit. Hence, the black hole entropy does not always go to zero at
zero temperature. This is an important difference between fluid
and black hole thermodynamics. The second difference is that black
hole thermodynamics is not extensive \cite{Landsberg:1980a},
meaning that one cannot increase the mass of the black hole in
such a way that all conjugate variables remain constant. The third
difference is the frequent absence of black hole thermodynamic
stability. The study of thermodynamic stability is conducted by
calculating heat capacity such that the positivity (negativity) of
heat capacity refers to a stable (unstable) state.  Negative heat
capacity is a fixture of gravitational thermodynamic problems. For
instance, the Kerr-Newman black hole thermodynamics is not stable
for any set of values of $(M, J, Q)$ \cite{Landsberg:1980b}.
However, there are black holes that can experience a stable state
e.g., BTZ black holes are thermodynamically stable for all of
their states. In some cases, the stability problem can be resolved
by restricting the number of fluctuating variables, adding an AdS
background, or altering the assumptions about the black hole's
topology. For example, Reissner-Nordstr\"{o}m and Kerr black holes
are not thermodynamically stable for any thermodynamic state.
However, in AdS spacetime both are stable for some values of black
hole parameters \cite{Banerjee:2011}.
\subsection{Thermodynamic curvature for black holes}\label{5-3}
As was already mentioned, black holes have well-defined
thermodynamic properties. Black hole thermodynamics leads
naturally to thermodynamic fluctuation theory and an information
metric
\begin{equation}
g_{\alpha\beta}=-\frac{\partial^{2}S}{\partial X^{\alpha}\partial
X^{\beta}}, \label{gab}
\end{equation}
where $ (M, J, Q,...)=(X^{1}, X^{2}, X^{3},...) $ play the role of
the conserved thermodynamic variables in black hole
thermodynamics. However, instead of $S=S(X^{1}, X^{2},
X^{3},...)$, one can use $M=M(Y^{1}, Y^{2}, Y^{3},...)$ where
$(Y^{1}, Y^{2}, Y^{3},...)=(S, J, Q,...)$. In this case, the
thermodynamic metric in the Weinhold energy form, with an
additional prefactor $ 1/T $ is obtained. The information metric
(\ref{gab}) produces the thermodynamic curvature $ \mathcal{R}$. The
concept of thermodynamic curvature of black holes was first
studied by Ferrara et al. They applied thermodynamic curvature to
calculate critical behavior in moduli spaces \cite{Ferrara:1997}.
Afterward, Cai and Cho \cite{Cai:1999} investigated a connection
between phase transitions in BTZ black holes to diverging $  \mathcal{R}$.
They also identified a correspondence with $  \mathcal{R}$ for the Takahashi
gas which suggests that an appropriate black hole statistical
model might be a system of hard rods. Then, Aman, Bengtsson, and
Pidokrajt \cite{Aman:2003} evaluated $ \mathcal{R} $ for Reissner-Nordstrom,
Kerr, and Kerr-Newman black holes with a nonzero cosmological
constant. An extensive review was also conducted in the context of
five-dimensional black holes and black rings by Arcioni and
Lozano-Tellechea \cite{Arcioni:2005} who connected phase
transitions to both diverging $ \mathcal{R} $ and diverging second
fluctuation moments. Afterward, Aman and Pidokrajt
\cite{Aman:2006} investigated thermodynamic curvature for
Reissner-Nordstrom and Kerr black holes in higher dimensions
spacetime and noticed that patterns in four dimensions continue to
higher dimensions. In Ref. \cite{Sarkar:2006}, authors also
studied thermodynamic curvature for a general class of BTZ black
holes, including quantum corrections to the entropy.

The thermodynamic curvature of black holes has been studied with
various thermodynamic criteria, each of which has been associated
with success and failure. We mentioned some of them here.

\subsubsection{Weinhold metric}
In black hole thermodynamics, as well as in the Weinhold method,
the mass of black holes can be considered as a thermodynamic
potential with appropriate extensive parameters, such as entropy $S$,
electric charge $Q$, and angular momentum $J$, along with their related
intensive quantities - temperature, electric potential, and
angular velocity.

For example, for a static charged black hole the denominator of
Weinhold Ricci scalar is obtained as
\begin{equation}
denom(\mathcal{R}_{W})=\left( M_{SS}M_{QQ}-M_{SQ}^{2}\right)
^{2}M^{2}\left( S,Q\right) ,  \label{denWein}
\end{equation}
where $M_{QQ}=\left( \frac{\partial ^{2}M }{\partial Q^{2}}
\right)_{S} $ and $M_{SQ}=\frac{\partial ^{2}M}{\partial S\partial
    Q} $. As we know, the heat capacity is given by
\begin{equation}
C_{Q}=\frac{M_{S}}{M_{SS}},  \label{heat}
\end{equation}%
where $M_{S}=\left( \frac{\partial M}{\partial S}\right) _{Q}$ and
$M_{SS}=\left( \frac{\partial ^{2}M}{\partial S^{2}}\right) _{Q}$
are regular functions. For consistency with the heat capacity
results, the roots of the Eq. (\ref{denWein}) should coincide with
divergencies of the heat capacity. As can be seen from Eq.
(\ref{denWein}), only in special case $M_{SQ}=0$ and nonzero
$M_{QQ}$, the divergence points of the heat capacity coincide with
divergencies of the Weinhold Ricci scalar. It is evident that for
the case of $M_{SS}=\frac{M_{SQ}^{2}}{M_{QQ}}$, extra divergencies
can be found for $\mathcal{R}_{W}$ which are not related to any
phase transition points of the heat capacity. Therefore, the
structure of this part of the denominator can provide extra
divergencies that do not correspond to any phase transition points
of the heat capacity.
A curious observation is that the Weinhold geometry of the Kerr
black holes is actually flat.

\subsubsection{Ruppeiner geometry}

Black hole thermodynamics exhibits unfamiliar features which are
generic to systems with long-range interactions and
self-gravitating systems, making Ruppeiner's arguments not
directly applicable to black holes. The first issue encountered is
negative specific heats, resulting in non-concave entropy
functions. The second issue is the absence of extensive variables,
causing the Ruppeiner metric not to be positive definite or have
any null eigenvectors. In other words, the canonical ensemble does
not exist, and choosing a physical dimension for the metric
becomes difficult. In spite of these challenges, it is believed that
the Ruppeiner geometry of black holes can provide interesting
insights in black hole physics. For further observations on the
role of the Ruppeiner metric in black hole physics, see Ref.
\cite{Ferrara:1997}. For background information on black hole
thermodynamics, see \cite{Davies;1977}.

Based on investigations conducted in the context of the
thermodynamic geometry of black holes, the Weinhold metric is
proportional to the metric on the moduli space for supersymmetric
extremal black holes with zero Hawking temperature, while the
Ruppeiner metric governing fluctuations naively diverges,
consistent with the argument that the thermodynamic description
breaks down near the extreme \cite{Preskill;1991,Kallosh;1992}.
The Ruppeiner approach has been used for RN black holes, BTZ black
holes, Kerr black holes, and Kerr-Newman black holes. It was found
that when considering the entropy as a function of mass and other
extensive variables, the Ruppeiner metric for RN and BTZ black
holes is always flat, with zero scalar curvature. In contrast, for
Kerr and Kerr-Newman black holes, the Ruppeiner metric is curved,
and the scalar curvature diverges at the extremal limit. This
reveals that the Ruppeiner metric for RN and BTZ black holes is
quite different from that of Kerr black holes. A remarkable point
is that in anti-de Sitter spacetime, the curvature scalar of the
Ruppeiner metric diverges for RN black holes. Here, we try to
review this case in more detail based on Ref. \cite{Aman:2003}.

The thermodynamics of these black holes is now defined by the
fundamental relation

\begin{equation}
M = \frac{\sqrt{S}}{2}\left( 1 + \frac{Q^2}{S}\right) \ .
\end{equation}
The Hawking temperature is
\begin{equation}
T  = \frac{1}{4\sqrt{S}}\left( 1 - \frac{Q^2}{S}\right).
\end{equation}
The extremal limit, beyond which the singularity becomes naked,
happens at
\begin{equation}
Q^2 = M^2 \hspace{5mm} \Leftrightarrow \hspace{5mm} \frac{Q^2}{S}
= 1 \ .
\end{equation}
In its natural coordinates the Weinhold metric becomes

\begin{equation} ds^2_W = \frac{1}{8S^{\frac{3}{2}}}\left( - \left(1 -
\frac{3Q^2}{S}\right)dS^2 - 8QdQdS + 8SdQ^2\right) \ .
\end{equation}

\begin{figure}[!htb]
    \centering
    {\includegraphics[width=0.50\textwidth]{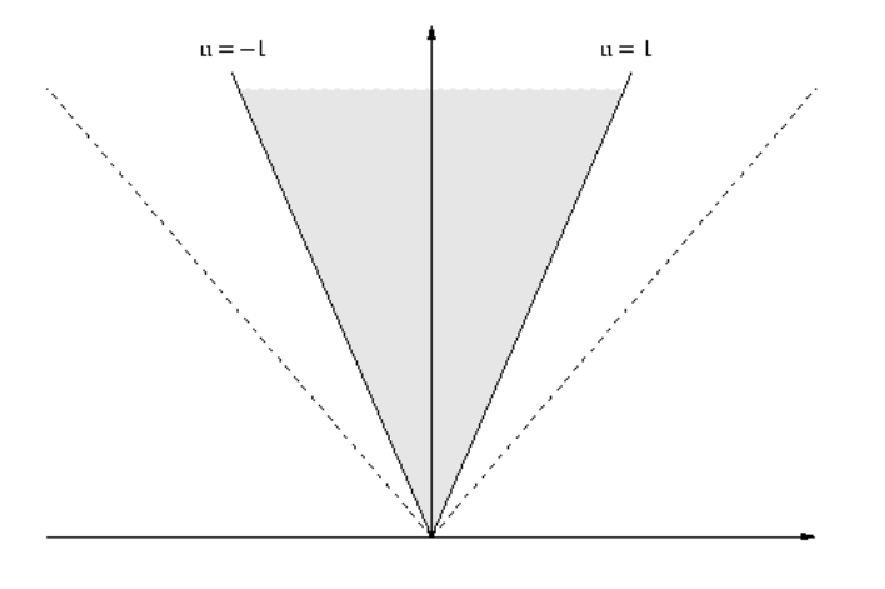}}
    \newline
    \caption{ The state space of the Reissner-Nordstr\"{o}m black holes shown as a wedge in a flat Minkowski space \cite{Aman:2003}.}
    \label{Fig2}
\end{figure}

Evidently, the component $g^W_{SS}$ vanishes and changes sign at $
\frac{Q^2}{S} = \frac{1}{3} $, indicating that the specific heat
$C_Q$ diverges and changes sign there. Introducing the new
coordinate
\begin{equation}
u = \frac{Q}{\sqrt{S}} \ ; \hspace{8mm} - 1 \leq u \leq 1 \ .
\end{equation}
the Weinhold metric has a diagonal form as
\begin{equation}
ds^2_W = \frac{1}{8S^{\frac{3}{2}}}\left( - (1-u^2)dS^2 +
8S^2du^2\right) \ .
\end{equation}
In these coordinates, the Ruppeiner metric is given by
\begin{equation}
ds^2 = \frac{1}{T}ds^2_W = - \frac{dS^2}{2S} +
4S\frac{du^2}{1-u^2} \ .
\end{equation}
One can also introduce new coordinates as
\begin{equation}
{\tau} = \sqrt{2S} \hspace{8mm} \sin{\frac{\sigma}{\sqrt{2}}} = u
\ ,
\end{equation}
which by using, the Ruppeiner metric takes the following form
\begin{equation}
ds^2 = - d{\tau}^2 + {\tau}^2d{\sigma}^2 \ ,
\end{equation}

which is flat and is recognizable as a timelike wedge in Minkowski
space when described by Rindler coordinates. The state space of
the Reissner-Nordstr\"{o}m black holes is depicted in Fig.
\ref{Fig2}.

In the presence of cosmological constant, the mass is given by
\begin{equation}
M = \frac{\sqrt{S}}{2}\left( 1 + \frac{S}{l^2} + \frac{Q^2}{S}
\right),
\end{equation}
and the Hawking temperature is obtained as
\begin{equation} T = \frac{1}{4\sqrt{S}}\left( 1 + \frac{3S}{l^2} - \frac{Q^2}{S}
\right) \ .
\end{equation}
where the extremal limit occurs when $ \frac{Q^2}{S} = 1 +
\frac{3S}{l^2} $. Using the above information, the Weinhold metric will be obtained as follows
\begin{equation}
ds^2_W = \frac{1}{8S^{\frac{3}{2}}}\left( - \left(1 -
\frac{3S}{l^2} - \frac{3Q^2}{S}\right) dS^2 - 8QdSdQ +
8SdQ^2\right) \ .
\end{equation}

\begin{figure}[!htb]
    \centering
    {\includegraphics[width=0.8\textwidth]{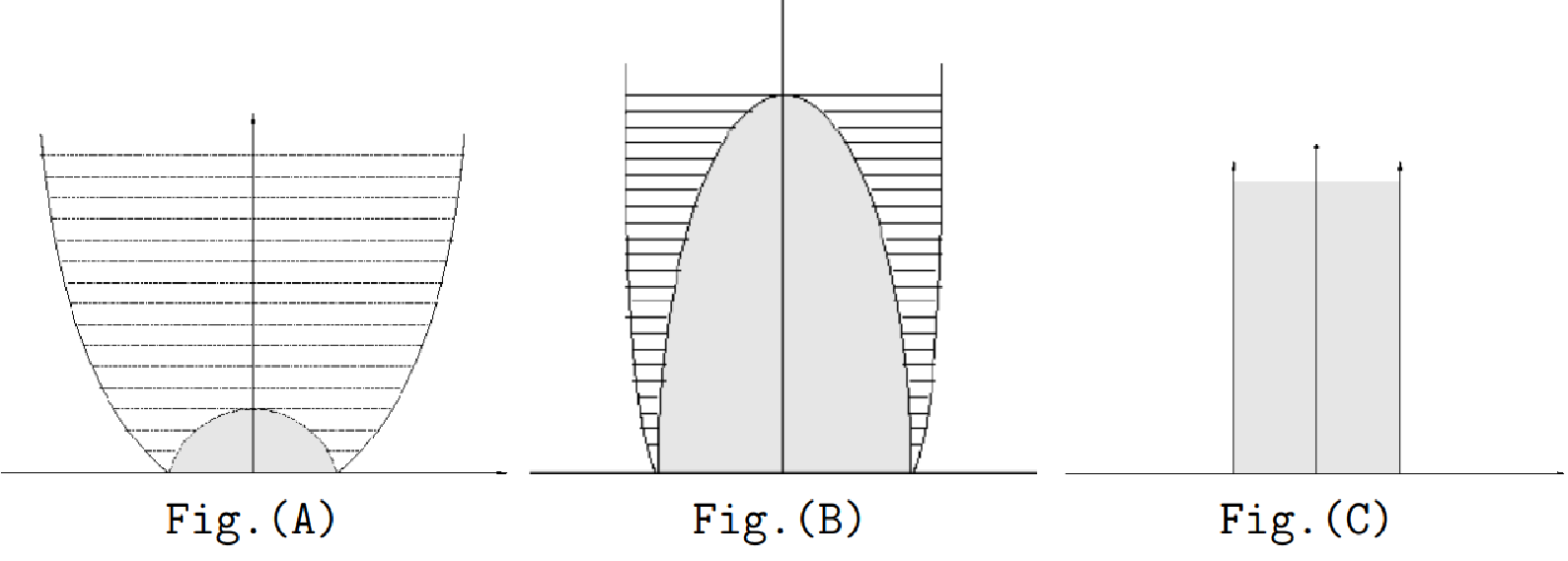}}
    \newline
    \caption{ The state space for Reissner-Nordstr\"{o}m black holes in the presence of a cosmological constant; the coordinates are $ u $ and $ S $ and $ \Lambda $ decreases from
$ A $ to $ C $ \cite{Aman:2003}.}
    \label{Fig3}
\end{figure}

Making use of the same coordinate transformation as above, one can
diagonalize the Weinhold metric and conformally relate it to the
Ruppeiner metric as
\begin{equation}
ds^2 = \frac{1}{1 + \frac{3{\tau}^2}{2l^2} - u^2}\left( - \left(1
- \frac{3{\tau}^2}{2l^2} - u^2\right) d{\tau}^2 + 2{\tau}^2du^2
\right) \ .
\end{equation}
In this case and with non-zero cosmological constant, the state
space will be as Fig. \ref{Fig3}. The curvature scalar of the
Ruppeiner metric is obtained as
\begin{equation}
R = \frac{9}{l^2}\frac{\left(\frac{3S}{l^2} + \frac{Q^2}{S}\right)
    \left( 1 - \frac{S}{l^2} - \frac{Q^2}{S}\right)}{\left( 1 - \frac{3S}{l^2} -
    \frac{Q^2}{S}\right)^2\left(1 + \frac{3S}{l^2} - \frac{Q^2}{S}\right) }
\ .
\end{equation}
As the above relation shows, the curvature scalar diverges both in the extremal limit and along
the curve where the metric changes signature,  i.e.,  where the
thermodynamic stability properties are changing.

It is worth pointing out that the same results were not obtained
in the investigation of the thermodynamic geometry of black holes
using the Weinhold and Ruppeiner methods. See table \ref{table1}
\cite{Aman:2003} for a detailed comparison. It is now well
established that these inconsistent results are a consequence of
the fact that Weinhold and Ruppeiner metrics are not Legendre
invariant, meaning that their properties depend on the
thermodynamic potential used in their construction. As we mentioned before, this property
make them inappropriate for describing the geometry of
thermodynamic systems.

\begin{table*}[htb!]
    \caption{The results of inconsistencies between the Weinhold and Ruppeiner methods for some black hole families.}
    \label{table1}\centering
    \begin{tabular}{|l|l|l|}
        \hline
        Black hole family & Ruppeiner & Weinhold \\ \hline
        RN & Flat & Curved, no Killing vectors \\
        RNadS & Curved, no Killing vectors & Curved, no Killing vectors \\
        Kerr & Curved, no Killing vectors & Flat \\
        BTZ & Flat & Curved, no Killing vectors \\
        Kerr--Newman & Curved & Curved \\ \hline
    \end{tabular}%
\end{table*}

\vspace{0.3cm} \noindent  $\bullet$ {\it{\textbf{Shortcoming of
Ruppeiner metric:}}}\quad
\vspace{0.2cm}

Although the Ruppeiner metric was able to describe phase transition and the nature of interactions between black hole microstructures for some types of black holes \cite{Ghosh:2020101,Ghosh:2020106,Ghosh:2302}, the investigations conducted in the context of thermodynamic
geometry showed that this approach cannot be a suitable
method for describing the phase transition of all black holes.
Because there are extra divergencies for Ruppeiner's Ricci scalar
which are not related to any phase transition of the heat
capacity. To achieve a better insight into understanding this
issue, we examine the denominator of the Ricci scalar of the
Ruppeiner metric for a charged case. The denominator of the Ricci
scalar for this case is
\begin{equation}
denom(\mathcal{R}_{R})=\left( M_{SS}M_{QQ}-M_{SQ}^{2}\right)
^{2}T(S,Q) M^{2}\left( S,Q\right) .  \label{denRupp}
\end{equation}

According to the above equation, the phase transitions of the heat
capacity correspond to the singularities of the Ruppeiner Ricci
scalar only when $M_{SQ}^{2}=0$ and $M_{QQ}\neq 0$. On the other
hand, due to the zeroes of $\left(M_{SS}M_{QQ}-M_{SQ}^{2}\right)$,
some divergence points of $\mathcal{R}_{R}$ may appear which
cannot coincide with phase transition points. In the case of
$M_{SQ}^{2}=0$, phase transition points of the heat capacity
($M_{SS}=0$) are covered by divergencies of the Ricci scalar of
the Ruppeiner metric. But in this case, one may encounter with
extra divergencies related to the roots of $M_{QQ}=0$ which are
not related to any phase transition points of the heat capacity.
Moreover, for the case of $M_{SQ}^{2} \neq 0$, the possible real
roots of $M_{QQ}=\frac{M_{SQ}^{2}}{M_{SS}}$ lead to the same extra
divergencies observed in the Weinhold metric.

\subsubsection{Fisher-Rao metric}
The Fisher-Rau (FR) metric is an alternative approach to classical
statistical mechanics that is used to analyze the geometry of
thermodynamic systems. This metric is a choice of Riemannian
metric in the space of probability distributions. In a statistical
mechanical context, the probability density distribution $
p(x|\vartheta) $, which is related to the partition function
($Z(\vartheta)$) of the corresponding system, takes the form of
the Gibbs measure \cite{Quevedo;2008cd}
\begin{equation}
p(x|\vartheta) = \exp\left[ - \vartheta_i H_i(x) - \ln Z(\vartheta)\right].
\, \label{prob}
\end{equation}

Where $H_i(x)$ represents Hamiltonian functions, $ x $ is a random
variable, and $\vartheta=(\vartheta_{1},...,\vartheta_{n})\in
R^{n}$ is an $n-$vector continuous parameter characterizing the
statistical model under consideration, the FR metric can be
derived from the information contained in the partition function.
A probability density function $ p(x) $ can be associated with a
vector in a real Hilbert space $ \mathcal{H} $ by taking the
square-root $ \psi (x)=\sqrt{P(x)} $, which is denoted by a vector
$ \psi^{a} $ in $ \mathcal{H} $ \cite{Brody:245}. Hence, the
Hilbert space embodies the state space of the system, and the
properties of the statistical model under consideration can be
described by embedding of $ p(x|\vartheta) $ in $ \mathcal{H} $.
The geometry resulting from the embedding has a natural Riemannian
metric, the FR metric in the classical case or the Fubini-Study
metric in the quantum case. For the classical one, Eq.
(\ref{prob}) takes the following form
\begin{equation}
g^{FR}= \frac{\partial^2 \ln Z(\vartheta)}{\partial \vartheta^\mu \partial
\vartheta^\nu} d\vartheta^\mu d\vartheta^\nu \ . \label{fisrao}
\end{equation}

The geometric features of the manifold described by metric
(\ref{fisrao}) were employed for various statistical models. For
the van der Waals fluid, we can consider the parameters as
$\vartheta^1 = 1/T$, and $\vartheta^2 = P/T$. With such a choice,
the internal energy $U$ (volume $V$) can be considered as the
corresponding Hamiltonian function to $\vartheta^1$
($\vartheta^2$). Hence, the partition function $Z(\vartheta)$ is a
function of temperature and pressure. The scalar curvature
obtained by this two-dimensional manifold diverges at the critical
points and the scaling exponent of the curvature near the
transition points coincides with that of the correlation volume.
The investigation in the context of the ideal gas indicated that
the scalar curvature vanishes in this case and the manifold is
flat. Such behavior has also been observed by Ruppeiner's and
Weinhold's geometry in these particular cases. As was shown in
Ref. \cite{jjk03}, these two metrics are related to the FR metric
by using Legendre transformations of the corresponding variables,
indicating that the FR metric is not invariant under Legendre
transformations.

For two well-known cases, RN and Kerr black holes, the components
of the FR metric $g^{FR}_{ij}(\vartheta)$ are usually chosen as
the "inverse" of the thermodynamic variables: $\vartheta^1=1/T$,
$\vartheta^2=P/T$, etc. Since the relations
$\theta^\mu=\theta^\mu(E^a)$ must allow the inverse
transformation, it can be shown that the FR metric is written as
$g^{FR}_{ab} = \partial^2 \ln Z(E)/\partial E^a\partial E^b$ for
the corresponding coordinates. The partition function is obtained
as
\begin{equation}
Z= \exp\left[ -\frac{1}{T}(M-TS - \Omega_H J - \phi Q)\right] \ .
\end{equation}

For RN black holes, the partition function is determined as
$Z=\exp(-M^{\prime}_{R}/T)$ and for Kerr black holes
is $Z=\exp(-M^{\prime}_{K}/T)$. The thermodynamic potentials $
M^{\prime}_{R} $ and $ M^{\prime}_{K} $ are related to the following mass
representations
\begin{eqnarray}
M^{\prime}_{R} & = & M - TS -\phi Q \ , \\
M^{\prime}_{K} & = & M - TS -\Omega_H  J . \nonumber
\label{tpotentials}
\end{eqnarray}
Thus, the components of the FR metric for black holes are
essentially given by
\begin{equation}
g^{FR}_{ab}= -\partial^2 (M/T) /\partial E^a \partial E^b.
\end{equation}
As already mentioned, the FR metric is not Legendre invariant for
black holes. Therefore, it is not a suitable method for solving
the problem contradictory results arising from the application of
Weinhold's and Ruppeiner's approaches.

Before ending this part, it is worth mentioning some points regarding the issue that how the representation based on Riemannian geometry, particularly as applied by Rao in 1945, which used entropy as a thermodynamic potential, relates to the information loss paradox, which often arises in the context of thermodynamics and statistical physics. Indeed, by utilizing Riemannian geometry, Rao established a framework where entropy serves as a thermodynamic potential. This approach allows for the quantification of statistical distances between probability distributions, providing a measure of information loss or retention during thermodynamic processes. In this regard, the Fisher-Rao metric and subsequent Hessian metrics play crucial roles in shedding light on the information loss paradox within thermodynamic geometry. The Fisher-Rao metric, derived from the Fisher information matrix, quantifies the distinguishability and separability between probability distributions. It enables the assessment of how much two system states differ in terms of their information content. This metric, together with Hessian metrics that capture the local curvature of the manifold, facilitates the analysis of stability, critical points, and phase transitions in thermodynamic potentials. By incorporating the Fisher-Rao and Hessian metrics into thermodynamic geometry, researchers gain valuable tools to quantify and characterize the distribution and transformation of information during thermodynamic processes. This, in turn, contributes to addressing the information loss paradox and deepening our understanding of complex systems in the realm of thermodynamics and statistical physics. \par

However, a key challenge of using Riemannian geometry, as applied by Rao in 1945 using entropy as a thermodynamic potential, lies in its practical implementation and applicability to diverse systems. The approach requires making assumptions about the underlying statistical distributions and their functional forms, which may not always accurately capture the complexity of real-world systems. Furthermore, accurately estimating the metric tensor in Riemannian geometry can be computationally demanding and may pose difficulties in high-dimensional spaces. On the other hand, the main limitation of the Fisher-Rao metric and subsequent Hessian metrics is their sensitivity to the choice of coordinate systems or parameterizations. Different parameterizations can lead to varying metric structures, potentially affecting the interpretation of distances or curvature. Additionally, in systems with limited available data, accurately estimating the metric tensors using finite samples may introduce uncertainties and biases in the computed metrics, which can impact the reliability of results obtained through thermodynamic geometry methods.

At the end of this discussion, it is necessary to mention the idea that remnants of black holes can be suggested as a resolution of the black hole information paradox. In one version of this idea \cite{Susskind:1231992}, the black hole stops evaporating when its dimension approaches the Planck scale. But this remnant quiescent object retains the vast amount of information that was stored in the black hole after its initial formation.
A common argument for remnants as information repositories is that they can hold large information despite their small dimension since they have a very large internal space in the form of a throat or horn. But Bekenstein proposed the existence of a universal bound on the entropy $ S $ of any object of maximal radius $ R $ and total energy
$E$ \cite{Bekenstein:1981}:
\begin{equation}
S\leq \frac{2\pi RE}{\hslash c}
\label{SGSL}
\end{equation}
and showed that if a remnant has an information capacity well above that specified by bound (\ref{SGSL}) in terms of its external dimensions, it will violate the GSL if it falls into a large black hole \cite{Bekenstein:19941912}. Indeed, the black hole remnants cannot resolve the information paradox (see Ref. \cite{Report} for more information in this regard).

\subsubsection{Quevedo metric}

First, we review the formulation of general relativistic black
hole thermodynamics in the language of geometrothermodynamics. As
we mentioned earlier, a rotating stationary black hole in general
relativity is completely described by three parameters $(M, Q,
J)$, which are related to each other through the first law of
black hole thermodynamics $dM = T dS + \phi dQ +\Omega_{H} dJ$.
For a given fundamental equation $M = M(S, J, Q)$, there are the
following conditions for thermodynamic equilibrium
\cite{Quevedo;2008cd}
\begin{equation}
T =\frac{\partial M}{\partial S} \ , \quad \Omega_H
=\frac{\partial M}{\partial J} \ , \quad \phi =\frac{\partial
M}{\partial Q} \ .
\end{equation}

Here, ${\cal T}$ is a 7-dimensional manifold characterized by
$Z^A=\{M,S,J,Q,T,\Omega_H ,\phi\}$, whereas the submanifold ${\cal
E}$ is 3-dimensional with coordinates $E^a = \{S,J,Q\}$, and is
defined by the following mapping
\begin{equation}
\varphi: \{S,J,Q\} \longmapsto \left\{ M(S,J,Q), S,J,Q,
\frac{\partial M}{\partial S}, \frac{\partial M}{\partial J},
\frac{\partial M}{\partial Q}\right\} \ .
\end{equation}

The mass  $M$ is considered as the thermodynamic potential dependent on the extensive variables $S$, $J$, and $Q$. It is
noteworthy that by using Legendre transformations, other
thermodynamic potentials can be also introduced as different
combinations of extensive and intensive variables. For instance
\begin{eqnarray}
M_1 & = & M - TS \ ,\nonumber \\
M_2 & = &  M - \Omega_H  J \ ,\nonumber \\
M_3 & = &  M - \phi Q \ ,\nonumber\\
M_4 & = &  M - TS -\Omega_H  J \ , \\
M_5 & = &  M - TS -\phi Q\ ,\nonumber\\
M_6 & = &  M - \Omega_H  J -\phi Q \ ,\nonumber\\
M_7 & = &  M - TS -\Omega_H  J -\phi Q \
.\nonumber \label{tpotentials}
\end{eqnarray}

Here, one can define the mapping $\varphi$ for each case
independent of the chosen thermodynamic potential. Since ${\cal
T}$ and ${\cal E}$ are invariant under Legendre transformations,
the properties of the underlying geometry for an ordinary
thermodynamic system do not depend on the thermodynamic potential.
This shows that it is consistent with standard thermodynamics.

In the mass representation, the fundamental Gibbs 1-form is given
by
\begin{equation}
\theta = dM - T dS - \Omega_H  d J - \phi d Q,
\end{equation}
whereas, in the entropy representation with the fundamental
equation $S=S(M,J,Q)$, it can be chosen as
\begin{equation}
\theta_S =  dS -\frac{1}{T} dM  +\frac {\Omega_H }{T} d J + \frac{
\phi}{T} d Q \ .
\end{equation}
In this case, the space of equilibrium states can be defined by
the following harmonic mapping
\begin{equation}
\varphi_S: \{M,J,Q\} \longmapsto \left\{ M, S(M,J,Q),J,Q,
T(M,J,Q), \Omega_H (M,J,Q), \phi(M,J,Q)\right\} \ ,
\end{equation}
with
\begin{equation}
\frac{1}{T} = \frac{\partial  S}{\partial M} \ ,\quad
\frac{\Omega_H }{T} = -\frac{\partial  S}{\partial J} \ ,\quad
\frac{\phi}{T} = -\frac{\partial  S}{\partial Q} \ .
\end{equation}

As has already been explained, components of Weinhold's metric
$g^W$ are given as the Hessian of the internal thermodynamic
energy (the mass representation), while the components of
Ruppeiner's metric $g^R$ are defined as the Hessian of the entropy
(the entropy representation). The simplest way to reach the
Legendre invariance for Weinhold's metric is to apply a conformal
transformation, with the thermodynamic potential as the conformal
factor. Hence, the simplest Legendre invariant generalization of
$g^W$ is as
\begin{equation}
g= M g^W = M \frac{\partial^2 M}{\partial E^a \partial E^b} d E^a
d E ^b \ , \label{gbhe}
\end{equation}
which can be written in terms of the
components of Ruppeiner’s metric as
\begin{equation}
g = M T g^R = - M \left(\frac{\partial S}{\partial M}\right)^{-1}
\frac{\partial^2 S}{\partial F^a \partial F^b} d F^a d F^b \ ,
\label{rup}
\end{equation}
where $E^a=\{S,J,Q\}$ and $ F^a = \{M,J,Q\}$. Using the mass
representation, the non-degenerate metric $G$ on the phase space
${\cal T}$ is defined as
\begin{equation}
G = (dM - TdS - \Omega_H  d J -\phi d Q)^2 + (TS+\Omega_H  J +
\phi Q) ( dT d S + d\Omega_H  d J + d\phi d Q) \ . \label{gbht}
\end{equation}

In Ref. \cite{Quevedo;2008cd}, Quevedo employed the Legendre
invariant generalizations of Weinhold's and Ruppeiner's metrics
and analyzed the geometry of the RN and Kerr black hole
thermodynamics, by using Legendre invariant thermodynamic metrics.
He noticed that the obtained results are geometrically consistent
for both cases. For the Reissner-Nordstr\"{o}m solution, there was
an agreement with the results of standard black hole
thermodynamics, whereas, for Kerr black holes, the simplest
Legendre invariant metrics could not reproduce the corresponding
phase transition structure.

\vspace{0.3cm} \noindent  $\bullet$ {\it{\textbf{Shortcoming of
Quevedo metric:}}}\quad
\vspace{0.2cm}

Although the divergencies of the Quevedo Ricci scalars coincide
with the divergencies of specific heat \cite{Banerjee:2017,Zhang:2018},
studies in this regard showed that the Quevedo metric cannot be a
suitable candidate to describe phase transition points of black
holes. It is due to the fact that they suffer from some extra
divergencies. For example, for charged black holes, the Quevedo
metric can be written as \cite{Quevedo}
\begin{equation}
ds_{Q}^{2}=\Omega \left( -M_{SS}dS^{2}+M_{QQ}dQ^{2}\right) ,
\label{Quev}
\end{equation}
where the (conformal) function $\Omega$ has one of the following
forms
\begin{equation}
\Omega =\left\{
\begin{array}{cc}
SM_{S}+QM_{Q}, & \text{case I} \\
SM_{S}, & \text{case II}%
\end{array}%
\right. .  \label{gabQ}
\end{equation}

To obtain the curvature singularity of the Quevedo metric, one
needs to calculate the Ricci scalar. Since the calculated results
of the Ricci scalar are too large, we avoid writing the Ricci
scalar here and just write down its denominator with the following
explicit forms
\begin{equation}
denom(\mathcal{R}_{Q1})=\left( SM_{S}+QM_{Q}\right)
^{3}M_{SS}^{2}M_{QQ}^{2}, \label{RQ1}
\end{equation}%
\begin{equation}
denom(\mathcal{R}_{Q2})=S^{3}M_{S}^{3}M_{SS}^{2}M_{QQ}^{2}.
\label{RQ2}
\end{equation}

Evidently, the divergencies of the heat capacity and Quevedo Ricci
scalars coincide with together due to the existence of $M_{SS}$ in
the denominator of both Quevedo Ricci scalars. It should be noted that
there exists an additional function $M_{QQ}^{2}$, and its roots
provides extra singular points for both cases of Quevedo Ricci
scalars. Another divergence point can also appear for nonzero
$M_{Q}$ and for a special choice of $M_{S}=-\frac{Q}{S}M_{Q}$.

According to Ref. \cite{Azreg:2014}, this problem comes from the
fact that in black hole thermodynamics,  the thermodynamic
potential is not a homogeneous first-order function of its natural
extensive variables contrary to classical thermodynamics. In fact,
the natural extrinsic thermodynamic variables expressing the first
law of thermodynamics are not the same variables in which
thermodynamic potentials are homogeneous. This is the difference
between black hole thermodynamics and classical ones. To work out
this problem, the modified extensive variables should be
considered instead of the natural ones (see Ref. \cite{Azreg:2014}
for more detail).

\subsubsection{HPEM metric}
Hendi et al. introduced a new metric, which is known as the HPEM metric, to eliminate the problems of
previous thermodynamical metrics by removing extra singular points
in the thermodynamical Ricci scalar that do not coincide
with phase transitions. HPEM metric is defined as \cite{Hendi:507}
\begin{equation}
ds_{HPEM}^{2}=S\frac{M_{S}}{M_{QQ}^{3}}\left(
-M_{SS}dS^{2}+M_{QQ}dQ^{2}\right) .  \label{newmetric}
\end{equation}
The remarkable point, here, is that the HPEM metric is defined in
the same way as the Quevedo metric but with a different conformal
function. In this metric, the total mass serves as a
thermodynamical potential with entropy and electric charge acting
as extensive parameters. The numerator and the denominator of the thermodynamical Ricci scalar
for this new metric is, respectively,
\begin{eqnarray}
num(\mathcal{R})
&=&6S^{2}M_{S}^{2}M_{QQ}M_{SS}^{2}M_{QQQQ}-6SM_{S}^{2}M_{QQ}^{2}M_{SS}M_{SSQQ}+2SM_{SQQ}^{2}M_{S}^{2}M_{QQ}M_{SS}
\notag \\
&&+2\left[SM_{S}M_{SSS}-\frac{1}{2}M_{SS}\left(SM_{SS}-M_{S}\right) \right]%
SM_{QQ}^{2}M_{S}M_{SQQ}-9S^{2}M_{QQQ}^{2}M_{S}^{2}M_{SS}^{2}  \notag \\
&&+4\left[ \frac{1}{4}M_{SQ}M_{SS}+M_{S}M_{SSQ}\right]
S^{2}M_{QQ}M_{S}M_{QQQ}+\left[
S^{2}M_{S}^{2}M_{SSQ}-S^{2}M_{SQ}M_{SS}M_{S}M_{SSQ}\right.  \notag \\
&&\left. SM_{QQ}M_{S}\left( SM_{SS}-M_{S}\right) M_{SS}-2\left(
S^{2}M_{SS}^{3}+M_{S}^{2}M_{SS}\right)
M_{QQ}+2S^{2}M_{SQ}^{2}M_{SS}^{2} \right] M_{QQ}^{2},
\label{NumNew}
\end{eqnarray}
and
\begin{equation}
denom(\mathcal{R})=S^{3}M_{S}^{3}M_{SS}^{2}.  \label{denNew}
\end{equation}
The denominator of the thermodynamical Ricci scalar (TRS), Eq. (\ref{denNew}), ensures that all the
phase transition points coincide with divergencies of the
mentioned thermodynamical Ricci scalar and there is no extra term that may provide extra
divergencies.

It should be noted that, in general, the derivatives of all orders
of $M$ (such as $M_{QQ}$, $M_{SS}$, $M_{SQ}$, $M_{SSQ}$ and so on)
are independent from each other. For the case of vanishing $M_{S}$
or $M_{SS}$, the numerator of TRS has nonzero value, but
denominator of TRS vanishes. In the case of $M_{S}=M_{SS}=0$, both
numerator and denominator vanish, but it should be noted that the
denominator approaches zero faster than the numerator. Therefore,
one concludes that when $M_{S}$ and/or $M_{SS}$ go to zero, the thermodynamical Ricci scalar
diverges. The validity of the HPEM metric was explored for
different types of black holes and was shown that this method is a
suitable one for studying the thermodynamic geometry of black
holes\cite{Hendi:528,Hendi:129,Hendi:296,Hendi:157,Hendi:64028,Sheykhi:124054,Hendi:769,wen:2017,Hendi:2016,Sheykhi:1650062,Sheykhi:4896,Hendi:5712016,Sheykhi:1045,Hendi:1750026,Hendi:767214,Jawad:2023}.

\subsubsection{Other approaches in the context of thermodynamic geometry}
A significant effort in the context of thermodynamic geometry was
made by Mansouri et al. They explored a formulation of
thermodynamic geometry for different types of black holes and
proved that their formulation can be applied to an arbitrary
thermodynamic system. In studying RN black holes, they noticed
that singularities of the scalar curvature $R(S,Q)$ match to phase
transition points of the heat capacity $ C_{\Phi} $ at a constant
electric potential, whereas phase transition points of the heat
capacity $ C_{Q} $ at a constant electric charge correspond to the
singularities of the scalar curvature $\overline{R}(S,\Phi)$.
Here, $\overline{R}(S,\Phi)$ is obtained by defining
$\overline{M}$ as a new conjugate potential of $M(S,Q)$ in
following form
\begin{equation}\label{114}
\overline{M}(S,\Phi )=M(S,Q)-\Phi Q\
\end{equation}

In fact, the scalar curvature $R(S, Q)$ is not able to explain the
properties of the phase transitions of $ C_{Q} $ but it can
describe phase transitions $ C_{\Phi} $. Regarding black holes
with three parameters (Kerr-Newman black holes), the singularities
of the scalar curvature $R(S, Q, J)$ coincide with divergencies of
heat capacity  at a constant electric potential and angular
velocity ${{C}_{\Phi ,\Omega }}$ and that also those of
$\overline{R}(S, \Phi, \Omega)$ correspond to the phase
transitions of $ C_{Q, J} $. In general, for a black hole with $n$
parameters, the singularities of $R(S,Q_{1},...,Q_{n})$ correspond
to the phase transition points of $C_{\Phi_{1},...,\Phi_{n}}$ and
the singularity points of
$\overline{R}(S,\Phi_{1},\Phi_{2},...,\Phi_{n})$ coincide with
divergencies of $C_{Q_{1},Q_{2},...,Q_{n}}$ (see Ref.
\cite{mirza1a} for more details). In Ref. \cite{mirza2a}, they
employed the Nambu brackets and obtained a simple representation
of the conformal transformations that connect different
thermodynamic metrics to each other. Using the bracket approach,
they studied the relationship between singularities of the scalar
curvature in various representations of thermodynamic geometry and
phase transition points of different heat capacities. They also
investigated the intrinsic, extrinsic, and total curvatures of a
certain hypersurface in the thermodynamic geometry of a physical
system and explained the relationship between them \cite{mirza3a}.
Their finding regarding KN-AdS black hole showed that by
considering a constant $ J $ hypersurface, the curvature scalar is
broken down into two parts. One is related to a zero intrinsic
curvature (the Ruppeiner curvature of the
Reissner-Nordstr\"{o}m-AdS black hole) and the other  is concerned
with an extrinsic part whose divergence points are the
singularities of a non-rotating Kerr-Newman-AdS black hole. In
Ref. \cite{mirza4a}, they constructed a new formalism
thermodynamic geometry by changing the coordinates of the
thermodynamic space by means of Jacobian matrices and showed that
the GTD metric is related to their new formalism of the
thermodynamic geometry with use of a singular conformal
transformation. It is worth mentioning that this new metric
removes all extra curvature singularities that appeared in GTD. In
fact, there is a one-to-one correspondence between the phase
transition points of a black hole and singularities of the
curvature associated with this new metric. In Ref. \cite{mirza5a},
they employed this new metric for black holes in pure Lovelock
gravity and found an exact correspondence between thermodynamic
Ricci scalar and specific heat $ C_{Q} $ at critical point.

The study of black hole criticality is done with two separate
approaches: one approach is related to the behavior of the heat
capacity which diverges at the critical point. Another case is
related to the AdS background with the cosmological constant,
which plays the role of thermodynamic pressure in the extended
phase space. In contrast to the usual thermodynamics,  $ P-V $
criticality, $T-S$ criticality, and also the $ Y-X $ criticality
(where $ Y $ is an external charge and $X$ is its corresponding
potential) can be found for the black holes. Some separate
attempts were done in this context to explain criticality in a
geometrical way based on the divergence point of heat capacity not
a point as the inflection point. In Ref. \cite{Bhattacharya:2017}
authors defined the relevant Legendre invariant thermogeometrics
corresponding to the two criticality conditions and illustrated
that the critical point refers to the divergency of the Ricci
scalars calculated from these metrics. For the first condition $
\left(\frac{\partial P}{\partial V} \right)_{T,Y}=0  $, the
Helmholtz free energy was considered as a proper thermodynamic
quantity to define a thermogeometrical metric. Whereas, for the
geometrical description of the second condition (the condition of
the point of inflection), the pressure was a proper
thermodynamical quantity (instead of $ F $). It is worth
mentioning that in Ref. \cite{Jafarzade:2021}, the authors
employed this approach to investigate the thermodynamic geometry
of charged accelerating black holes and found that the Ricci scalar included an extra divergence point at the critical temperature and was not able to describe the critical point. In other words, this method cannot be a suitable candidate for this kind of black hole. According to their analysis, only the HPEM’s metric can provide an appropriate picture of phase transition for accelerating AdS black holes in both extended and non-extended phase space.

\section{Conclusion}
\label{5-4a}
In this study, we conducted a comprehensive examination of various
facets pertaining to geometrical thermodynamics. Our analysis
encompassed the exploration of two distinct approaches involving
Riemannian metrics: namely, the utilization of {\it{Hessian
metrics}} and {\it{Legendre invariant metrics}} to develop a
geometric framework for describing equilibrium thermodynamics. The
first approach involves the incorporation of metrics into the
equilibrium space, wherein the components of these metrics align
with the Hessian matrix associated with specific thermodynamic
potentials. This approach enables a direct connection between the
geometry of the equilibrium space and thermodynamic properties.
Alternatively, the second approach establishes the thermodynamic
phase space as a Riemannian contact manifold through a purely
geometric perspective. Within this framework, the class of
Legendre invariant metrics has been introduced in the formalism of
geometric thermodynamics. These metrics serve the purpose of
accounting for the fundamental observation that the physical
properties of a system in equilibrium thermodynamics remain
unchanged regardless of the chosen thermodynamic potential, from a
geometric standpoint.

After discussing the two previously mentioned approaches, an
examination of the geometric representation of thermodynamics was
undertaken for various ordinary systems, including ideal gas and
real gas. Additionally, a comprehensive evaluation of these
approaches in relation to the issue of black holes was provided.
It was noted that black holes can be regarded as thermodynamic
systems by equating the parameters of the black hole (mass,
surface gravity, and area) to the thermodynamic system's energy,
temperature, and entropy, respectively. The four laws of black
hole mechanics, which correspond to the four laws of ordinary
thermodynamics, were also reviewed. These laws are summarized in
the table \ref{table6}
\begin{table}[H]
\begin{center}
\def\arraystretch{1.5}
\begin{tabular}{| c | m{.35\textwidth} | m{.35\textwidth} |}
    \hline
    \textbf{Law}  & ~~~~~~~~~~~~~~\textbf{Thermodynamics} &~~~~~~~~~~ \textbf{Black Hole Mechanics} \\
     \hline
    Zeroth Law & The temperature $ T $ is constant over a system in thermal
equilibrium & The surface gravity $ \kappa $ is constant on the
event horizon
of a stationary black hole  \\
    \hline
    First Law & \begin{equation}dE=TdS+ work ~terms \nonumber \end{equation} & \begin{equation}dM=\frac{\kappa}{8\pi G} + work~ terms \nonumber\end{equation} \\
    \hline
    Second Law & \begin{equation}dS_{TD}\geq 0\nonumber\end{equation} & \begin{equation}d(S_{outside}+\frac{A}{4})\geq 0\nonumber\end{equation} \\
    \hline
    Third Law & \begin{equation} dS_{TD}\rightarrow 0 \;\text{as}\; T\rightarrow 0\nonumber \end{equation} & It is impossible to achieve $\kappa = 0$ within a finite number of steps. \\
    \hline
\end{tabular}
\caption{Summary of the laws of thermodynamics and their black
hole mechanical counterparts.} \label{table6}
\end{center}
\end{table}

In addition, we explored the utilization of the thermodynamic
metrics mentioned above in the context of black hole
thermodynamics. We demonstrated that the Weinhold, Ruppeiner, and
Fisher-Rao metrics yield inconsistent outcomes when applied to the
same black holes due to their lack of Legendre invariance. While
the Quevedo metric resolves the issue of conflicting results from
previous metrics, it contains additional divergencies in its Ricci
scalar that do not correspond to phase transition points.

To address these challenges, Hendi et al. introduced a novel
metric that eliminates the problems encountered with previous
thermodynamic metrics by eliminating extra singular points.
Furthermore, we examined alternative approaches within the realm
of thermodynamic geometry for black holes, which can be applied to
any thermodynamic system. One such approach, proposed by Mansouri
et al., established a connection between singularities of the
scalar curvature at constant charge and phase transition points of
the heat capacity at a constant electric potential.

Another approach focused on Legendre invariant thermogeometrics,
which encompassed the two criticality conditions. Although this
approach yielded consistent results for many black holes, it was
not deemed suitable for studying accelerating black holes.

In conclusion, while geometrical thermodynamics approaches have achieved interesting results, this paper highlights several ongoing challenges and open questions that require further investigation. Some of these challenges are outlined below:

1. Incorporating contact geometry and GTD into systems with strong interactions may pose challenges due to the complexity of the underlying dynamics. The nonlinearity and intricate interaction patterns may hinder the identification and calculation of relevant geometric quantities, potentially limiting the applicability of GTD.

2. Extending GTD to non-equilibrium conditions can be intricate, as it requires considering dissipative processes and time-dependent phenomena. Capturing the dynamics and temporal evolution of the system within the geometric framework may require the development of novel mathematical tools and techniques.

3. In systems exhibiting emergent behaviors, where collective phenomena arise from the interactions of individual components, the applicability of GTD may be limited. The geometric framework traditionally relies on the notion of equilibrium and may not fully capture the emergent properties of the system, demanding the formulation of new concepts and theories.

4. Generalizing GTD to incorporate quantum and relativistic effects in a consistent manner represents a significant challenge. Quantum systems may involve noncommutative geometry, entanglement, and uncertainty principles, while relativistic systems require accounting for curved spacetime and the interplay between geometry and dynamics. Bridging GTD with these theories requires careful consideration and potential modifications to accommodate these fundamental principles.

\section*{Acknowledgements}

We are grateful to the anonymous referee for the insightful comments. S. Mahmoudi and  Kh. Jafarzade are grateful to the Iran Science
Elites Federation for the financial support.

\appendix

\section{Riemannian geometry}\label{one}
Riemannian geometry is widely used in general relativity to
describe spacetime. In fact, based on the geometry we are dealing
with, a length element or metric can be defined for the
investigated spacetime. Regarding thermodynamic systems, a
thermodynamic coordinate surface can be defined using the system's
thermodynamic parameters, and for a system with two independent
variables, we will have a two-dimensional surface. In what
follows, we will review the main elements of Riemannian geometry.

Regarding the Riemannian geometry in two dimensions, manifold or surface is the first constituent element. Based on physics topics, physical quantities can be defined by the points placed on the manifold which typically involve more than one
coordinate. In the case of thermodynamic processes, these points
denote the thermodynamic states.

The second main constituent part of Riemannian geometry provides a line element's rule for two adjacent points with
coordinate difference $\Delta x^{\alpha}$, which is
expressed as follows
\begin{equation}
\Delta \ell^2\equiv g_{\mu\nu}\Delta x^\mu \Delta x^\nu,\,\,\,\,\,\,\,\,\mu,\nu=1,2
\label{line element}
\end{equation}
in which $g_{\alpha\beta}$ is a positive-definite matrix representing the metric elements. A manifold whose distance between points is defined in Eq. \eqref{line element} is called
a Riemannian manifold. It is notable that being independent of the choice of the coordinate system is a requirement for defining the distance between two points.
 For instance,
Euclidean planes are familiar Riemannian surfaces on which the distance is defined by Cartesian coordinates
as follows
\begin{equation}
     \Delta \ell^2= (\Delta x^{0})^2+(\Delta x^1)^2. \label{cartizian}
     \end{equation}
Moreover, the above distance can be defined in terms of polar coordinates,
\begin{equation}
     \Delta \ell^2= \Delta r^2+ r^2\,\Delta\theta^2, \label{polar}
     \end{equation}
     where $r$ indicates the radial coordinate and $\theta$ is the angular coordinate.
Another example of two-dimensional Riemannian manifold is a sphere of radius $a$ in spherical coordinates $(\theta,\phi)$
\begin{equation}
\Delta \ell^2=a^2 \Delta\theta^2+ a^2\,\sin^2
\theta\,\Delta\phi^2. \label{spherical}
\end{equation}

Now, let us find the rule of transformation for the metric elements. To this end, one can consider some different coordinates
$({{x}^{\prime}}^1,{{x}^{\prime}}^2)$ in which the metric elements are
${{g}^{\prime}}_{\alpha\beta}$. Since
\begin{equation}
 \Delta x^\alpha=\frac{\partial x^{\alpha}}{\partial {{x}^{\prime}}^\mu} \Delta {{x}^{\prime}}^\mu,\label{transformed metric}
\end{equation}
we get
\begin{equation}\Delta \ell^2=g_{\alpha\beta}\frac{\partial x^{\alpha}}{\partial \tilde{x}^\mu}\frac{\partial x^{\beta}}{\partial {{x}^{\prime}}^\nu}\Delta {{x}^{\prime}}^\mu \Delta {{x}^{\prime}}^\nu\equiv {{g}^{\prime}}_{\mu\nu} \Delta {{x}^{\prime}}^\mu \Delta {{x}^{\prime}}^\nu.\label{1030} \end{equation}
Therefore, the new metric elements read
\begin{equation}
{{g}^{\prime}}_{\mu\nu}=g_{\alpha\beta}\frac{\partial
x^{\alpha}}{\partial {{x}^{\prime}}^\mu}\frac{\partial
x^{\beta}}{\partial {{x}^{\prime}}^\nu}, \label{transformation rule}
\end{equation}
which is the transformation law
for a $2^{nd}$ order tensor \cite{Arfken, Laugwitz1965}. \par

The third essential elements of a Riemannian geometry which is Riemannian
curvature examines the issue that
\textit{if the curvatures at given points of two different coordinate systems are the same, they will describe the same manifold.}
In order to find this scalar invariant, one can employ
the following conventions
\begin{eqnarray}
\Gamma^{\sigma}_{\;\;\mu\nu}&=&\frac{1}{2}g^{\sigma\rho}
(\partial_\nu g_{\rho\mu} + \partial_\mu g_{\rho\nu} - \partial_\rho g_{\mu\nu}), \\
R^{\sigma}_{\;\;\rho\mu\nu}&=& \partial_\nu
\Gamma^{\sigma}_{\;\;\rho\mu}
- \partial_\mu \Gamma^{\sigma}_{\;\;\rho\nu}+\Gamma^{\delta}_{\;\;\rho\mu}\Gamma^{\sigma}_{\;\;\delta\nu}-\Gamma^{\delta}_{\;\;\rho\nu}\Gamma^{\sigma}_{\;\;\delta\mu}, \\
R_{\mu\nu}&=&R^{\sigma}_{\;\;\mu\sigma\nu}, \\
\mathcal{R}&=& g^{\mu\nu}R_{\mu\nu},
\end{eqnarray}
where $g^{\mu\nu}g_{\mu\nu}=1$.
For a 2-dimensional space, the Riemannian scalar curvature is
obtained as follows
\begin{eqnarray}\label{Ricci}
\mathcal{R}=&-&\frac{1}{\sqrt{g}}\bigg[\frac{\partial}{\partial x^{0}}\left(\frac{g_{01}}{g_{00}\sqrt{g}}\frac{\partial g_{00}}{\partial x^{1}}-\frac{1}{\sqrt{g}}\frac{\partial g_{11}}{\partial x^{0}}\right)\nonumber\\
&+&\frac{\partial}{\partial
x^{1}}\left(\frac{2}{\sqrt{g}}\frac{\partial g_{01}}{\partial
x^{1}}-\frac{1}{\sqrt{g}}\frac{\partial g_{00}}{\partial
x^{1}}-\frac{g_{01}}{g_{00}\sqrt{g}}\frac{\partial
g_{00}}{\partial x^{0}}\right)\bigg],
\end{eqnarray}
which for a diagonal one, it will be simplified to
\begin{equation}
\mathcal{R}=\frac{1}{\sqrt{g}}\left[\frac{\partial}{\partial
x^{0}}\left(\frac{1}{\sqrt{g}}\frac{\partial g_{11}}{\partial
x^{0}}\right)+\frac{\partial}{\partial
x^{1}}\left(\frac{1}{\sqrt{g}}\frac{\partial g_{00}}{\partial
x^{1}}\right)\right].\label{scalar}
\end{equation}
\section{Contact manifolds}\label{two}
In this appendix, we aim to review the main properties of contact
geometry.\par

Consider a $(2n+1)-$dimensional differential manifold ${\cal T}$
and its tangent manifold $T({\cal T})$. It can be shown that for
an arbitrary family of hyperplanes ${\cal D}\subset T({\cal T})$
there exists a non-vanishing differential one-form $\theta$ on the
cotangent manifold $T^*({\cal T})$ such that
\begin{equation}  \label{family}
    \cal{D} = \mathrm{ker}(\theta).
\end{equation}
The Frobenius integrability condition $\theta\wedge d \theta =0$
requires that the hyperplanes ${\cal D}$ is completely integrable.
By contrast, ${\cal D}$ is non-integrable if  $\theta \wedge d
\theta \neq 0$. In the limiting case which $\theta$ satisfies the
condition
\begin{equation}\label{limit}
    \theta \wedge (d \theta)^n \neq 0,
\end{equation}
the hyperplane ${\cal D}$ becomes maximally non-integrable and a
contact structure can be defined on ${\cal T}$. Introducing a set
of local coordinates $\{\phi, x_a, y^a\}$ with $a=1,...,n$, it is
possible to express $\theta$ in its canonical form,
\begin{equation} \label{etadarboux}
    \theta = d \phi - x_a d y^a. \,\,\,\,\,\,\,\text{(Darboux theorem)}
\end{equation}
The pair $({\cal T},{\cal D})$ determines a contact manifold
\cite{handbook} and it is often denoted as $({\cal T},\theta)$ to
emphasize the role of the contact form $\theta$. In addition, the
set $({\cal T}, \theta, G)$ defines a Riemannian contact manifold
in which $G$ is a non-degenerate Riemannian metric on ${\cal T}$.



\section{Legendre transformations}\label{three}
The Legendre transformation is a convenient procedure in theoretical
physics that acts an important role in classical mechanics
\cite{CCC} as well as statistical mechanics and thermodynamics
\cite{SM, JWC}. As two examples of the initial and important
applications of these transformations in physics, the following
can be mentioned:

\vspace{0.3cm}
\begin{itemize}
 \item{
    In classical mechanics, where it yields the relation between the Hamiltonian $\mathcal{H}(p)$ and the Lagrangian $\mathcal{L}(\dot{q})$ to switch from Hamiltonian to Lagrangian dynamics, and conversely.}
 \item{In statistical mechanics where it provides connection between the internal energy and the various sorts of the thermodynamic potentials like enthalpy, Gibbs and Helmholtz free energies.}
\end{itemize}

\vspace{0.3cm} In a nutshell, a Legendre transformation aims to convert a function of one group of variables (such as velocity, pressure, or temperature) to another function of a {\it{conjugate}} set of variables (momentum, volume, and entropy, respectively). This transformation can be performed for a function with any number of
variables. For simplicity, we explain its concept for a two
variable function.\par

 Consider a function $f(x,y)$ with the differential
 \begin{equation}
    df=\frac{\partial f}{\partial x}\Big|_{y} dx+ \frac{\partial f}{\partial y}\Big|_{x} dy
 \end{equation}
  in which $x$ and $y$ are two independent variables.
Making use of the definition $\xi=\frac{\partial f}{\partial
x}\Big|_{y}$ and $\eta=\frac{\partial f}{\partial y}\Big|_{x}$,
the above relation can be written as follows
\begin{equation}
    df=\xi dx+\eta dy.
\end{equation}
Based on the aforementioned definitions, $x$ and $\xi$ as well as
$y$ and $\eta$ are conjugate pair of variables. Herein, the
Legendre transformation with respect to the variable $x$ is
defined as
\begin{equation}
    h=f-x\frac{\partial f}{\partial x}=f-\xi x,
\end{equation}
and hence
\begin{equation}
    dh=df-xd\xi-\xi dx=-xd\xi+\eta dy.
\end{equation}
As can be seen $h$ is a function of two independent variables
$\xi$ and $y$, i.e. $h=h(\xi,y)$. Moreover, we have
\begin{eqnarray}
    \frac{\partial h}{\partial \xi}\Big|_{y}=-x, \,\,\,\,\,\,\,\,\,\,\,\,\,\,\,\,\,\,\,\,\,\,\,\,\,\,\,\,\,\,\,\,\,\,\,\,   \frac{\partial h}{\partial y}\Big|_{\xi}=\eta.
\end{eqnarray}
Therefore, the Legendre transformation exchange the role of one
variable with its conjugate, along with a minus sign. Indeed,
for a function with two variables, there are 4 possible kinds
of  the transformed function. If instead we have 3 independent
variables, there are 8 different transformed functions. In
general, there are $2^n$  transformed function for a function of
$n$ independent variables, as each variable can be represented by
either member of a conjugate pair.\par

In geometric language, Legendre transformations are a special case
of contact transformations which leave invariant the contact
structure of $\cal T$. To explain this point in more detail, let us consider the class of maps with the feature of leaving the contact structure invariant. A transformation $f: \mathcal T \to \mathcal T$ is a
diffeomorphism of the contact manifold if $f$ maintains the
contact structure, i.e
\begin{equation}
    f^* (\theta) = \Omega\, \theta\,=\,\bar\theta\in [\,\theta\,]\,\,\quad\,\text{where}\,\quad\Omega\neq0.
\end{equation}
where $f^{\star}$ is the pullback of $f$. In this case, $f$ is
called a contact transformation or a {\it{contactomorphism}}
\cite{contactomorphism}. A Legendre transformation belongs to the
special class of contactomorphisms in which $f^* (\theta) = \theta$. Hence, it represents a
symmetry of the contact structure and is defined by the following
$2n + 1$ equations between the sets of coordinates
$\{\phi,x_a,y^a\}$ and $\{\bar\phi,\bar{x}_a,\bar{y}^a\}$,
\begin{equation} \label{legendretransf}
\bar\phi = \phi -  y^i x_i, \quad \bar{y}^i = -x_i, \quad
\bar{y}^j = y^j, \quad \bar{x}_i = y^i, \quad \bar{x}_j = x_j\,,
\end{equation}
where $i\cup j$ is any disjoint decomposition of the set of
indices $\{1, ..., n\}$. A direct calculation shows that
\begin{equation} \label{etacov}
\theta= d \phi - x_a \ d y^a = d \bar\phi - \bar{x}_a d
\bar{y}^a\,.
\end{equation}
It is notable that since this transformation only exchanges the
$i$th pair of coordinates, it is called a {\it{partial Legendre
transformation}}. However, the transformation with the property of exchanging each pair of coordinates is called the {\it{total Legendre transformation}}.\par

In the context of the thermodynamics, the importance of the
Legendre transformation stems from the fact that  it changes the
dependence of the energy  function from an extensive variable to
its conjugate intensive variable, which can usually be controlled
more easily in a physical experiment. In equilibrium
thermodynamics, a system with $n$ degrees of freedom is fully
described by $n$ extensive variables $E_i$ along with their
corresponding conjugate intensive variables $I_i$ and a
thermodynamic potential $\Phi$ which relates them to each other.
Taking into account what was mentioned before, the Legendre
transformations, which correspond to a redefinition of the
thermodynamic potential, can be expressed as follows
\begin{eqnarray}
\tilde{\Phi}_{(i)}&\equiv& \Phi_{(i)}-I_{(i)}E^{(i)}\,\,\,\,\,\,\text{(No sum over $i$)}\label{Legendre1}\\
\tilde{I}_{(i)}&\equiv& E^{(i)}\label{Legendre2}\\
\tilde{E}^{(i)}&\equiv& - I_{(i)}\label{Legendre3}
\end{eqnarray}
where $I_{i}=\frac{\partial \Phi}{\partial E^{i}}$. Besides, for $j\neq i$, we have $\tilde{I}_{(j)}=I_{(j)}$ and $\tilde{E}^{(j)}=E^{(j)}$.



\begin{thebibliography}{99}

    \bibitem{frankel}
     T. Frankel, {\em The Geometry of Physics: An Introduction} (Cambridge University Press, Cambridge, UK, 1997).
\bibitem{ym53}
 C. N. Yang and R. L.  Mills, Phys. Rev. {\bf 96}, 191 (1954).
\bibitem{Ruppeiner1995}
G. Ruppeiner,
  \textit{Rev. Mod. Phys.} \textbf{67}, 605 (1995); {\bf 68}, 313(E) (1996).
\bibitem{Rup91}
 G. Ruppeiner, Phys. Rev. A. {\bf 44}, 3583 (1991).
\bibitem{Rup2}
G. Ruppeiner,
 American Journal of Physics
\textbf{78,} 1170 (2010).

\bibitem{Rup3}
 G. Ruppeiner,
 Phys. Rev. E\textbf{\ 86,} 021130
(2012).

\bibitem{Rup4}
H. O. May, P. Mausbach and G. Ruppeiner,
    Phys. Rev. E\textbf{\ 88,} 032123 (2013).
\bibitem{Ruppeiner:2011gm}
G.~Ruppeiner, A.~Sahay, T.~Sarkar and G.~Sengupta,
Phys. Rev. E \textbf{86}, 052103 (2012).
\bibitem{Rup2015}
G. Ruppeiner, M. Mausbach, and H. O. May, Phys. Lett. A
{\bf{379}}, 646 (2015).


\bibitem{gibbs}
 J. Gibbs, {\it The collected works}, Vol. 1, Thermodynamics (Yale
University Press, 1948).
\bibitem{car}
C. Charatheodory,
 Gesammelte Mathematische Werke, Band 2 (Munich, 1995).
\bibitem{Rao}
C. R. Rao, {{Bull. Calcutta Math. Soc.}}, {\bf{37}} (1945).
\bibitem{amari85}
S. Amari, {\it Differential-Geometrical Methods in Statistics}
(Springer-Verlag, Berlin, 1985).
\bibitem{Wein1975}
F. Weinhold, J. Chem. Phys. {\bf{63}}, 2479, 2484, 2488, 2496
(1975).
\bibitem{Wein75}
F. Weinhold, J. Chem. Phys. {\bf{65}}, p. 558 (1975).
\bibitem{Rupp79}
G. Ruppeiner, Phys. Rev. A {\bf{20}}, 1608 (1979).
\bibitem{nul85}
J. Nulton and P. Salamon, {\it Geometry of the  ideal gas}, Phys.
Rev. A {\bf 31},  2520 (1985).
\bibitem{san04}
 M. Santoro,
 J. Chem. Phys. {\bf 121},  2932 (2004).
\bibitem{san05a}
 M. Santoro,
Chem. Phys. {\bf 310},  269 (2005).
\bibitem{san05b}
 M. Santoro,
 Chem. Phys. {\bf 313},  331 (2005).
\bibitem{san05c} M. Santoro and Serge Preston, arXiv:math-ph/0505010
\bibitem{jjk03}
 D.A. Johnston, W. Janke, and R. Kenna,
 Acta Phys. Polon. B {\bf 34},  4923 (2003).
\bibitem{jan04}
 W. Janke, D.A. Johnston, and R. Kenna,
 Physica A {\bf 336},  181 (2004).




\bibitem{Hermann}
R. Hermann, {\it{Geometry, physics and systems}} (Marcel Dekker,
New York, 1973)
\bibitem{Mrugala1}
R. Mrugala,
 Rep. Math. Phys. {\bf{14}}, 419 (1978).
\bibitem{Mrugala2}
R. Mrugala,
 Rep. Math. Phys. {\bf{21}}, 197 (1985).
\bibitem{Arnold}
V. I. Arnold, {\it{Mathematical Methods of Classical Mechanics}}
(Springer, New York, 1980).
\bibitem{Quevedo}
H.~Quevedo,
J. Math. Phys. \textbf{48}, 013506 (2007).
\bibitem{RT1}
R. Mrugala, J.D. Nulton, J.C. Schon, P. Salmon, Rep. Math. Phys. \textbf{29}, 109 (1991).
\bibitem{RT3}
Hernandez, G.; Lacomba, E.A.,
J. Differ. Geom. Appl. {\bf{8}}, 205 (1998).
\bibitem{RT2}
R. Mrugala, 
  Rep. Math. Phys. {\bf{46}}, 461 (2000).
\bibitem{CM1}
M. De Leon, C. Sardon, {\it   A geometric approach to solve time
dependent and dissipative Hamiltonian systems},
arXiv:1607.01239.
\bibitem{CM2}
 A. Bravetti,  H. Cruz, D. Tapias,
   Ann. Phys. {\bf{376}}, 17 (2017).
\bibitem{SM1}
A. Bravetti, D. Tapias,
    Phys. Rev. E  {\bf{93}}, 22139 (2016).

\bibitem{Ghosh:2019100}
A. Ghosh, C. Bhamidipati, Phys. Rev. D \textbf{100}, 126020 (2019).

\bibitem{Rupp05}
G. Ruppeiner, Phys. Rev. E {\bf{72}}, 016120 (2005).


\bibitem{Bardeen}
J. M. Bardeen, B. Carter, and S. W. Hawking, Math. Phys. vol. \textbf{31}, 170 (1973).
 \bibitem{Bekenstein:1973}
 J.~D. Bekenstein,
 Phys. Rev. D
 {\bf 7}, 2333 (1973).
 \bibitem{Hawking:1975vcx}
 S.~W. Hawking,
 {Commun. Math. Phys.
    {\bf43}, 199 (1975)}. [Erratum: Commun.Math.Phys. 46, 206 (1976)]
 \bibitem{FN}
 V. Frolov and I. Novikov,
 \textit{Black hole physics: basic concepts and new developments}, Fundam. Theor. Phys. \textbf{96} (1998).
 \bibitem{Wald-book}
  R.M. Wald, General Relativity, University of Chicago Press (1984).
\bibitem{Davies:1978zz}
P.~C.~W.~Davies,
Rept. Prog. Phys. \textbf{41}, 1313 (1978).

 \bibitem{Hut:1977zx}
 P. Hut, Monthly Notices of the Royal Astronomical Society, 180, Issue \textbf{3}, 379 (1977).

\bibitem{Sokolowski:1980uva}
L.~M.~Sokolowski and P.~Mazur,
J. Phys. \textbf{13}, A1113-1120 (1980).
\bibitem{Cai:1997cs}
R.~G.~Cai,
Phys. Rev. Lett. \textbf{78}, 2531 (1997).
\bibitem{Shen:2005nu}
J.~y.~Shen, R.~G.~Cai, B.~Wang and R.~K.~Su,
Int. J. Mod. Phys. A \textbf{22}, 11 (2007).
\bibitem{David Kubiznak12}
 D. Kubiznak, R. B. Mann,
     JHEP \textbf{07}, 033 (2012).
\bibitem{Rong-Gen Cai13}
 R. G. Cai, L. M. Cao, L. Li, R. Q. Yang,
   JHEP \textbf{09}, 005 (2013).
\bibitem{Sharmila Gunasekaran12}
 S. Gunasekaran, D. Kubiznak, R. B. Mann,
  JHEP \textbf{11}, 110 (2012).
\bibitem{Antonia M. Frassino14}
 A. M. Frassino, D. Kubiznak, R. B. Mann, F. Simovic,
 JHEP \textbf{09}, 080 (2014).
\bibitem{Natacha Altamirano14}
 N. Altamirano, D. Kubiznak, R. B. Mann, Z. Sherkatghanad,
 Class. Quant. Grav. \textbf{31}, 042001 (2014).
\bibitem{Natacha Altamirano13}
 N. Altamirano, D. Kubiznak, R. B. Mann,
  Phys. Rev. D  \textbf{88}, 101502 (2013).
\bibitem{Joy Das Bairagya20}
 J. D. Bairagya, K. Pal, K. Pal, T. Sarkar,
   Physics Letters B \textbf{805}, 135416 (2020).
\bibitem{X. H. Ge15}
X. H. Ge, Y. Ling, C. Niu, S. J. Sin,
 Phys. Rev. D \textbf{92}, 106005 (2015).
\bibitem{Rong-Gen Cai16}
 R. G. Cai, S. M. Ruan, S. J. Wang, R. Q. Yang, R. H. Peng,
  JHEP \textbf{09}, 161 (2016).
\bibitem{Jia-Lin Zhang15b}
J. L. Zhang, R. G. Cai, H. W. Yu,
  Phys. Rev. D \textbf{91}, 044028 (2015).
\bibitem{Ren Zhao13}
 R. Zhao, H. H. Zhao, M. S. Ma, L. C. Zhang,
   Eur. Phys. J. C \textbf{73}, 2645 (2013).
\bibitem{Ren Zhao13b}
R. Zhao, M. S. Ma, H. F. Li, L. C. Zhang,
 Advances in High Energy Physics \textbf{1155}, 371084 (2013).
\bibitem{Shao-Wen Wei09}
 S. W. Wei, Y. X. Liu, Y. Q. Wang,
  \emph{Dynamic properties of thermodynamic phase transition for five-dimensional neutral Gauss-Bonnet AdS black hole on free energy landscape, arXiv:2009.05215}
\bibitem{S H Hendi17}
S. H. Hendi, R. B. Mann, S. Panahiyan, B. Eslam Panah,
  Phys. Rev. D  \textbf{95}, 021501 (2017).
\bibitem{A. Dehghani20}
 A. Dehghani, S. H. Hendi, R. B. Mann,
  Phys. Rev. D \textbf{101}, 084026 (2020).
\bibitem{Rabin Banerjee20}
 H. Ranjbari, M. Sadeghi, M. Ghanaatian, Gh. Forozani,
  Eur. Phys. J. C \textbf{80}, 17 (2020).
\bibitem{M. Chabab2019}
M. Chabab, H. El Moumni, S. Iraoui, K. Masmar,
 Eur. Phys. J. C \textbf{79}, 342 (2019).
\bibitem{M. Chabab2016}
 M. Chabab, H. El Moumni, S. Iraoui, K. Masmar,
  Eur. Phys. J. C \textbf{76}, 676 (2016).
\bibitem{Meng-Sen Ma17b}
 S. Z. Han, J. Jiang, M. Zhang,  W. B. Liu,
    Communications in Theoretical Physics \textbf{72}, 10 (2020).
\bibitem{Dehyadegari2020}
 A. Dehyadegari, A. Sheykhi,
  Phys. Rev. D \textbf{102}, 064021 (2020).
\bibitem{Daniela Mago2020}
 D. Mago, N. Breton,
   Phys. Rev. D \textbf{102}, 084011 (2020).
\bibitem{Sajadi2019}
 S. N. Sajadi, N. Riazi, S. H. Hendi,
 Eur. Phys. J. C \textbf{79}, 775 (2019).
\bibitem{Hendi2019}
S. H. Hendi, A. Dehghani,
  Eur. Phys. J. C \textbf{79}, 227 (2019).
\bibitem{Hendi2018}
 S. H. Hendi, Z. S. Taghadomi, C Corda,
  Phys. Rev. D \textbf{97}, 084039 (2018).
\bibitem{Eslam}
 S. H. Hendi, B. Eslam Panah, S. Panahiyan, M. S. Talezadeh,
   Eur. Phys. J. C \textbf{77}, 133 (2017).
\bibitem{Hendi2016a}
 S. H. Hendi, R. Moradi, Z. Armanfard, M. S. Talezadeh,
   Eur. Phys. J. C \textbf{76}, 263 (2016).
\bibitem{Hendi2016b}
 S. H. Hendi, S. Panahiyan, B. Eslam Panah,
  Eur. Phys. J. C  \textbf{76}, 396 (2016).
\bibitem{Zou17}
D. C. Zou, R. H. Yue, M. Zhang,
 Eur. Phys. J. C \textbf{77}, 256 (2017).
\bibitem{Jianfei15}
 J. F. Xu, L. M. Cao, Y. P. Hu,
  Phys. Rev. D \textbf{91}, 124033 (2015).
\bibitem{Mahmoudi:2022hqq}
S.~Mahmoudi, K.~Jafarzade and S.~H.~Hendi,
JHEP \textbf{12}, 009 (2022).
\bibitem{Hendi:2022opt}
S. H. Hendi,  S. Hajkhalili, S. Mahmoudi,
 Fortschritte der Physik, {\bf{2200101}} (2023).
\bibitem{Hendi:2021yii}
S.~H.~Hendi, S.~Hajkhalili, M.~Jamil and M.~Momennia,
Eur. Phys. J. C \textbf{81}, 1112 (2021).
\bibitem{Dehghani:2020kbn}
A.~Dehghani and S.~H.~Hendi,
Phys. Rev. D \textbf{104},  024025 (2021).
\bibitem{Hendi:2020mhg}
S.~H.~Hendi, F.~Azari, E.~Rahimi, M.~Elahi, Z.~Owjifard and
Z.~Armanfard,
Annalen Phys. \textbf{532}, 2000162 (2020).
\bibitem{Hendi:2020knv}
S.~H.~Hendi, S.~N.~Sajadi and M.~Khademi,
Phys. Rev. D \textbf{103}, 064016 (2021).
\bibitem{Yuichi-2006}
 Y. Sekiwa.
 Phys. Rev. D {\bf{73}}, 084009 (2006).
\bibitem{Miho-2009}
 M. Urano, A. Tomimatsu, and H. Saida.
  Class. Quant. Grav., {\bf{26}}, 105010 (2009).
\bibitem{Saoussen-2019}
 S. Mbarek, R. B. Mann,
 Reverse Hawking-Page Phase Transition in de Sitter Black Holes,
  JHEP {\bf{02}}, 103 (2019).
\bibitem{David-2016}
 D. Kubiznak, F. Simovic,
 transitions, Classical and Quantum Gravity {\bf{33}}, 245001 (2016).
\bibitem{Simovic-2019}
 F. Simovic, R. B. Mann,
  JHEP {\bf{05}}, 136 (2019).
\bibitem{Sumarna-2020}
S. Haroon, R. A. Hennigar, R. B. Mann, F. Simovic,
 Phys. Rev. D {\bf{101}}, 084051 (2020).
\bibitem{Simovic-202020}
 F. Simovic, D. Fusco, R. B. Mann,
  Thermodynamics of de Sitter Black Holes with Conformally Coupled Scalar Fields, arXiv:2008.07593 [gr-qc]
\bibitem{Chabab-202020}
 M. Chabab, H. El Moumni, J. Khalloufi,
  On Einstein-non linear-Maxwell-Yukawa de-Sitter black hole thermodynamics, arXiv:2001.01134 [hep-th]
\bibitem{Brian-2013}
B. P. Dolan, D. Kastor, D. Kubiznak, R. B. Mann, J. Traschen,
 Phys. Rev. D. {\bf{87}}. 104017 (2013).
\bibitem{Bhattacharya-2016}
S. Bhattacharya,
  Eur. Phys. J. C {\bf{76}}, 112 (2016).
\bibitem{James-2016}
 J. McInerney, G. Satishchandrana, J. Traschena,
   Class. Quantum Grav. {\bf{33}}, 105007 (2016).
\bibitem{Kanti-2017}
 P. Kanti, T. Pappas,
 Phys. Rev. D {\bf{96}}, 024038 (2017).
\bibitem{Romans-1992}
 L. J. Romans,
  Nucl.Phys. B {\bf{383}}, 395 (1992).
\bibitem{zhang-2016}
L. C. Zhang, R. Zhao, M. S. Ma,
 Phys. Lett. B {\bf{761}}, 74 (2016).
\bibitem{zhang-2019}
 L. C. Zhang, R. Zhao,
 Physics Lett. B {\bf{797}}, 134798 (2019).
\bibitem{ma-2020}
 Y. B. Ma, Y. Zhang, L. C. Zhang, L. Wu, Y. M. Huang, Y. Pan,
    Eur. Phys. J. C  {\bf{80}}, 213 (2020).
\bibitem{guoxiong-2020}
 X. Y. Guo, H. F. Li, L. C. Zhang, R.  Zhao,
 Commun. Theor. Phys. {\bf{72}}, 085403 (2020).
\bibitem {IN-Strominger:1996sh}
A.~Strominger and C.~Vafa,
Phys.\ Lett.\ B \textbf{379}, 99 (1996)
\bibitem {IN-Callan:1996dv}
C.~G.~Callan and J.~M.~Maldacena,
 Nucl.\ Phys.\ B \textbf{472}, 591
(1996)
\bibitem {IN-Emparan:2006it}
R.~Emparan and G.~T.~Horowitz,
\ Phys.\ Rev.\ Lett.\ \textbf{97}, 141601 (2006).
\bibitem{wei-2019}
 S. W. Wei, Y. X. Liu, R. B. Mann,
   Phys. Rev. Lett. {\bf{123}}, 071103 (2019).
\bibitem{wei-2020}
 S. W. Wei, Y. X. Liu,
  Phys. Rev. D {\bf{101}}, 104018 (2020).
\bibitem{zou-2020}
R. Zhou, Y. X. Liu, S. W. Wei,
 Phys. Rev. D {\bf{102}}, 124015 (2020).
\bibitem{wei-202011}
 S. W. Wei, Y. X. Liu,
 hole, Phys. Lett. B {\bf{803}}, 135287 (2020).
\bibitem{miao-2018}
Y. G. Miao and Z. M. Xu,
 Phys. Rev. D {\bf{98}}, 044001 (2018).
\bibitem{miao-2019}
 Y. G. Miao and Z. M. Xu,
   Nucl. Phys. B {\bf{942}}, 205 (2019).
\bibitem{guo-2020}
X. Y. Guo, H. F. Li, L. C. Zhang, R. Zhao,
  Eur. Phys. J. C  {\bf{80}}, 168 (2020).
\bibitem{guo-2019}
 X. Y. Guo, H. F. Li, L. C. Zhang, R. Zhao,
  Phys. Rev. D {\bf{100}}, 064036 (2019).
\bibitem{mann-2021}
 G. A. Marks, F. Simovic, R. B. Mann,
 Phase Transitions in 4D Gauss-Bonnet-de Sitter Black Holes. arXiv: 2107.11352.
\bibitem{Volovik-2021}
G. E. Volovik, Effect of the inner horizon on the black hole
thermodynamics: Reissner-Nordstr\"om black hole and Kerr black
hole, arXiv: 2107.11193.
\bibitem{Mishin}
 Y. Mishin,
  Annals of Physics {\bf{363}}, 48 (2015).



\bibitem{Ruppeiner:2013yca}
G.~Ruppeiner,
Springer Proc. Phys. \textbf{153}, 179 (2014).

\bibitem{Landau}
 L. D. Landau and E. M. Lifshitz, {\it Statistical Physics} (Pergamon, New York, 1977).
\bibitem{Pathria}
R. K. Pathria, {\it Statistical Mechanics} (Butterworth-Heinemann,
Oxford, 1996).
\bibitem{Graham}
R. Graham,
Z. Phys. B {\bf 26}, 397 (1977).
\bibitem{Call}
H. B. Callen,
 {\it Thermodynamics and an Introduction to Thermostatistics} (John Wiley \& Sons, New York, 1985).

\bibitem{conformal}
P. Salamon, J. D. Nulton, and E. Ihrig., J. Chem. Phys. {\bf 80},
436 (1984).
\bibitem{Liu:2010sz}
H.~Liu, H.~Lu, M.~Luo and K.~N.~Shao,
JHEP \textbf{12}, 054 (2010).
\bibitem{Salamon;1980}
P. Salamon, R. S. Berry and B. Andresen, J. Chem. Phys. \textbf{73}, 1001
(1980).




\bibitem{Gilmore;1981}
R. Gilmore, J. Chem. Phys. \textbf{75}, 5964 (1981).

\bibitem{Feldman;1985}
T. Feldman, B. Andersen, A. Qi and P. Salamon, Chem. Phys. \textbf{83},
5849 (1985).





\bibitem{Torres;1993}
 G. F. Torres del Castillo and M. Montesinos-Velasquez, Rev. Mex. FÃ­s. \textbf{39}, 194 (1993).



\bibitem{condensed1}
F. Weinhold,
  J. Chem. Phys. , {\bf{63}}, 6, (1975)
  \bibitem{condensed2}
  F. Weinhold,
   J. Chem. Phys. {\bf{65}}, 2, (1976).
\bibitem{condensed3}
H. Janyszek and R. Mrugala,
Phys. Rev. A {\bf{39}}, 6515, (1989).
\bibitem{condensed5}
B.~P.~Dolan,
Proc. Roy. Soc. Lond. A \textbf{454}, 2655 (1998).
\bibitem{condensed6}
B.~P.~Dolan, D.~A.~Johnston and R.~Kenna,
J. Phys. A \textbf{35}, 9025 (2002).
\bibitem{condensed7}
W.~Janke, D.~A.~Johnston and R.~Kenna,
Phys. Rev. E \textbf{67}, 046106 (2003).

\bibitem{condensed8}
H.~Quevedo and S.~A.~Zaldivar, ``A geometrothermodynamic approach
to ideal quantum gases and Bose-Einstein condensates,''
[arXiv:1512.08755 [gr-qc]].
\bibitem{condensed9}
H.~Quevedo, F.~Nettel, C.~S.~Lopez-Monsalvo and A.~Bravetti,
J. Geom. Phys. \textbf{81}, 1-9 (2014).
\bibitem{condensed10}
H.~Quevedo, A.~Sanchez, S.~Taj and A.~Vazquez,
Gen. Rel. Grav. \textbf{43}, 1153 (2011).

\bibitem{Quevedo;2008}
H. Quevedo, A. Vazquez, AIP Conf. Proc. \textbf{977},165,   (2008).

\bibitem{Mirza:059}
B. Mirza and M. Zamaninasab, JHEP \textbf{0706}, 059 (2007).

\bibitem{Medved:2149}
A. J. M. Medved, Mod. Phys. Lett. A \textbf{23}, 2149 (2008).

\bibitem{Aman:2003}
J. E. Aman, I. Bengtsson, and N. Pidokrajt, Gen. Relativ. Gravit.
\textbf{35}, 1733 (2003).

\bibitem{Quevedo;2008cd}
H.Quevedo, Gen. Rel. Grav. \textbf{40}, 971 (2008).


\bibitem{Belgiorno:2002iw}
F.~Belgiorno,
J. Math. Phys. \textbf{44}, 1089 (2003)

\bibitem{Qeovedo1}
H. Quevedo, M. N. Quevedo, and A. Sanchez,  Eur. Phys. J. C
{\bf{79}}, 1 (2018).

\bibitem{Burke;1987}
W. L. Burke, Applied differential geometry, Cambridge University
Press, Cambridge, UK, 1987.



\bibitem{Laugwitz1965}
 D. Laugwitz: \textit{Differential and Riemannian Geometry} (Academic, New York 1965)
 
 
 \bibitem{Wei:2019yvs}
 S.~W.~Wei, Y.~X.~Liu and R.~B.~Mann,
 Phys. Rev. D \textbf{100}, 124033 (2019).

\bibitem{Oshima}
H. Oshima, T. Obata, and H. Hara,
J. Phys. A: Math. Gen. \textbf{3}, 6373 (1999).

\bibitem{Quevedo:2012jg}
H.~Quevedo and A.~Ramirez,
[arXiv:1205.3544 [math-ph]].

\bibitem{Rupp81} G. Ruppeiner,
  Phys. Rev. A {\bf 24}, 488 (1981).

\bibitem{Rupp90B} G. Ruppeiner and J. Chance,
  J. Chem. Phys. {\bf 92}, 3700 (1990).

\bibitem{Brody95} D. Brody and N. Rivier,
 Phys. Rev. E {\bf 51}, 1006(1995).



\bibitem{Jan02} W. Janke, D. A. Johnston, and R. P. K. C. Malmini,
 Phys. Rev. E {\bf 66}, 056119 (2002).


\bibitem{Brody03} D. C. Brody and A. Ritz,
 J. Geom. Phys. {\bf 47}, 207 (2003).




\bibitem{Weinberg1972}
S. Weinberg: \textit{Gravitation and Cosmology} (Wiley, New York
1972)




\bibitem{Mirza1}
B. Mirza and  Z. Talaei,
   Phys. Lett. A {\bf{377}} 513 (2013).

 \bibitem{Mirza2}
   B. Mirza and H. Mohammadzadeh,
    J. Phys. A: Math. Theor. {\bf{44}}  475003 (2011).

   \bibitem{Mirza3}
   B. Mirza and H. Mohammadzadeh,
     Phys Rev E. {\bf{84}}. 031114 (2010).

   \bibitem{Mirza4}
   B. Mirza and H. Mohammadzadeh,
    Phys Rev E. {\bf{82}}, 031137(2010).

   \bibitem{Mirza5}
   B. Mirza and H. Mohammadzadeh, Phys. Rev. E {\bf{80}}, 011132 (2009).

   \bibitem{Mirza6}
   B. Mirza and H. Mohammadzadeh, Phys. Rev. E {\bf{78}}, 021127 (2008).

\bibitem{Halperin;1984} 
B. I. Halperin, Phys. Rev. Lett. \textbf{52}, 1583 (1984).

\bibitem{Misner:1973}
C. W. Misner, K. S. Thorne, J. A. Wheeler, Gravitation (Freeman,
San Francisco 1973).

\bibitem{Lewis:1931}
G. N. Lewis,
  J. Am.
Chem. Soc. \textbf{53}, 2578 (1931).

\bibitem{Adamson:edu}
R. B. Mann, Black Holes: Thermodynamics, Information, and Firewalls, Springer (2015).

\bibitem{Kittel:1980}
C. Kittel, and H. Kroemer, Thermal Physics, 2nd ed. New York: W. H. Freeman, 1980.


\bibitem{Hawking;1972hk}
S. W. Hawking, Phys. Rev. Lett. \textbf{26}, 1344 (1971); Comm. Math. Phys.
\textbf{25}, 152 (1972).


\bibitem{Jacobson:2mn}
T. Jacobson, Introductory Lectures on Black Hole Thermodynamics,
Institute for Theoretical Physics University of Utrecht.

\bibitem{Bardeen:1973abc}
J. Bardeen, B. Carter, and S. W. Hawking, "The four laws of black hole
mechanics", Commun. Math. Phys. \textbf{31}, 161 (1973).

\bibitem{Gibbons:1977ab}
G.W. Gibbons, S.W. Hawking, Phys. Rev. D \textbf{15}, 2752 (1977).

\bibitem{Bekenstein:92111}
J. D. Bekenstein, Prodeedings of the Eight Marcel Grossmann Meeting, T. Piran and R. Ruffini, eds. (World Scientific Singapore 1999), pp. 92-111.

\bibitem{Hod:1998}
S. Hod, Phys. Rev. Lett. \textbf{81}, 4293 (1998).

\bibitem{Hod:1999}
S. Hod, Gen. Rel. Grav. \textbf{31}, 1639 (1999, Fifth Award at Gravity Research
Foundation).

\bibitem{Bekenstein:124052}
J. D. Bekenstein, Phys. Rev. D \textbf{91}, 124052 (2015).

\bibitem{Sakalli:2015}
I. Sakalli, Eur. Phys. J. C \textbf{75}, 144 (2015).

\bibitem{Maggiore:2008}
M. Maggiore, Phys. Rev. Lett. \textbf{100}, 141301 (2008).


\bibitem{Bekenstein:3292}
J. D. Bekenstein, Phys. Rev. D \textbf{9}, 3292 (1974).

\bibitem{Page;2032005}
Don N. Page, New J. Phys.\textbf{7}, 203 (2005).

\bibitem{Wald;62001}
R. M. Wald, Living Rev.Rel. \textbf{4}, 6 (2001).

\bibitem{Heusler:1996}
M. Heusler, Black Hole Uniqueness Theorems, Cambridge University
Press (Cambridge, 1996).

\bibitem{Carter:1973a}
B. Carter, "Black Hole Equilibrium States" in Black Holes, ed. by C. DeWitt and B.S. DeWitt, 57-214, Gordon and Breach (New York, 1973).







\bibitem{Ohanian:923}
R. M. Wald and H. C. Ohanian, American Journal of Physics,  \textbf{53}, 923 (1985).

\bibitem{Bardeen:170161}
J. M. Bardeen, B. Carter, and S. W. Hawking, \textbf{170}, 161 (1973).

\bibitem{Israel:1971nm}
W. Israel, Phys. Rev. Lett. \textbf{57}, 397 (1971).



\bibitem{Landsberg:1980a}
P. T. Landsberg, D. Tranah: Thermodynamics of non-extensive
entropies. I. Collective Phenomena 3, 73 (1980).

\bibitem{Landsberg:1980b}
D. Tranah, P. T. Landsberg: Thermodynamics of non-extensive
entropies. II. Collective Phenomena 3, 81 (1980).

\bibitem{Banerjee:2011}
R. Banerjee, S. Kumar Modak, S. Samanta, Phys.Rev.D \textbf{84}, 064024 (2011).

\bibitem{Ferrara:1997}
S. Ferrara, G. W. Gibbons, and R. Kallosh, Nuc. Phys. B \textbf{500}, 75
(1997).


\bibitem{Cai:1999}
R. G. Cai and J. H. Cho, Phys. Rev. D \textbf{60}, 067502 (1999).



\bibitem{Arcioni:2005}
G. Arcioni and E. Lozano-Tellechea, Phys. Rev. D \textbf{72}, 104021
(2005).

\bibitem{Aman:2006}
J. E. Aman and N. Pidokrajt, Phys. Rev. D \textbf{73}, 024017 (2006).
\bibitem{Sarkar:2006}
T. Sarkar, G. Sengupta, and B. N. Tiwari, J. High Energy Phys. \textbf{11},
015 (2006).


\bibitem{Davies;1977}
P. C. W. Davies, Proc. R. Soc. Lond. A \textbf{353}, 499 (1977).

\bibitem{Preskill;1991}
J. Preskill, P. Schwarz, A. D. Shapere, S. Trivedi and F. Wilczek,
Mod. Phys. Lett. A \textbf{6}, 2353 (1991).


\bibitem{Kallosh;1992}
R. Kallosh, A. D. Linde, T. Ortin, A. W. Peet and A. Van Proeyen,
Phys. Rev. D \textbf{46}, 5278 (1992).































\bibitem{Ghosh:2020101}
A. Ghosh, C. Bhamidipati, Phys. Rev. D \textbf{101}, 046005 (2020).
\bibitem{Ghosh:2020106}
A. Ghosh, C. Bhamidipati, Phys. Rev. D \textbf{101}, 106007 (2020).
\bibitem{Ghosh:2302}
A. Ghosh, C. Bhamidipati, Front. Phys. \textbf{11}, 1132712 (2023).




\bibitem{Brody:245}
D. C. Brody and L. P. Hughston, Phys. Lett. A \textbf{245}, 73
(1998).

\bibitem{Susskind:1231992}
 L. Susskind and L. Thorlacius, Nucl. Phys. B \textbf{382}, 123 (1992); T. Banks, A. Dabholkar, M.R. Douglas, and M. OLoughlin,
Phys. Rev. D \textbf{45}, 3607 (1992); T. Banks, M. OLoughlin and A. Strominger, Phys. Rev. D \textbf{47}, 4476 (1993).

\bibitem{Bekenstein:1981}
 J. D. Bekenstein, Phys. Rev. D \textbf{23}, 287 (1981).
\bibitem{Bekenstein:19941912}
J. D. Bekenstein, Phys.Rev. D \textbf{49}, 1912 (1994). 

\bibitem{Report} P. Chen, Y. C. Ong, D.-h. Yeom, Phys. Report \textbf{603},
1 (2015).

\bibitem{Banerjee:2017}
R. Banerjee, B. Ranjan Majhi, S. Samanta,   Phys. Lett. B \textbf{767}, 25 (2017).

\bibitem{Zhang:2018}
M. Zhang, X-Y. Wang, W-B. Liu, Phys. Lett. B \textbf{783}, 169 (2018).

\bibitem{Azreg:2014}
M. Azreg-Aïnou, Eur. Phys. J. C \textbf{74}, 2930 (2014).

\bibitem{Hendi:507}
 S. H. Hendi, S. Panahiyan, B. Elam Panah, M. Momennia, Eur. Phys. J. C \textbf{75} (2015) 507.
\bibitem{Hendi:528}
S. H. Hendi, S. Panahiyan, B. Eslam Panah, M. Momennia, Ann. Phys.
(Berlin) \textbf{528}, 819 (2016).

\bibitem{Hendi:129}
S. H. Hendi, S. Panahiyan, B. Eslam Panah,  JHEP \textbf{01} 129
(2016).

\bibitem{Hendi:296}
S. H. Hendi, Mir Faizal, B. Eslam Panah, S. Panahiyan, Eur. Phys.
J. C \textbf{76}, 296 (2016).

\bibitem{Hendi:157}
S. H. Hendi, B. Eslam Panah, S. Panahiyan,  JHEP \textbf{11}, 157
(2015).

\bibitem{Hendi:64028}
S. H. Hendi, A. Sheykhi, S. Panahiyan, B. Eslam Panah,  Phys. Rev.
D \textbf{92}, 064028 (2015).

\bibitem{Sheykhi:124054}
A. Sheykhi, F. Naeimipour, S. M. Zebarjad, Phys. Rev. D
\textbf{92}, 124054 (2015).

\bibitem{Hendi:769}
S. H. Hendi, B. Eslam Panah, S. Panahiyan, Phys. Lett. B
\textbf{769}, 191 (2017).

\bibitem{wen:2017}
W-Y. Wen, Int. J. Mod. Phys. D \textbf{10}, 1750106 (2017).

\bibitem{Hendi:2016}
S. H. Hendi, B. Eslam Panah, S. Panahiyan,  JHEP \textbf{05}, 029
(2016).

\bibitem{Sheykhi:1650062}
A. Sheykhi, S. Hajkhalili, Int. J. Mod. Phys. D \textbf{25},
1650062  (2016).

\bibitem{Sheykhi:4896}
A. Sheykhi, F. Naeimipour, S.M. Zebarjad, Gen. Rel. Grav.
\textbf{48}, 96  (2016).

\bibitem{Hendi:5712016}
S. H. Hendi, G-Q. Li, J-X. Mo, S. Panahiyan, B. Eslam Panah, Eur.
Phys. J. C \textbf{76}, 571 (2016).

\bibitem{Sheykhi:1045}
A. Sheykhi, S. H. Hendi, F. Naeimipour, S. Panahiyan, B. Eslam
Panah, Can. J. Phys. \textbf{94}, 1045 (2016).

\bibitem{Hendi:1750026}
S. H. Hendi, S. Panahiyan, M. Momennia, B. Eslam Panah, Int. J.
Mod. Phys. D \textbf{26}, 1750026 (2016).


\bibitem{Hendi:767214}
S. H. Hendi, B. Eslam Panah, S. Panahiyan, A. Sheykhi, Phys. Lett.
B \textbf{767}, 214 (2017).

\bibitem{Jawad:2023}
A. Jawad, M. Hussain, S. Rani, Universe \textbf{9}, 87 (2023).
\bibitem{mirza1a}
S. A. Hosseini Mansoori, B. Mirza,  Eur. Phys. J. C \textbf{74}
2681(2014).
\bibitem{mirza2a}
S. A. Hosseini Mansoori, B. Mirza, M. Fazel, JHEP \textbf{04}, 115
(2015).

\bibitem{mirza3a}
S. A. Hosseini Mansoori, B. Mirza, E. Sharifian, Phys. Lett. B
\textbf{759}, 298  (2016).


\bibitem{mirza4a}
S. A. Hosseini Mansoori, B. Mirza, Phys. Lett. B \textbf{799}, 135040 (2019).

\bibitem{mirza5a}
M. Ebrahimi Khuzani, B. Mirza, M. Tavakoli Kachi, Int. J. Mod. Phys. D \textbf{13}, 2250097 (2022).


\bibitem{Bhattacharya:2017}
K. Bhattacharya, B. R. Majhi, Phys. Rev. D \textbf{95}, 104024
(2017).

\bibitem{Jafarzade:2021}
Kh. Jafarzade, J. Sadeghi, B. Eslam Panah, S. H. Hendi, Annals of
Physics \textbf{432}, 168577 (2021).


\bibitem{Arfken}
 G. B. Arfken and H. J. Weber, {\it Mathematical Methods for Physicists} (Academic Press, New York, 2001).

\bibitem{handbook}
F. Dillen and L. Verstraelen, {\it Handbook of Differential
Geometry} (Elsevier B. V., Amsterdam, 2006).

\bibitem{CCC}
C.-C. Cheng, ``Maxwell's equations in dynamics,'' Am. J.
Phys. \textbf{66}, 622 (1966); A. L. Fetter and J. D. Walecka, \textit{%
    Theoretical Mechanics of Particles and Continua} (McGraw-Hill, New York,
1980).
\bibitem{SM}
 K. Huang, \textit{Statistical Mechanics} (John Wiley \& Sons,
1987), 2nd ed.; H. S. Robertson, \textit{Statistical
Thermophysics} (Prentice Hall, 1997).
\bibitem{JWC}
J. W. Cannon, ``Connecting thermodynamics to students' calculus,''
Am. J. Phys. \textbf{72}, 753 (2004).

\bibitem{contactomorphism}
A. Banyaga, {\it{The structure of classical diffeomorphism
groups}} (Kluwer Academic Publishers, Dordrecht, The Netherlands,
1997).

\end{thebibliography}
\end{document}